\documentclass[aps,prl,10pt,twocolumn,floatfix,superscriptaddress,longbibliography,nofootinbib]{revtex4-2}

\usepackage{graphicx}
\usepackage{amssymb,amsmath,amsfonts,gensymb}
\usepackage{bm,hyperref}
\usepackage{color}
\usepackage{ulem}
\usepackage{bbm}
\usepackage{dcolumn}
\usepackage{mathtools}
\usepackage{pifont}

\newcommand{\ket}[1]{\vert #1 \rangle}

\newcommand{\eV}{{\text{eV}}}

\newcommand{\bk}{\mathbf{k}}

\newcommand{\bd}{\mathbf{d}}
\newcommand{\btk}{\widetilde{\mathbf{k}}}
\newcommand{\btq}{\widetilde{\mathbf{q}}}
\newcommand{\br}{\mathbf{r}}

\newcommand{\I}{\uppercase\expandafter{\romannumeral 1}}
\newcommand{\II}{\uppercase\expandafter{\romannumeral 2}}
\newcommand{\III}{\uppercase\expandafter{\romannumeral 3}}
\newcommand{\IV}{\uppercase\expandafter{\romannumeral 4}}
\newcommand{\V}{\uppercase\expandafter{\romannumeral 5}}
\newcommand{\VI}{\uppercase\expandafter{\romannumeral 6}}
\newcommand{\VII}{\uppercase\expandafter{\romannumeral 7}}

\newcommand{\p}{\mathbf{p}}
\newcommand{\ta}{\tilde{\mathbf{a}}}

\newcommand{\bfk}{\mathbf{k}}
\newcommand{\G}{\mathbf{G}}

\newcommand{\B}{\mathbf{B}}

\def\k{\mathbf{k}}

\makeatletter
\def\@ssect@ltx#1#2#3#4#5#6[#7]#8{%
	\def\H@svsec{\phantomsection}%
	\@tempskipa #5\relax
	\@ifdim{\@tempskipa>\z@}{%
		\begingroup
		\interlinepenalty \@M
		#6{%
			\@ifundefined{@hangfroms@#1}{\@hang@froms}{\csname @hangfroms@#1\endcsname}%
			{\hskip#3\relax\H@svsec}{#8}%
		}%
		\@@par
		\endgroup
		\@ifundefined{#1smark}{\@gobble}{\csname #1smark\endcsname}{#7}%
	}{%
		\def\@svsechd{%
			#6{%
				\@ifundefined{@runin@tos@#1}{\@runin@tos}{\csname @runin@tos@#1\endcsname}%
				{\hskip#3\relax\H@svsec}{#8}%
			}%
			\@ifundefined{#1smark}{\@gobble}{\csname #1smark\endcsname}{#7}%
			\addcontentsline{toc}{#1}{\protect\numberline{}#8}%
		}%
	}%
	\@xsect{#5}%
}%
\renewcommand{\fnum@figure}{\textbf{Fig.\thefigure}}
\renewcommand{\fnum@table}{\textbf{Table \thetable}}
\makeatother

\hypersetup{hidelinks}

\begin{document}

% Main manuscript (revision colors removed).
\title{Transdimensional Chern insulators in rhombohedral multilayer graphene superlattices}
	\author{Min Li}
\affiliation{School of Physical Science and Technology, ShanghaiTech Laboratory for Topological Physics, State Key Laboratory of Quantum Functional Materials, ShanghaiTech University, Shanghai 201210, China}

\author{Xin Lu}
\affiliation{School of Physical Science and Technology, ShanghaiTech Laboratory for Topological Physics, State Key Laboratory of Quantum Functional Materials, ShanghaiTech University, Shanghai 201210, China}

\author{Lei Wang}
\affiliation{National Laboratory of Solid-State Microstructures, Collaborative Innovation Center of Advanced Microstructures, School of Physics, Nanjing University, Nanjing, China}

\author{Jianpeng Liu}
\email{liujp@shanghaitech.edu.cn}
\affiliation{School of Physical Science and Technology, ShanghaiTech Laboratory for Topological Physics, State Key Laboratory of Quantum Functional Materials, ShanghaiTech University, Shanghai 201210, China}
\affiliation{Liaoning Academy of Materials, Shenyang 110167, China}

	\bibliographystyle{apsrev4-2}

\begin{abstract}
Recent experiments on lightly electron-doped rhombohedral multilayer graphene (RMG) in the ``transdimensional'' regime revealed anomalous Hall hysteresis under both perpendicular and in-plane magnetic fields, pointing to the emergence of transdimensional orbital magnetism in an interaction-driven ``Fermi-lune'' metal. Here, we demonstrate that, by coupling to the moir\'e potential, this transdimensional orbital ferromagnetic metal can evolve into quantized Chern insulator phases in twisted RMG/hBN moir\'e superlattices. Using unrestricted Hartree--Fock calculations, we find that interaction-driven reconstruction isolates a spin-valley-polarized $|C|=1$ Chern band, leading to a ``transdimensional Chern insulator'' (TDCI) state exhibiting both in-plane and out-of-plane orbital magnetizations. A minimal two-band model is introduced to describe the unique properties of TDCI, which unveils the intimate connection between non-reciprocal velocity and in-plane orbital magnetization.
We further analyze the collective excitations of TDCIs, and identify soft charge and spin modes that may lead to density wave instabilities in certain cases. An in-plane magnetic field can harden the soft density-wave mode and may drive a transition from density-wave state to TDCI. Finally, when the $|C|=1$ band is fractionally filled, the many-body ground state may be a fractional Chern insulator with broken rotation and mirror symmetries. %
\end{abstract}

\maketitle

\begin{figure}[tbp]
    \centering
    \includegraphics[width=\columnwidth]{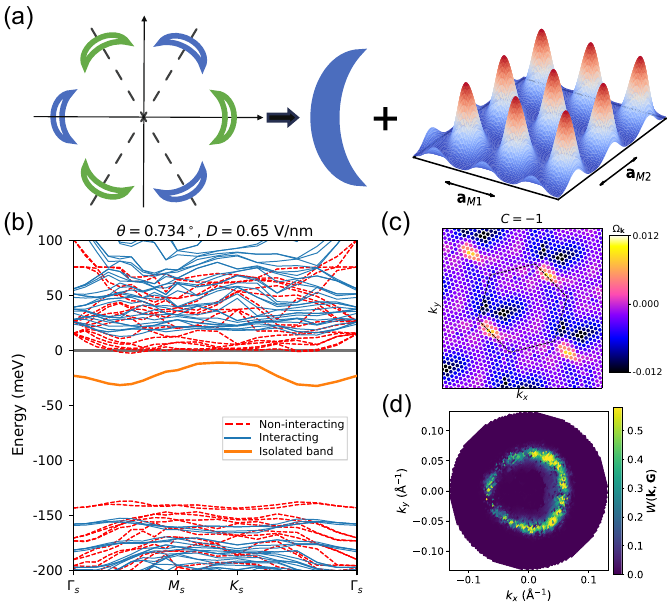}
    \caption{\textbf{From transdimensional orbital magnetism to a Chern insulating state.} (a) Interaction-driven Fermi-lune states in pristine RMG. The blue and green crescents denote the $K$ and $K'$ valleys. Within each valley, the three lune orientations are related by $C_3$ rotations, giving six symmetry-related Fermi-lune domains when both valleys are included. Interactions spontaneously select one domain, represented by the filled crescent. (b) Noninteracting bands, self-consistent Hartree--Fock bands, and the isolated occupied TDCI band of twisted octalayer RMG/hBN at $\theta=0.734^{\circ}$ and $D=0.65\,\mathrm{V/nm}$. (c) Berry-curvature distribution of the isolated TDCI band, showing strong anisotropy. (d) Unfolded spectral weight of the isolated band in the Brillouin zone of pristine RMG, revealing the Fermi-lune-like contour and broken $C_3$ symmetry.}
    \label{fig1}
\end{figure}

\textit{Introduction}
Rhombohedral multilayer graphene (RMG) has emerged as a versatile platform for correlated and topological phases in an electrically tunable two-dimensional electron system. Its chiral low-energy bands become progressively flatter with the increase of layer number and are strongly reshaped by a perpendicular displacement field~\cite{min-chiral-prb08,koshino-abc-prb09,zhang-abc-prb10,pierucci-flat-acsnano15}, enhancing electron-electron interaction effects and enabling symmetry-broken metals~\cite{zhou-halfmetal_R3G-nature-2021,liu-symbreakR4G-natnano-2024,arp-ivc-natphys24,huang-spin-orbital-prb23,kpgra}, superconductivity~\cite{zhou-superconductivity_R3G-nature-2021,han-chiral-sc-nature25,ghazaryan-annular-sc-prl21,qin-frg-sc-prl23,you-ivc-sc-prb22,patterson-spin-canting-nature25,yang-soc-sc-natmat25,seo-field-boosted-sc-nature26,okounkova-anomalous-metal-nature26,deng-hexalayer-sc-arxiv25}, Chern insulators~\cite{han-symbreakR5G-natnano-2024,winterer-ferroelectric-natphys24,zhang-spontaneous-qh-prl11}, and novel orbital magnetic order~\cite{ju-orbitalmulti-nature23,comment_exp,das-multiferroic-prb25} etc. Alignment with hBN introduces a long-wavelength moir\'e potential that defines commensurate miniband fillings~\cite{koshino-hBN-prb14,jung-hbn-graphene-prb14,jung-multilayer-arxiv23,bernevig-fci-plg-arxiv23}. Integer and fractional Chern insulator states and competing charge-ordered phases have consequently been observed across several RMG/hBN systems~\cite{fqah-ju-nature24,ju-eqah-nature25,aronson-fci-cdw-prx25,xie-tunable-fci-natmat25,zheng-switchable-r6g-prl25,liu-heptalayer-qah-prl26,guo-prb24,lixiao-fci-prb24,senthil-fci-prl24,ashvin-fci-prl24,zhou-moireless-prl24,tan-parent-berry-prx24,soejima-ahc-model-prb24,dong-ahc-stability-prb24}.

A distinct route to correlated topology starts from interaction-driven Fermi-surface reconstruction in pristine RMG. In the transdimensional regime where the sample thickness is comparable to electrons' vertical coherence length, RMG may enter a low-symmetry metallic state with a crescent-shaped Fermi contour under slight electron doping, termed ``Fermi lune''~\cite{fermilune}. This state spontaneously breaks time-reversal symmetry $\mathcal{T}$, threefold rotation $C_3$, and crystalline mirror symmetries. In addition to the spontaneous current loop within the two dimensional plane as in the case of a conventional orbital ferromagnet, this state also carries coherent interlayer current loops and thus generates both out-of-plane and in-plane orbital magnetizations. Consequently, its anomalous Hall response exhibits hysteresis under both perpendicular and in-plane magnetic fields ~\cite{comment_exp}, revealing a transdimensional orbital magnetic character. Unlike moir\'e Chern insulator phases, however, the Fermi-lune state remains metallic and its Hall response is not quantized.

This raises a central question: can the symmetry breaking and mixed orbital magnetization of the Fermi-lune metal be inherited by a quantized Chern insulating phase? A moir\'e superlattice provides the essential ingredients by introducing a new translational scale. A metallic Fermi lune state, after being folded to the superlattice Brillouin zone, may form isolated moir\'e Chern bands, leading to TDCI. We demonstrate this mechanism in realistic twisted RMG/hBN moir\'e superlattices, where interaction-driven reconstruction and moir\'e potential transform the symmetry-broken metallic state into isolated Chern bands carrying transdimensional orbital magnetization. Band unfolding reveals that such a Chern insulating state inherits the momentum-space anisotropy and reduced rotational symmetry of its Fermi-lune parent.  
The unique properties of TDCI can be captured by a minimal two-band model, which unveils the subtle connection between non-reciprocal velocity and in-plane orbital magnetization. Beyond the static mean-field treatment, we further investigate neutral collective excitations to distinguish collectively stable TDCIs from regimes with competing finite-momentum density-wave instabilities, and argue that an in-plane magnetic field may drive transitions between TDCI and density-wave instabilities. Finally, exact diagonalization calculations demonstrate a fractional Chern insulator obtained from fractional filling of the symmetry-breaking $|C|=1$ flat band.

\textit{Fermi-lune parent state}
At low electron densities, increasing the displacement field drives the noninteracting Fermi surface of pristine RMG from a circular pocket to an annulus. For the annulus Fermi surface, electrons are more spread across the entire sample rather than being exponentially localized at one surface. Coulomb interactions can then stabilize a crescent-shaped Fermi-lune metal, which exhibits spontaneous out-of-plane orbital current loops and in-plane orbital magnetizations \cite{fermilune}. For a fixed spin, the three $C_3$-related orientations and time-reversal operations give six symmetry-related magnetic domains. For a representative magnetic domain with in-plane magnetization pointing to the crystalline $y$ direction, the state breaks $\mathcal{T}$, $C_3$, and $\mathcal{M}_y$, while preserving $\mathcal{M}_y\mathcal{T}$, allowing orbital magnetization $\mathbf{M}=(0,M_y,M_z)$. This reduced symmetry and mixed orbital magnetization provide the defining fingerprints inherited by the insulating TDCI.

\textit{TDCI in RMG/hBN moir\'e superlattices}
\label{sec:moire}
We next investigate whether the symmetry-breaking orbital character of the Fermi-lune parent state can survive in a realistic moir\'e environment and become incorporated into a quantized Chern insulator phase. For twisted RMG/hBN, the noninteracting continuum Hamiltonian near valley $\mu=\pm1$ is
\begin{equation}
	H^{0,\mu}=H_{\mathrm{RMG}}^{0,\mu}+V_{\mathrm{hBN}},
\end{equation}
where $H_{\mathrm{RMG}}^{0,\mu}$ is the continuum model of RMG, and $V_{\mathrm{hBN}}$ represents the hBN moir\'e potential ~\cite{koshino-hBN-prb14,jung-hbn-graphene-prb14,jung-multilayer-arxiv23,bernevig-fci-plg-arxiv23}. Electron-electron interactions are described by the double-gate-screened Coulomb interaction
$V^{\rm dg}(\mathbf{q})=e^2\tanh(|\mathbf{q}|d_s)/(2\epsilon_{\rm BN}\epsilon_0|\mathbf{q}|)$, with $d_s=40\,\mathrm{nm}$ and relative background dielectric constant $\epsilon_{\rm BN}=4$. Details of the continuum model and Hartree--Fock implementation are provided in the Supplemental Material~\cite{tdci-supplemental}.

Figure~\ref{fig1}(b) shows the noninteracting (red dashed) and self-consistent Hartree--Fock (blue solid) single-particle energy spectra of twisted octalayer RMG/hBN at twist angle $\theta=0.734^{\circ}$ and displacement field $D=0.65\,\mathrm{V/nm}$. The hBN moir\'e potential folds the non-interacting conduction bands (with an annular Fermi surface with six lune sectors) into the superlattice Brillouin zone, generating a narrow conduction-band manifold; however, the noninteracting state remains metallic at filling 1 due to spin-valley degeneracy. Once Coulomb interactions are included, electrons spontaneously occupy one spin-valley flavor, opening a global gap and isolating an occupied Chern band with $|C|=1$.  Importantly, the resulting insulating state breaks rotation, mirror and time-reversal symmetry, exhibiting the mixed orbital current-loop characters inherited from the Fermi-lune parent state. In particular, the out-of-plane current loop gives rise to a sizable in-plane orbital magnetization of $M_y\sim2.5\mu_B$ per electron (see Supplemental Material~\cite{tdci-supplemental} for the calculation methods). The isolated band also exhibits an anisotropic Berry curvature distribution across the moir\'e Brillouin zone [Fig.~\ref{fig1}(c)]. To establish the connection between the insulating TDCI and its metallic parent state, we unfold the isolated band into the Brillouin zone of pristine RMG, clearly revealing a Fermi-lune-like contour breaking all $C_3$, $\mathcal{M}_y$ and $\mathcal{T}$ symmetries [Fig.~\ref{fig1}(d)]. For comparison, an ordinary Chern insulator without transdimensional orbital magnetization retains $C_3$ symmetry and exhibits a largely annular unfolded spectral-weight distribution [Fig.~S1(b) in the Supplemental Material~\cite{tdci-supplemental}]. These results motivate the definition of the TDCI as an interaction-induced Chern insulator that simultaneously carries transdimensional orbital magnetization, thus exhibiting magnetic hysteresis to both in-plane and out-of-plane magnetic fields. The persistence of mixed orbital magnetization establishes its transdimensional character.

\begin{figure}[tbp]
    \centering
    \includegraphics[width=\columnwidth]{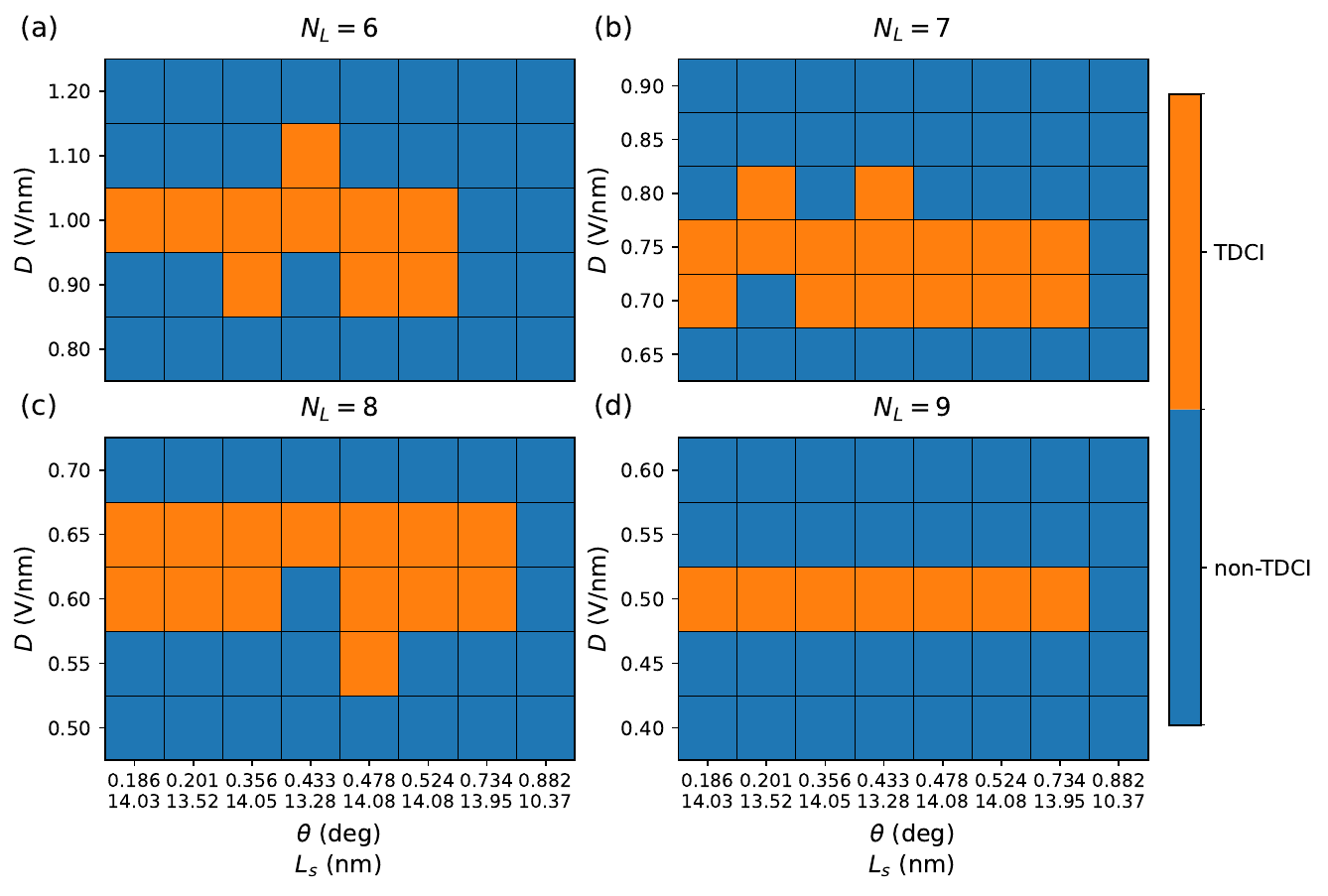}
    \caption{\textbf{Hartree–Fock phase diagrams of the TDCI in RMG/hBN moir\'e superlattices.} TDCI phases for $N_L=6$--$9$ as functions of displacement field $D$ and twist angle $\theta$. Orange cells denote TDCI states; blue cells denote other states, including band insulators, Chern insulators without transdimensional orbital magnetization, and metallic states. The TDCI regions cluster around a moir\'e period of approximately $14\,\mathrm{nm}$ and shift toward smaller $D$ with increasing layer number.}
    \label{fig_moire_phase}
\end{figure}
We further present the Hartree--Fock phase diagrams of TDCI state in RMG/hBN moir\'e superlattices in the parameter space of displacement field $D$ and twist angle $\theta$. Here, we highlight representative twist angles that host self-consistent TDCI mean-field solutions across multiple layer numbers $N_L=6$--$9$. Across the layer numbers considered, the TDCI regions
consistently cluster around a moir\'e period of $L_s\sim 13\rm{-}14\,\mathrm{nm}$, suggesting this length scale as a particularly favorable regime for realizing TDCI. We identify the TDCI phase through the self-consistent coexistence of a spin-valley-polarized insulating state, a nonzero Chern number, and in-plane orbital magnetization. Its emergence results from the interplay among the moir\'e potential, displacement field, carrier filling, and Coulomb interactions. With increasing layer number, the TDCI regions systematically shift toward smaller displacement fields. For $N_L<6$, the sample thickness becomes substantially smaller than vertical coherence length, and the TDCI is disfavored because such a state gains less exchange energy for smaller $N_L$ with weaker in-plane orbital magnetization.  The robustness of TDCI over multiple layer numbers and finite parameter windows establishes RMG/hBN moir\'e superlattices as a realistic and tunable platform for exploring transdimensional topology.

\textit{Two-band effective model}
To capture the essential orbital character of the TDCI, we introduce a minimal two-band Hamiltonian on a hexagonal moir\'e lattice for a fixed spin and valley $\mu=\pm1$,
\begin{equation}
\begin{aligned}
 h_{\mu i}(\bk)={}& t \hat{\mathbf n}_i\!\cdot\!\mathbf F(\bk)\,\sigma_0
 -\mu v F_x(\bk)\sigma_x\\
 &+v F_y(\bk)\sigma_y+m(\bk)\sigma_z .
\end{aligned}
\label{eq:tdci_two_band}
\end{equation}
Here $\sigma_0$ and $\sigma_{x,y,z}$ are the identity and Pauli matrices of two effective layer orbitals centered at heights $\pm\ell/2$; $v$ describes interorbital hybridization, $m(\bk)=m+r G(\bk)$ controls band inversion, and $t$ is the signed tilt amplitude. All four parameters $v,m,r, t$ have units of energy. The domain orientation is specified by $\hat{\mathbf n}_i=\hat{\mathbf M}_i\times\hat{\mathbf z}$, where $\hat{\mathbf M}_i$ is the unit vector along the in-plane orbital magnetization of domain $i=1,\ldots,6$, and $\hat{\mathbf z}$ is the layer normal. Thus, the first domain $\hat{\mathbf M}_1=\hat{\mathbf y}$ gives the energy tilt term $t F_x(\bk)$.  The other five magnetic domains are obtained by applying $C_3$, $\mathcal{T}$ and/or mirror operations to the first one. 
The dimensionless lattice functions are  $\mathbf F(\bk)=\frac{2}{3}\sum_{j=1}^{3} \frac{\boldsymbol\delta_j}{L_s}\sin(\bk\!\cdot\!\boldsymbol\delta_j)$, $ G(\bk)=\sum_{j=1}^{3}[1-\cos(\bk\!\cdot\!\boldsymbol\delta_j)]$,  with $\boldsymbol\delta_j=L_s(\cos\theta_j,\sin\theta_j)$, $\theta_j=2\pi(j-1)/3$, and $L_s$ the moir\'e period; $\bk$ lies in the moir\'e Brillouin zone. For $v\neq0$, $r>0$ and $-4r<m<0$, the lower band has $C=\mu$, provided the indirect gap remains open.

Selecting a valley and a tilt orientation breaks $\mathcal T$, $C_3$, and the valley-exchanging mirror symmetry. The $\hat{\mathbf M}_1=\hat{\mathbf y}$ domain preserves $\mathcal M_y\mathcal T$, with $\mathcal M_y:x\mapsto-x$, allowing both $M_y$ and $M_z$; time reversal sends $(\mu,\hat{\mathbf n}_i)$ to $(-\mu,-\hat{\mathbf n}_i)$. Writing the lower-band dispersion as $\varepsilon_{-,i}=t\hat{\mathbf n}_i\cdot\mathbf F-\mathcal E(\bk)$, with $\mathcal E(\bk)=\sqrt{v^2|\mathbf F(\bk)|^2+m(\bk)^2}$, its group velocity $\mathbf v_{-,i}=\hbar^{-1}\nabla_{\bk}\varepsilon_{-,i}$ has a non-reciprocal component
\begin{equation}
\begin{aligned}
 \delta\mathbf v_i(\bk)&\equiv
 \frac{\mathbf v_{-,i}(\bk)+\mathbf v_{-,i}(-\bk)}{2}\\
 &=\frac{t}{\hbar}\nabla_{\bk}
 [\hat{\mathbf n}_i\cdot\mathbf F(\bk)].
\end{aligned}
\label{eq:tdci_nonreciprocal_velocity}
\end{equation}
Using the layer-position operator $\hat z=\ell\sigma_z/2$ and straight-bond Peierls coupling, the in-plane orbital magnetization per unit area of the filled lower band is
\begin{equation}
 \mathbf M_{\parallel,i}
 =\frac{e\ell}{2}\int_{\mathrm{mBZ}}\frac{d^2k}{(2\pi)^2}
 \frac{m(\bk)}{\mathcal E(\bk)}\,
 \hat{\mathbf z}\times\delta\mathbf v_i(\bk),
\label{eq:tdci_magnetization_velocity}
\end{equation}
where $-e$ is the electron charge and $\mathrm{mBZ}$ denotes the moir\'e Brillouin zone. This expression connects the in-plane orbital moment to the non-reciprocal velocity weighted by the layer polarization $\langle\sigma_z\rangle_-=-m(\bk)/\mathcal E(\bk)$. Such layer-resolved counterflow can carry an in-plane orbital moment even though the net equilibrium current of the filled band vanishes. Equation~\eqref{eq:tdci_magnetization_velocity} links in-plane orbital magnetization to non-reciprocal group velocities. This connection provides insight into the coexistence of transdimensional anomalous Hall effect and non-reciprocal longitudinal transport in the parent Fermi-lune metal.
We emphasize that this simplified two-band model is intended to capture key qualitative features of the TDCI, rather than to quantitatively reproduce the Hartree--Fock states obtained for RMG/hBN moir\'e superlattices.

\begin{figure*}[tp]
    \centering
    \includegraphics[width=6.5in]{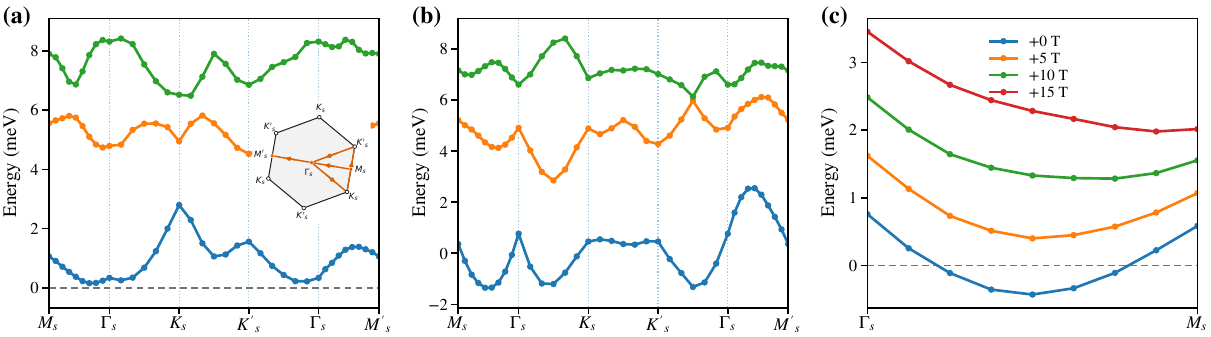}
    \caption{\textbf{Collective charge excitations of the TDCI.} (a) Stable heptalayer RMG/hBN at $\theta=0.604^{\circ}$ and $D=0.70\,\mathrm{V/nm}$, where the positive-norm charge branches remain positive. (b) An unstable TDCI at $\theta=0.356^{\circ}$ and $D=0.75\,\mathrm{V/nm}$, where the finite-momentum minimum of a positive-norm charge branch crosses zero and becomes negative, demonstrating an energetic instability toward charge density wave order. (c) Hardening of the lowest charge mode with increasing positive in-plane field $B_y$ for octalayer RMG/hBN at $\theta=0.186^{\circ}$ and $D=0.60\,\mathrm{V/nm}$. Spin and valley collective modes are presented in Fig.~S2 of the Supplemental Material~\cite{tdci-supplemental}.}
    \label{fig_bse}
\end{figure*}

\textit{Collective modes and competing orders}
\label{sec:collective}
We further study the collective neutral excitations of the TDCI by solving the Bethe--Salpeter equation for particle--hole fluctuations about the self-consistent Hartree--Fock state~\cite{negele-rmp82,uria-xatu-cpc24,wu-tbg-collective-prl20,kumar-collective-prb21,kwan-fci-prb25,froese-excitons-prr25,xie-tmd-magnons-2026,alavirad-sau-prb20,desrochers-ahc-instability-prl26}. We track the positive-norm excitation branches. A mode that softens to zero and becomes negative signals an energetic instability of the reference Hartree--Fock state in the corresponding fluctuation channel~\cite{alavirad-sau-prb20,kumar-collective-prb21}. The detailed formulation, interaction kernels, and mode normalization are provided in the Supplemental Material~\cite{tdci-supplemental}.

Taking the occupied spin-valley flavor as $(+,\uparrow)$, we consider charge $[(+,\uparrow), (+,\uparrow)]$, spin $[(+,\downarrow), (+,\uparrow)]$, and spin-preserving valley $[(-,\uparrow), (+,\uparrow)]$ excitation channels. The charge channel probes density fluctuations within the occupied flavor, whereas the spin and valley channels probe transverse spin and intervalley fluctuations. Figure~\ref{fig_bse}(a) shows the charge excitation spectrum of a representative stable TDCI in heptalayer RMG/hBN at $\theta=0.604^{\circ}$ and $D=0.70\,\mathrm{V/nm}$. All charge branches remain positive throughout the moir\'e Brillouin zone, but there is a clear minimum near a finite wavevector between $\Gamma_s$ and $M_s$ points [Fig.~\ref{fig_bse}(a)]. As shown in Figs.~S2(a,b) of the Supplemental Material~\cite{tdci-supplemental}, this state exhibits a gapless spin Goldstone mode at $\mathbf{Q}=0$ and a gapped valley mode, similar to those in a conventional orbital ferromagnet, completing the collective-mode signature of a locally stable spin-valley-polarized TDCI mean-field state.

Some self-consistent TDCI solutions exhibit an instability under finite-momentum charge fluctuations. In these cases, the charge collective spectrum develops a softened minimum at finite momentum, which can reach zero energy and eventually become negative, signaling an energetic instability towards charge density wave order. A representative example is found in heptalayer RMG/hBN at $\theta=0.356^{\circ}$ and $D=0.75\,\mathrm{V/nm}$ [Fig.~\ref{fig_bse}(b)], where a charge branch softens below zero energy. In the corresponding spin spectrum [Fig.~S2(c) in the Supplemental Material~\cite{tdci-supplemental}], a positive-norm branch also becomes negative around the same wavevector, indicating a competing spin density wave instability. 

The unstable charge/spin mode can be further tuned by an in-plane magnetic field. For octalayer RMG/hBN at $\theta=0.186^{\circ}$ and $D=0.60\,\mathrm{V/nm}$, increasing $B_y$ hardens the lowest finite-momentum charge excitation [Fig.~\ref{fig_bse}(c)]. This field dependence reflects coupling to the in-plane orbital magnetization of the TDCI and provides a direct collective-mode manifestation of its transdimensional magnetic character. It also implies a possible transition from density-wave state to TDCI tuned by in-plane magnetic fields. In contrast, an ordinary Chern insulators exhibit gapped charge modes (with the gap $\sim $meV), as shown in Fig.~S3 of the Supplemental Material~\cite{tdci-supplemental}.

The two-band model in Eq.~\eqref{eq:tdci_two_band} illustrates a mechanism for finite-momentum charge-mode softening. For a fixed spin and valley $\mu$, the energy cost of an unbound particle--hole pair excitation with momentum $\mathbf Q$ in magnetic domain $i$ is
\begin{equation}
\begin{aligned}
 E^{\rm ph}_i(\bk,\mathbf Q)={}&
 \mathcal E(\bk+\mathbf Q)+\mathcal E(\bk)\\
 & + t \hat{\mathbf n}_i\!\cdot\!
 [\mathbf F(\bk+\mathbf Q)-\mathbf F(\bk)] .
\end{aligned}
\label{eq:tdci_ph_softening}
\end{equation}
The tilt term can shift the conduction-band minimum and valence-band maximum to different momenta, lowering the continuum particle-hole excitation threshold $\min_{\bk}E^{\rm ph}_i(\bk,\mathbf Q)$ at finite $\mathbf Q$. Residual particle--hole interactions can further lower a collective branch below this continuum and drive it to zero, while the single-particle gap remains open. The non-reciprocal dispersion thus provides a kinematic route to a finite-momentum charge-density-wave instability, whereas the actual soft wavevector and instability threshold additionally depend on the interaction kernel and orbital form factors. Further details are provided in the Supplemental Material~\cite{tdci-supplemental}.

\begin{figure}[tbp]
	\centering
	\includegraphics[width=\columnwidth]{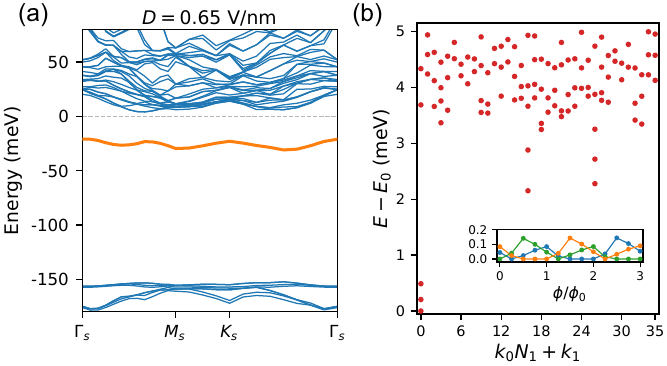}
	\caption{\textbf{From TDCI to TDFCI in twisted RMG/hBN.}
		(a) Hartree--Fock band structure of twisted octalayer RMG/hBN at $\theta=0.621^{\circ}$ and $D=0.65\,\mathrm{V/nm}$. The isolated occupied band has $C=1$.
		(b) Many-body spectrum at electron filling $\nu_{\rm band}=1/3$ for $6	imes 6$ cluster, showing a threefold low-energy manifold. The inset shows spectral flow at the same filling on a $4\times6$ cluster.}
	\label{fig4}
\end{figure}

\textit{Transdimensional fractional Chern insulator}
Having established the integer TDCI and its collective stability, we next ask whether fractional filling of its isolated Chern band can support a fractional Chern insulator (FCI). We consider twisted octalayer RMG/hBN at $\theta=0.621^{\circ}$ and $D=0.65\,\mathrm{V/nm}$. At moir\'e filling $\nu=1$, interactions open an insulating gap and isolate an occupied conduction band with Chern number $|C|=1$ [Fig.~\ref{fig4}(a)]. This band retains the mixed orbital character of the integer TDCI and provides a natural parent band for its fractionalized descendant. Technically, we project the screened Coulomb interaction onto this isolated Hartree--Fock Chern band and perform exact diagonalization calculations, following previous studies of FCIs in rhombohedral graphene~\cite{guo-prb24,lixiao-fci-prb24,senthil-fci-prl24}. More details are presented in Supplementary Materials~\cite{tdci-supplemental}. 

At electron filling $\nu_{\rm band}=1/3$, corresponding to hole filling $\nu_h=2/3$ with respect to filling 1, the $36$-site ($N_s=36$) cluster at 1/3 electron filling (2/3 hole filling) exhibits three nearly degenerate low-energy states separated from the higher-energy spectrum by a finite gap [Fig.~\ref{fig4}(b)]. This threefold torus degeneracy is consistent with a Laughlin-type $1/3$ FCI in a $C=1$ band. The topological character of this manifold is further supported by spectral flow under flux insertion~\cite{neupert-fci-prl11,regnault-fci-prx11}. On the $N_s=24$ cluster at the same filling, the three low-energy states cyclically evolve into one another and remain separated from the higher-energy continuum [inset of Fig.~\ref{fig4}(b)]. The spectrum returns to its initial configuration after three flux quanta, as expected for a threefold topological ground-state manifold. Because this fractional state is formed within the same orbital-magnetic $|C|=1$ band that realizes the integer TDCI, it combines fractional Chern topology with the transdimensional orbital character of its parent state. Thus such a fractional state is expected to exhibit in-plane orbital magnetization and magnetic hysteresis to in-plane magnetic field. 

\textit{Discussion and outlook}
Our results establish a route from symmetry-broken orbital magnetism in a metallic parent state to quantized Chern phases in RMG moir\'e systems. In twisted RMG/hBN, the moir\'e potential and Coulomb interactions reconstruct the Fermi-lune metal into a spin--valley-polarized Chern insulator with an isolated $|C|=1$ band. The resulting TDCI retains the reduced symmetry and coexisting in-plane and out-of-plane orbital magnetizations of its metallic parent. A minimal two-band model links the in-plane orbital magnetization to non-reciprocal group velocities and illustrates how the asymmetric dispersion can favor low-energy particle--hole excitations at finite momentum. Our collective-mode calculations reveal competing  density-wave instabilities in certain parameter regimes. An in-plane magnetic field can harden the soft charge mode, suggesting a means of stabilizing the TDCI against these fluctuations and potentially driving transitions between the TDCI and competing density-wave states.

Fractional filling of the isolated $|C|=1$ band further supports an FCI that is expected to inherit the transdimensional orbital character of its parent. Experimentally, quantized anomalous Hall transport accompanied by Hall hysteresis under both perpendicular and in-plane magnetic fields would provide a characteristic signature of the integer TDCI. Enhanced finite-momentum charge or spin fluctuations could reveal nearby density-wave instabilities. The reduced symmetry also allows edge modes on opposite boundaries to have different group velocities, offering further signatures of the underlying orbital order (see Fig.~S4 of the Supplemental Material~\cite{tdci-supplemental}). Finally, proximity to a superconductor may provide a route to chiral topological superconductivity with in-plane-field tunability, provided that effective pairing can be induced in the spin--valley-polarized bands.

% Independent main-text references, embedded in this file.
\def\bibsection{\section*{References}}
\makeatother

% Supplementary Information with independently numbered references.
\clearpage
\onecolumngrid
\phantomsection
\pdfbookmark[0]{Supplementary Information}{supplementary-information}
\begin{center}
{\large\bfseries Supplementary Information}\par\medskip
{\large Transdimensional Chern insulators in rhombohedral multilayer graphene superlattices}\par\medskip
Min Li$^{1}$, Xin Lu$^{1}$, Lei Wang$^{2}$ and Jianpeng Liu$^{1,3,*}$\par\medskip
$^{1}$\textit{School of Physical Science and Technology, ShanghaiTech Laboratory for Topological Physics, State Key Laboratory of Quantum Functional Materials, ShanghaiTech University, Shanghai 201210, China}\par
$^{2}$\textit{National Laboratory of Solid-State Microstructures, Collaborative Innovation Center of Advanced Microstructures, School of Physics, Nanjing University, Nanjing, China}\par
$^{3}$\textit{Liaoning Academy of Materials, Shenyang 110167, China}\par
\medskip$^*$\href{mailto:liujp@shanghaitech.edu.cn}{\nolinkurl{liujp@shanghaitech.edu.cn}}
\end{center}
\medskip
\renewcommand{\thefigure}{S\arabic{figure}}
\renewcommand{\thetable}{S\arabic{table}}
\renewcommand{\theequation}{S\arabic{equation}}
\setcounter{figure}{0}
\setcounter{table}{0}
\setcounter{equation}{0}
\setcounter{section}{0}
\setcounter{secnumdepth}{3}
\renewcommand{\theHequation}{S.\arabic{equation}}
\renewcommand{\theHfigure}{S.\arabic{figure}}
\renewcommand{\theHtable}{S.\arabic{table}}
\renewcommand{\theHsection}{S.\arabic{section}}
\renewcommand{\theHsubsection}{\theHsection.\arabic{subsection}}
\renewcommand{\theHsubsubsection}{\theHsubsection.\arabic{subsubsection}}
\def\a{\mathbf{a}}
\def\b{\mathbf{b}}
\def\p{\mathbf{p}}
\def\ta{\tilde{\mathbf{a}}}
\def\L{\mathbf{L}}
\def\br{\mathbf{r}}

\section{Methods}
\subsection{Non-interacting continuum model of rhombohedral multilayer graphene}
In our theoretical study, we adopt the continuum model derived by Moon and Koshino \cite{SI-koshino-hBN-prb14} to twisted rhombohedral multilayer graphene/hBN moir\'e superlattice for $N_L=6$--$9$. The corresponding Hamiltonian can be decomposed into three distinct components: the rhombohedral multilayer graphene Hamiltonian $H_{\rm{RMG}}$, the intralayer hBN  Hamiltonian $H_{\rm{hBN}}$ and the moir\'e coupling between them. 

The non-interacting single-particle properties of rhombohedral multilayer graphene can  be properly described by an effective low-energy continuum model, which is derived from a widely used Slater-Koster tight-binding model for multilayer graphene with carbon's $p_z$-like Wannier orbitals. Details of the Slater-Koster hopping parameters are given in Ref.~\cite{SI-koshino-hBN-prb14}. We further perform a Fourier transform on the tight-binding model to $\mathbf{k}$ space and expand the resulting Hamiltonian around the  Dirac points of monolayer graphene $\mathbf{K}^{\mu}=-\mu4\pi/3a_0(1,0)$, where $\mu=\pm 1$ denotes the valley index. For a rhombohedral graphene stack with $N_L$ layers, the low-energy continuum Hamiltonian for valley  $\mu$ is thus obtained as
\begin{align}
	H^{0,\mu}_{\rm{RMG}}
	=    \begin{pmatrix}
		h_{\text {intra }}^{0, \mu} & \left(h_{\text {inter }}^{0, \mu}\right)^{\dagger} & 0 & 0&... & 0 \\
		h_{\text {inter }}^{0, \mu} & h_{\text {intra }}^{0, \mu} & \left(h_{\text {inter }}^{0, \mu}\right)^{\dagger} & 0&... & 0 \\
		0 & h_{\text {inter }}^{0, \mu} & h_{\text {intra }}^{0, \mu} & \left(h_{\text {inter }}^{0, \mu}\right)^{\dagger}  & ... & 0\\
		\vdots  & ...& ...& ... & ... & \vdots \\
		0 & 0 & 0 & ...& h_{\text {inter }}^{0, \mu} & h_{\text {intra }}^{0, \mu}
	\end{pmatrix}  
	\label{eq:Htot_RMG}
\end{align}
where $H^{0,\mu}_{\rm{RMG}}(\mathbf{k})$ is a $2N_L\times 2N_L$ matrix at each $\mathbf{k}$. $h_{\text {intra }}^{0, \mu}=-\hbar v_F\mathbf{k}\cdot\mathbf{\sigma}_{\mu}$ is the intralayer Hamiltonian of monolayer graphene, 
with $\hbar v_F^0 \approx 5.253 \mathrm{eV}\,\text{\AA}$ being the non-interacting Fermi velocity of Dirac fermions \cite{SI-kpgra}. The Pauli matrix  $\sigma_\mu=\left(\mu \sigma_x, \sigma_y\right)$ is defined in sublattice space. The interlayer coupling term $h_{\text {inter }}^{0, \mu}$ takes the form
\begin{equation}
	h_{\text {inter }}^{0, \mu}=
	\begin{pmatrix}
		\hbar v_{\perp}\left(\mu k_x+i k_y\right) & t_{\perp} \\
		\hbar v_{\perp}\left(\mu k_x-i k_y\right) & \hbar v_{\perp}\left(\mu k_x+i k_y\right)\;,
	\end{pmatrix}
\end{equation}
where $t_{\perp}=0.34 \mathrm{eV}, \hbar v_{\perp}=0.335 \mathrm{eV}\,\text{\AA}$.

The hBN also possesses a honeycomb lattice structure with a nominal lattice constant $a_{\rm{hBN}}\simeq1.016a_0$, where $a_0$ denotes the lattice constant of graphene. For each twist angle, we choose a nearby ratio of integers for $a_{\rm{hBN}}/a_0$ to satisfy the commensurability condition. The two inequivalent sublattices of hBN (denoted as $A$ and $B$) are occupied by the two distinct atomic species, boron (B) and nitrogen (N), respectively, or in the reversed atomic assignment. For the rhombohedral multilayer graphene-hBN heterostructure investigated in this work, the stacking geometry is constructed based on a non-rotational alignment configuration. Specifically, a $B$ ($A$) sublattice site of graphene is superimposed onto a boron (nitrogen) site of hBN at the in-plane coordinates $(0,0)$ ($(0,a_0/\sqrt{3})$), respectively, such that the in-plane $A$-$B$ bonds are parallel to each other.

After rotation, the lattice vectors of hBN are given by
\begin{align}
	\ta_{1,2} = M R(\theta) \a_{1,2}
\end{align}
where $M = a_{\rm{hBN}}/a_0$ is the lattice constant ratio and $R(\theta)$ is the two-dimensional rotation matrix. The mismatch between the two lattices generates a moir\'e pattern with the moir\'e lattice vectors $\L_i$ and reciprocal lattice vectors $\G_i$ ($i=1,2$)
\begin{align}
	\L_{i} &= (1-R(-\theta) M^{-1})^{-1} \a_i \ , \\     
	\G_{i} &= (1- M^{-1}R(\theta)) \b_i \  . 
\end{align}
Following the derivation in \cite{SI-koshino-hBN-prb14} and considering only the hopping between bottom-layer graphene and hBN, the moir\'e interlayer hopping terms from the bottom-layer graphene to hBN are
\begin{align}
	U = u_0 \left[ \begin{pmatrix}
		1 & 1 \\
		1 & 1
	\end{pmatrix}+ \begin{pmatrix}
		1 & e^{-i\mu\frac{2 \pi}{3}} \\
		e^{i\mu\frac{2 \pi}{3}} & 1
	\end{pmatrix} e^{i \mu \G_1 \cdot \br}+ \begin{pmatrix}
		1 & e^{i\mu\frac{2 \pi}{3}} \\
		e^{-i\mu\frac{2 \pi}{3}} & 1
	\end{pmatrix} e^{i \mu (\G_1+\G_2) \cdot \br}
	\right]
	\label{eq:u-moire}
\end{align}
where $u_0=0.152\,\eV$. Since the band edges of insulating hBN and its gap are much larger than the long-wavelength moir\'e coupling at small twist angle, we can safely neglect completely the dispersion in hBN by dropping the $\k$ dependence in $H_{\rm{hBN}}$, namely
\begin{align}
	H_{\rm{hBN}} = 
	\begin{pmatrix}
		V_{\rm{N}} & 0 \\
		0 & V_{\rm{B}}
	\end{pmatrix}
\end{align}
with $V_{\rm{N}} = -1.40\,\eV$ and $V_{\rm{B}}=3.34\,\eV$. For the same reason, the low-energy physics of rhombohedral multilayer graphene-hBN heterostructure is dominated by the graphene's side so that the interlayer moir\'e coupling (Eq.~\eqref{eq:u-moire}) can be integrated out into an effective moir\'e superlattice potential $V_{\rm{hBN}}$ acting directly on Layer 1 of rhombohedral multilayer graphene, namely
\begin{align}
	V_{\rm{hBN}} = - U^\dagger H_{\rm{hBN}}^{-1} U = V^{\rm{eff}}(\br) + M^{\rm{eff}}(\br) \sigma_z + e v_F \mathbf{A}^{\rm{eff}}(\br) \cdot \bm{\sigma}_\mu.
\end{align}
We classify different terms in the effective potential by their sublattice structure. Simple algebra calculations give
\begin{subequations}
	\begin{align}
		V^{\rm{eff}}(\br) &= V_0 - V_1 \sum_{j=1}^{3} \cos \alpha_j(\br) \\
		M^{\rm{eff}}(\br) &= \sqrt{3} V_1 \sum_{j=1}^{3} \sin \alpha_j(\br) \\
		e v_F \mathbf{A}^{\rm{eff}}(\br) &= 2 \mu V_1 \sum_{j=1}^{3} \begin{pmatrix}
			\cos [2\pi (j+1)/3] \\
			\sin [2\pi (j+1)/3]
		\end{pmatrix}\cos \alpha_j(\br) \\
		\alpha_j (\br) &=  \G_j \cdot \br + \psi + \frac{2 \pi}{3} \quad \rm{with} \quad \G_3 = -\G_1 - \G_2
	\end{align}
\end{subequations} 
where $V_0=0.0289\,\eV$, $V_1 = 0.0210\,\eV$ and $\psi = -0.29$\,rad. Finally, the continuum model for twisted rhombohedral multilayer graphene/hBN heterostructure is readily written as
\begin{align}
	H^{0,\mu}
	=    \begin{pmatrix}
		h_{\text {intra }}^{0, \mu}+ V_{\rm{hBN}} & \left(h_{\text {inter }}^{0, \mu}\right)^{\dagger} & 0 & 0&... & 0 \\
		h_{\text {inter }}^{0, \mu} & h_{\text {intra }}^{0, \mu} & \left(h_{\text {inter }}^{0, \mu}\right)^{\dagger} & 0&... & 0 \\
		0 & h_{\text {inter }}^{0, \mu} & h_{\text {intra }}^{0, \mu} & \left(h_{\text {inter }}^{0, \mu}\right)^{\dagger}  & ... & 0\\
		\vdots  & ...& ...& ... & ... & \vdots \\
		0 & 0 & 0 & ...& h_{\text {inter }}^{0, \mu} & h_{\text {intra }}^{0, \mu}
	\end{pmatrix}  
	\label{eq:Htot_all}
\end{align}
In the following calculations,  we always consider $e v_F \mathbf{A}^{\rm{eff}}(\br)$ as a whole, instead of treating $\mathbf{A}^{\rm{eff}}(\br)$ separately. %
Also, the constant energy shift $V_0$ can be absorbed in chemical potential so we will omit it in the following.

To incorporate the effects of an applied out-of-plane electric field, we introduce layer-dependent on-site energies that increase linearly with the layer index. In experiments, the typical strength of the out-of- plane electric field D is around $1\,\text{V/nm}$. For a dielectric constant $\epsilon_r=4$ and an interlayer spacing between adjacent graphene sheets of $\sim d_0$, the electric-field-induced energy difference between adjacent layers is estimated to be on the order of $\sim 80\,\text{meV}$.
We further consider the application of an in-plane magnetic field .$\mathbf{B}=\left(B_x, B_y, 0\right)$, which can be described by a $z$-dependent in-plane vector potential $\mathbf{A}(\mathbf{z})=\B \times \mathbf{z}$. The effect of the in-plane field is incorporated into the Bloch Hamiltonian by modifying its $\bfk$ dependent matrix elements\cite{SI-lee-mag-inplane}. Specifically, the matrix element $H_{l_1, l_2}(\bfk, \B)$ between Layers $l_1$ and $l_2$ ($l_1, l_2=1,2,\cdots,N_L$) is given by:
\begin{equation}
	H_{l_1, l_2}(\bfk, \B)=H_{l_1, l_2}\left(\bfk+\frac{e}{\hbar} \frac{(l_1+l_2-2) d_0}{2} \B \times \mathbf{e}_z\right)
\end{equation}
where $d_0=3.35 \, \text{\AA}$ is the interlayer distance, and $\mathbf{e}_z$ is the unit vector in the $z$ direction. For $l_1 = l_2$, this corresponds to the intralayer $h_{\text {intra }}^{0, \mu}$ part, for $l_1 = l_2 \pm 1$ corresponds to the interlayer $h_{\text {inter }}^{0, \mu}$ part in $H^{0,\mu}_{\rm{RMG}}$.

\subsection{Renormalization of continuum model parameters}
Our work focuses on the low-energy electronic properties twisted rhombohedral multilayer graphene/hBN heterostructure at filling factor $\nu=1$. We define a low-energy window with an energy cutoff $E_C^*$, 
where electron-electron interactions are treated using a fully unrestricted self-consistent Hartree--Fock approach as detailed in the following subsection. 
Outside this window, electron-electron interactions are incorporated perturbatively as corrections to the single-particle Hamiltonian, partitioning electrons into
``low-energy'' electrons (within $E_C^*$) and ``remote-band" ones (outside $E_C^*$). 
While our emphasis lies on low-energy physics, the occupied remote-band electrons play a critical role: their long-range Coulomb interactions mediate interband correlations that renormalize low-energy properties. This renormalization fundamentally alters the effective description of the system, amplifying parameters in the low-energy Hamiltonian compared to non-interacting models. For instance, graphene’s Fermi velocity near the Dirac points is notably enhanced due to screening by filled Dirac sea states—a phenomenon validated by theory and experiments.

To systematically account for these correlation effects, we employ a perturbative renormalization group (RG) framework, which has been applied to monolayer graphene, magic-angle twisted bilayer graphene and RMG-hBN moir\'e superlattice to obtain the interaction-renormalized low-energy parameters. This methodology progressively integrates out high-energy remote-band degrees of freedom, encoding their influence into scale-dependent renormalizations of low-energy parameters. Rather than reproducing the full derivation, we adopt established expressions for renormalized parameters derived in Ref.~\cite{SI-guo-prb24}
\begin{subequations}
	\begin{align}
		v_F^* &= v_F^0 \left(1 + \frac{e^2}{16 \pi \epsilon_0 \epsilon_r \hbar v_F^0} \log \left( \frac{L_s}{n_{\text{cut}} a_0} \right) \right) \\
		M^{\rm{eff},*} &= M^{\rm{eff}} \left(1 + \frac{e^2}{16 \pi \epsilon_0 \epsilon_r \hbar v_F^0} \log \left( \frac{L_s}{n_{\text{cut}} a_0} \right) \right)^2 \\
		t_\perp^* &= t_\perp \left(1 + \frac{e^2}{16 \pi \epsilon_0 \epsilon_r \hbar v_F^0} \log \left( \frac{L_s}{n_{\text{cut}} a_0} \right) \right) \\
		e v_F^* \mathbf{A}^{\rm{eff},*} &= e v_F^0 \mathbf{A}^{\rm{eff}} \left(1 + \frac{e^2}{16 \pi \epsilon_0 \epsilon_r \hbar v_F^0} \log \left( \frac{L_s}{n_{\text{cut}} a_0} \right) \right).
	\end{align}
	\label{eq:rg}
\end{subequations}
Here $L_s\equiv|\L_i|$ is the moir\'e period, and the ratio $L_s/(n_{\text{cut}}a_0)$ plays the role of $E_C/E_C^*$. In the Hartree--Fock calculations, we use $n_{\text{cut}}=8$, retaining eight valence and eight conduction bands per spin and valley. $\hbar$ is the reduced Planck constant and $\epsilon_0$ is the vacuum permittivity. For the phase-diagram and collective-mode calculations, we retain two valence and eight conduction bands per spin and valley. %

\subsection{Hartree--Fock calculations}
\label{sec:hf_methods}

We project the renormalized continuum Hamiltonian in Eq.~\eqref{eq:rg} onto the low-energy window $E_C^*$, retaining $n_{\rm cut}=8$ valence and eight conduction bands per spin and valley. For the phase-diagram and collective-mode calculations, we retain two valence and eight conduction bands per spin and valley. The transformation to this band basis is
\begin{equation}
 \hat c_{\sigma\mu\alpha\mathbf G}(\btk)
 =\sum_n C_{\mu\alpha\mathbf G,n}(\btk)\hat c_{\sigma\mu,n}(\btk),
 \label{eq:transform}
\end{equation}
where $C_{\mu\alpha\mathbf G,n}$ is the Bloch eigenvector of the renormalized continuum Hamiltonian, $\btk$ lies in the moir\'e Brillouin zone, and $\alpha$ includes the layer and sublattice components. The projected interaction is
\begin{align}
 \hat V^{\rm intra}
 &=\frac{1}{2N_s}\sum_{\btk\btk'\btq}
 \sum_{\substack{\sigma\sigma'\mu\mu'\\nm n'm'}}\sum_{\mathbf G_0}
 V(\btq+\mathbf G_0)
 \Omega^{\mu,\mu'}_{nm,n'm'}(\btk,\btk',\btq,\mathbf G_0)\nonumber\\
 &\quad\times\hat c^{\dagger}_{\sigma\mu,n}(\btk+\btq)
 \hat c^{\dagger}_{\sigma'\mu',n'}(\btk'-\btq)
 \hat c_{\sigma'\mu',m'}(\btk')\hat c_{\sigma\mu,m}(\btk),
 \label{eq:Hintra-band}
\end{align}
with the form factor
\begin{align}
 \Omega^{\mu,\mu'}_{nm,n'm'}(\btk,\btk',\btq,\mathbf G_0)
 &=\sum_{\alpha\alpha'\mathbf G\mathbf G'}
 C^*_{\mu\alpha(\mathbf G+\mathbf G_0),n}(\btk+\btq)
 C^*_{\mu'\alpha'(\mathbf G'-\mathbf G_0),n'}(\btk'-\btq)\nonumber\\
 &\quad\times C_{\mu\alpha\mathbf G,m}(\btk)
 C_{\mu'\alpha'\mathbf G',m'}(\btk').
 \label{eq:hf_form_factor}
\end{align}
Each band-index pair labels the outgoing and incoming states on one electron line. The interaction conserves spin and valley along each electron line. Here $N_s$ is the number of moir\'e unit cells, equal to the number of momenta in the uniform Brillouin-zone discretization. We use the cell-area-normalized, double-gate-screened interaction
\begin{equation}
 V(\mathbf q)=\frac{e^2\tanh(|\mathbf q|d_s)}
 {2\epsilon_0\epsilon_r\Omega_0|\mathbf q|},
 \qquad d_s=40\,\mathrm{nm},\qquad\epsilon_r=4,
 \label{eq:bse_screening}
\end{equation}
where $\Omega_0$ is the moir\'e unit-cell area. Thus $V$ has units of energy and $V(0)=e^2d_s/(2\epsilon_0\epsilon_r\Omega_0)$.

The interaction in Eq.~\eqref{eq:Hintra-band} is decoupled into Hartree and Fock terms and solved self-consistently on a $24\times24$ momentum mesh. We use 128 symmetry-breaking initial states characterized by $\hat\tau_a\hat l_b\hat\sigma_c\hat s_{0,z}$, with $a,b,c=0,x,y,z$. The matrices $\hat\tau$, $\hat l$, $\hat\sigma$, and $\hat s$ act in valley, layer, sublattice, and spin space, respectively; $\hat l$ acts on the top and bottom layers.

For the spin--valley-polarized solutions used in the excitation calculation, the Hartree--Fock Hamiltonian is diagonal in spin and valley. Its band eigenvectors $U^{\sigma\mu}_{jn}(\btk)$ define
\begin{equation}
 \hat c_{\sigma\mu,j}(\btk)
 =\sum_n U^{\sigma\mu}_{jn}(\btk)\hat d_{\sigma\mu,n}(\btk),
 \qquad
 C^{\rm HF}_{\sigma\mu\alpha\mathbf G,n}(\btk)
 =\sum_j C_{\mu\alpha\mathbf G,j}(\btk)U^{\sigma\mu}_{jn}(\btk).
 \label{eq:hf_eigenbasis}
\end{equation}
We denote the corresponding energies and occupations by $\varepsilon^{\rm HF}_{\sigma\mu,n}(\btk)$ and $f_{\sigma\mu,n}(\btk)$. Replacing each $C$ in Eq.~\eqref{eq:hf_form_factor} by $C^{\rm HF}$ gives $\Omega^{\rm HF;\sigma\mu,\sigma'\mu'}_{nm,n'm'}$, the interaction form factor in the self-consistent band basis.

\clearpage
\subsection{Unfolding the Hartree--Fock miniband into graphene momentum space}
\label{sec:SI_band_unfolding}

To connect the insulating TDCI with its metallic parent state, we unfold the isolated Hartree--Fock miniband into the momentum space of pristine RMG. The unfolded weight resolves the graphene momenta contributing to the interacting moir\'e eigenstate. Using the coefficients defined in Eq.~\eqref{eq:hf_eigenbasis}, the Hartree--Fock Bloch state in a fixed spin--valley sector is
\begin{equation}
 \ket{\psi^{\rm HF}_{\sigma\mu,n}(\btk)}
 =\sum_{\alpha\mathbf G}C^{\rm HF}_{\sigma\mu\alpha\mathbf G,n}(\btk)
 \ket{\mathbf K_\mu+\btk+\mathbf G,\alpha;\sigma},
 \label{eq:SI_unfold_basis}
\end{equation}
where $\mathbf K_\mu$ is the graphene valley center and $\alpha$ includes the layer and sublattice components. The unfolded weight of this band is
\begin{equation}
 W_{\sigma\mu,n}(\btk,\mathbf G)
 =\sum_\alpha\left|C^{\rm HF}_{\sigma\mu\alpha\mathbf G,n}(\btk)\right|^2,
 \qquad \sum_{\mathbf G}W_{\sigma\mu,n}(\btk,\mathbf G)=1.
 \label{eq:SI_unfold_weight}
\end{equation}
We plot this weight at the valley-centered graphene momentum $\btk+\mathbf G$, suppressing the fixed spin, valley, and band labels in $W$. At filling $\nu=1$, the selected band is the isolated Hartree--Fock miniband filled relative to charge neutrality.

\begin{figure}[htbp]
 \centering
 \includegraphics[width=\textwidth]{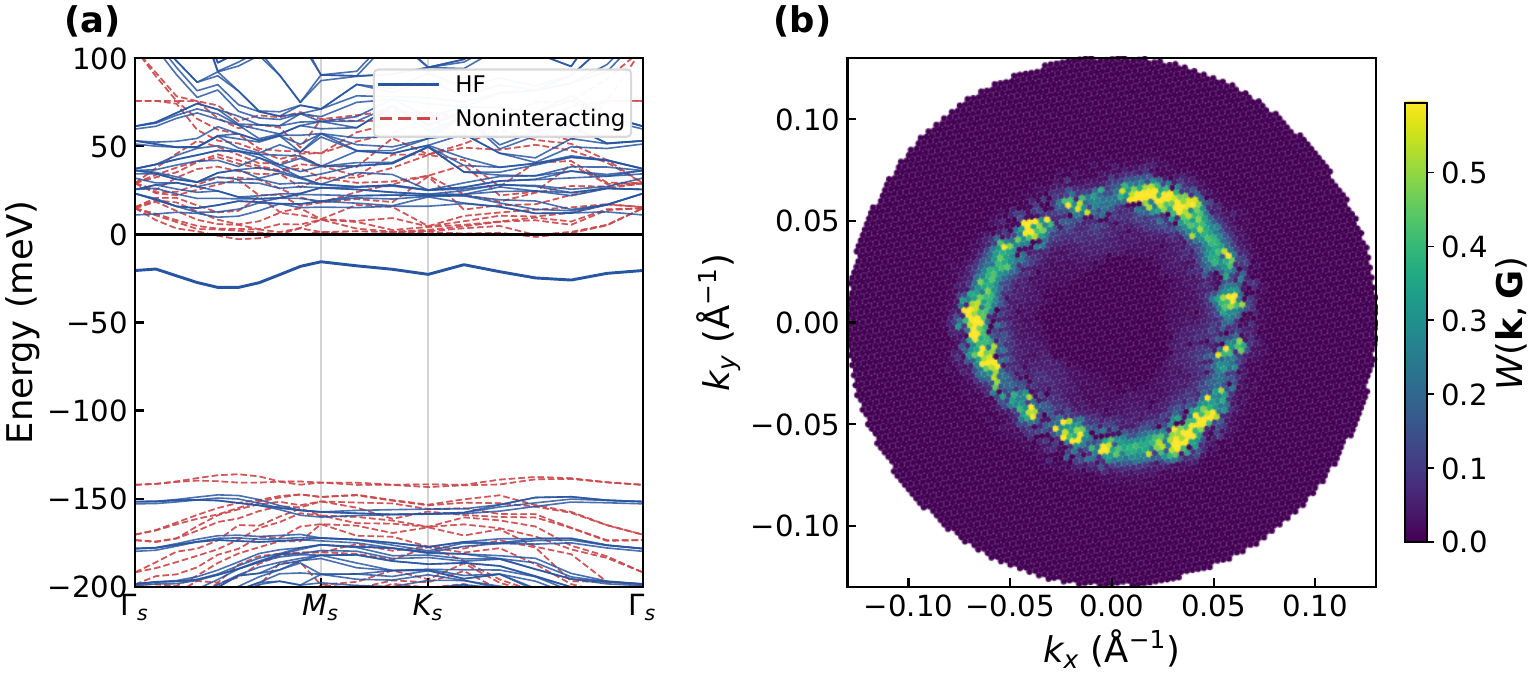}
 \caption{Ordinary Chern-insulator reference state preserving $C_3$ symmetry and without transdimensional orbital magnetization in octalayer RMG/hBN at $\theta=0.356^\circ$ and $D=0.65\,\mathrm{V/nm}$. (a) Hartree--Fock bands (blue solid lines) and noninteracting moir\'e bands (red dashed lines) along $\Gamma_s$--$M_s$--$K_s$--$\Gamma_s$, with energies referenced to their respective $\nu=1$ chemical potentials. (b) Unfolded momentum weight of the isolated Hartree--Fock miniband occupied at $\nu=1$, calculated from Eq.~\eqref{eq:SI_unfold_weight}. The momentum axes are measured relative to the graphene valley center, and the color scale gives $W(\btk,\mathbf G)$. The largely annular distribution contrasts with the anisotropic Fermi-lune-like weight of the TDCI in Fig.~1(d) of the main text.}
 \label{fig:SI_C3_reference}
\end{figure}

Figure~\ref{fig:SI_C3_reference} shows an ordinary Chern insulator in octalayer RMG/hBN at $\theta=0.356^\circ$ and $D=0.65\,\mathrm{V/nm}$. This reference state preserves $C_3$ symmetry and has no transdimensional orbital magnetization. Its unfolded distribution is largely annular, with an approximately threefold-symmetric weight pattern. It provides a comparison with the $\theta=0.734^\circ$ TDCI at the same displacement field in Fig.~1(d) of the main text, whose strongly anisotropic Fermi-lune-like weight reflects the broken $C_3$ and $\mathcal M_y$ symmetries.

\clearpage

\subsection{Bethe--Salpeter equation for neutral collective excitations}
\label{sec:bse_methods}

\subsubsection{Particle--hole basis and interaction kernel}
We calculate neutral collective excitations about the self-consistent Hartree--Fock ground state $\ket{\mathrm{HF}}$. For transferred momentum $\mathbf Q$, the particle--hole basis is
\begin{equation}
 \ket{I;\mathbf Q}
 =\hat P_I^{\dagger}(\mathbf Q)\ket{\mathrm{HF}},\qquad
 \hat P_I^{\dagger}(\mathbf Q)
 =\hat d^{\dagger}_{\sigma_2\mu_2,n}(\btk+\mathbf Q)
 \hat d_{\sigma_1\mu_1,m}(\btk),
 \label{eq:bse_pair_basis}
\end{equation}
where $m$ is occupied, $n$ is empty, and $I=(n,m,\btk;\sigma_2\mu_2,\sigma_1\mu_1)$. We denote the set of these forward transitions by $\mathcal T_{\rm ph}(\mathbf Q)$. The excitation energy of an unbound pair is
\begin{equation}
 \Delta\varepsilon_I(\mathbf Q)
 =\varepsilon^{\rm HF}_{\sigma_2\mu_2,n}(\btk+\mathbf Q)
 -\varepsilon^{\rm HF}_{\sigma_1\mu_1,m}(\btk).
 \label{eq:bse_pair_energy}
\end{equation}
The Bethe--Salpeter equation (BSE) couples these pairs through the residual interaction. In the Hartree--Fock basis, its resonant block consists of $\Delta\varepsilon_I$ and the direct and exchange electron--hole kernels, $K=K^{\rm d}+K^{\rm x}$~\cite{SI-uria-xatu-cpc24}. The ground-state self-energies are already contained in $\varepsilon^{\rm HF}$; the residual kernel is obtained from variations of the same projected Hartree--Fock functional~\cite{SI-negele-rmp82,SI-kumar-collective-prb21}.

Using the form factor defined in Eq.~\eqref{eq:hf_form_factor}, the direct kernel for a fixed spin--valley transfer is
\begin{align}
 K^{{\rm d};\sigma_2\mu_2,\sigma_1\mu_1}_{nm,n'm'}(\btk,\btk';\mathbf Q)
 ={}&-\frac{1}{N_s}\sum_{\mathbf G_0}V(\btk-\btk'+\mathbf G_0)\nonumber\\
 &\times\Omega^{{\rm HF};\sigma_2\mu_2,\sigma_1\mu_1}_{nn',m'm}
 (\btk'+\mathbf Q,\btk,\btk-\btk',\mathbf G_0).
 \label{eq:bse_fock_kernel}
\end{align}
It describes electron--hole attraction and is generated by variation of the Fock self-energy. The interaction transfers relative momentum $\btk-\btk'+\mathbf G_0$, conserving spin and valley on the electron and hole lines separately. The same projected-interaction structure appears in neutral-exciton calculations for moir\'e TMD heterobilayers~\cite{SI-froese-excitons-prr25}.

The exchange kernel couples transitions within spin--valley sectors $(\sigma,\mu)$ and $(\sigma',\mu')$:
\begin{align}
 K^{{\rm x};\sigma\mu,\sigma'\mu'}_{nm,n'm'}(\btk,\btk';\mathbf Q)
 ={}&\frac{1}{N_s}\sum_{\mathbf G_0}V(\mathbf Q+\mathbf G_0)\nonumber\\
 &\times\Omega^{{\rm HF};\sigma\mu,\sigma'\mu'}_{nm,m'n'}
 (\btk,\btk'+\mathbf Q,\mathbf Q,\mathbf G_0).
 \label{eq:bse_hartree_kernel}
\end{align}
This term is generated by the Hartree-potential variation and represents the exchange, or local-field, contribution in BSE terminology. It depends on the transferred momentum $\mathbf Q+\mathbf G_0$. At $\mathbf Q=0$, orthogonality eliminates the $\mathbf G_0=0$ interband matrix element, while finite-$\mathbf G_0$ terms remain. Both kernels use the interaction $V$, normalization $1/N_s$, and active-band space of the ground-state calculation. Momenta outside the moir\'e Brillouin zone are folded back with the corresponding relabeling of plane-wave indices.

\subsubsection{Full resonant and antiresonant BSE}
We retain both excitation and de-excitation amplitudes. A collective-mode operator is written in the quasiboson approximation as
\begin{equation}
 \hat O_{\nu}^{\dagger}(\mathbf Q)
 =\sum_I X_I^{\nu}(\mathbf Q)\hat P_I^{\dagger}(\mathbf Q)
 -\sum_J Y_J^{\nu}(\mathbf Q)\hat P_J(-\mathbf Q),
 \label{eq:bse_mode_operator}
\end{equation}
where $J$ labels forward transitions at $-\mathbf Q$. For a mode with definite spin--valley transfer, these transitions belong to the conjugate transfer channel, so that $\hat P_J(-\mathbf Q)$ carries the same momentum and spin--valley transfer as the $X$ terms. Evaluating the equation of motion in the Hartree--Fock Slater determinant gives the full BSE, equivalently the linear-response time-dependent Hartree--Fock (TDHF) equation~\cite{SI-negele-rmp82,SI-kumar-collective-prb21,SI-kwan-fci-prb25},
\begin{equation}
 \mathcal L(\mathbf Q)
 \binom{X^{\nu}(\mathbf Q)}{Y^{\nu}(\mathbf Q)}
 =\Omega_{\nu}(\mathbf Q)\binom{X^{\nu}(\mathbf Q)}{Y^{\nu}(\mathbf Q)},\qquad
 \mathcal L(\mathbf Q)=
 \begin{pmatrix}
 A(\mathbf Q)&B(\mathbf Q)\\
 -B^*(-\mathbf Q)&-A^*(-\mathbf Q)
 \end{pmatrix}.
 \label{eq:bse_momentum_blocks}
\end{equation}
The matrix elements are
\begin{equation}
 A_{IJ}(\mathbf Q)=\Delta\varepsilon_I(\mathbf Q)\delta_{IJ}+K_{IJ}(\mathbf Q),
 \qquad B_{IJ}(\mathbf Q)=K_{I\bar J}(\mathbf Q),
 \label{eq:bse_ab_elements}
\end{equation}
where $\bar J$ is the occupied--empty coherence obtained by reversing the particle and hole orbitals of $J$. It has momentum $\mathbf Q$ and the reversed spin--valley transfer. The kernels in Eqs.~\eqref{eq:bse_fock_kernel} and \eqref{eq:bse_hartree_kernel} apply to both occupation branches. Thus $A$ scatters particle--hole pairs, whereas $B$ couples resonant and antiresonant transitions. The interaction symmetries imply $A^\dagger(\mathbf Q)=A(\mathbf Q)$ and $B^*(-\mathbf Q)=B^\dagger(\mathbf Q)$. The lower blocks are the complex conjugates of the forward blocks at $-\mathbf Q$, with the corresponding conjugate-channel labels~\cite{SI-kumar-collective-prb21}.

Setting $B=0$ and retaining only the resonant block gives the Tamm--Dancoff approximation (TDA), $AX=\Omega X$, used in the optical-exciton formulation of Ref.~\cite{SI-uria-xatu-cpc24}.

\subsubsection{Collective channels and stability}
For the reference state polarized in the $(\sigma,\mu)=(\uparrow,+)$ sector, the forward transitions considered in the main text are
\begin{equation}
 \begin{array}{c|cc}
  \text{channel}&(\sigma_1,\mu_1)&(\sigma_2,\mu_2)\\\hline
  \text{charge}&(\uparrow,+)&(\uparrow,+)\\
  \text{spin flip}&(\uparrow,+)&(\downarrow,+)\\
  \text{intervalley}&(\uparrow,+)&(\uparrow,-)
 \end{array}
 \label{eq:bse_channels}
\end{equation}
Spin-flip transitions have zero charge-density vertex, so their exchange kernel $K^{\rm x}$ vanishes. The same holds for intervalley transitions within the long-wavelength valley-diagonal interaction in Eq.~\eqref{eq:Hintra-band}, which neglects microscopic valley-transfer vertices. Both transverse channels therefore contain only $K^{\rm d}$. For charge transitions, $K^{\rm x}$ also couples different diagonal spin--valley sectors; retaining all such sectors gives the total-charge response, while projection onto $(\uparrow,+)$ gives the intraflavor charge channel discussed in the main text. The antiresonant basis in each case contains all occupied--empty coherences with the same spin--valley transfer as the resonant fluctuation.

For a specified channel, the particle--hole continuum starts at
\begin{equation}
 E_{\rm cont}(\mathbf Q)=\min_{I\in\mathcal T_{\rm ph}(\mathbf Q)}\Delta\varepsilon_I(\mathbf Q).
 \label{eq:bse_continuum_threshold}
\end{equation}
Here $\mathcal T_{\rm ph}(\mathbf Q)$ is restricted to the forward transitions in that channel. Interaction-bound collective branches can lie below this threshold and soften while the single-particle Hartree--Fock gap remains finite, as in collective-mode studies of TBG, rhombohedral graphene, and moir\'e TMD systems~\cite{SI-wu-tbg-collective-prl20,SI-kwan-fci-prb25,SI-froese-excitons-prr25,SI-xie-tmd-magnons-2026}.

The quasiboson commutator $\langle[\hat P_I,\hat P_J^\dagger]\rangle_{\rm HF}=\delta_{IJ}$ gives the mode norm
\begin{equation}
 \mathcal N_{\nu}=X^{\nu\dagger}X^\nu-Y^{\nu\dagger}Y^\nu.
 \label{eq:bse_mode_norm}
\end{equation}
For real-energy solutions of Eq.~\eqref{eq:bse_momentum_blocks}, we select positive-norm branches and normalize them to $\mathcal N_\nu=1$. In a locally stable Hartree--Fock state these branches have nonnegative excitation energies, with zero modes along continuous-symmetry directions. A real positive-norm branch that softens through zero and becomes negative signals an energetic instability in the corresponding channel~\cite{SI-alavirad-sau-prb20,SI-kumar-collective-prb21}. Hermitian conjugation generates a partner at opposite momentum and reversed spin--valley transfer with energy $-\Omega_\nu$ and opposite norm. Such negative-frequency, negative-norm partners are distinct from positive-norm negative-energy instabilities.

The purely orbital in-plane magnetic field considered here preserves spin SU(2) while changing the Hartree--Fock energies and form factors entering the BSE. For the field-dependent spectra in Fig.~3(c) of the main text, we solve the Hartree--Fock equations self-consistently at each $B_y$ and fixed filling. The resulting field-dependent energies $\varepsilon^{\rm HF}(B_y)$ and Bloch coefficients $C^{\rm HF}(B_y)$ are then used to reconstruct both $A(\mathbf Q;B_y)$ and $B(\mathbf Q;B_y)$ and solve the full BSE.

\begin{figure}[htbp]
 \centering
 \includegraphics[width=0.86\textwidth]{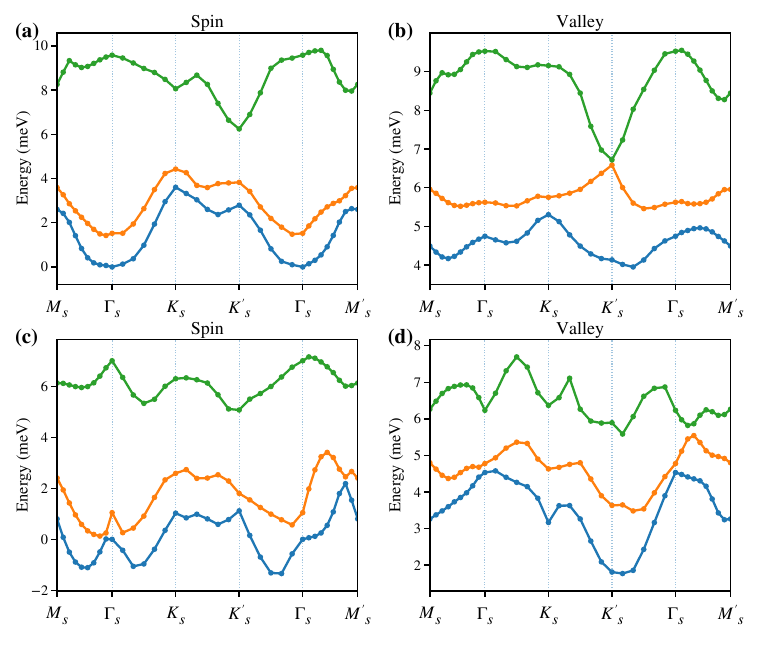}
 \caption{Spin and valley collective modes of heptalayer RMG/hBN. (a,b) Spin-flip and intervalley spectra, respectively, of the locally stable TDCI at $\theta=0.604^\circ$ and $D=0.70\,\mathrm{V/nm}$. The spin spectrum contains a Goldstone mode at $\Gamma_s$ ($\mathbf Q=0$), while the valley spectrum is gapped. (c,d) Corresponding spectra of the TDCI at $\theta=0.356^\circ$ and $D=0.75\,\mathrm{V/nm}$. A positive-norm spin branch becomes negative at finite momentum, indicating a spin-density-wave instability competing with the charge instability shown in Fig.~3(b) of the main text; the displayed valley modes remain gapped. The spectra follow the high-symmetry path indicated on the horizontal axes.}
 \label{fig:spin_valley_modes}
\end{figure}

\clearpage
Figure~\ref{fig:spin_valley_modes} supplements the charge spectra in Fig.~3 of the main text with the spin-flip and intervalley modes of the same self-consistent TDCI states. At $\theta=0.604^\circ$ and $D=0.70\,\mathrm{V/nm}$, the lowest spin branch is the gapless Goldstone mode at $\mathbf Q=0$ associated with spontaneously broken spin SU(2) symmetry~\cite{SI-wu-tbg-collective-prl20,SI-kumar-collective-prb21}. The valley modes remain gapped, consistent with the absence of an exact valley-SU(2) symmetry in the continuum Hamiltonian~\cite{SI-kwan-fci-prb25}. Together with the positive charge-mode energies throughout the moir\'e Brillouin zone reported in the main text, these spectra establish the collective-mode signature of a locally stable spin--valley-polarized TDCI.

At $\theta=0.356^\circ$ and $D=0.75\,\mathrm{V/nm}$, a positive-norm spin branch develops negative energies at finite momentum, including the region between $\Gamma_s$ and $M_s$ where the charge mode also becomes negative. This identifies a competing spin-density-wave instability alongside the charge-density-wave instability of the uniform TDCI. The displayed valley branches remain gapped along the high-symmetry path.

\begin{figure}[htbp]
 \centering
 \includegraphics[width=\textwidth]{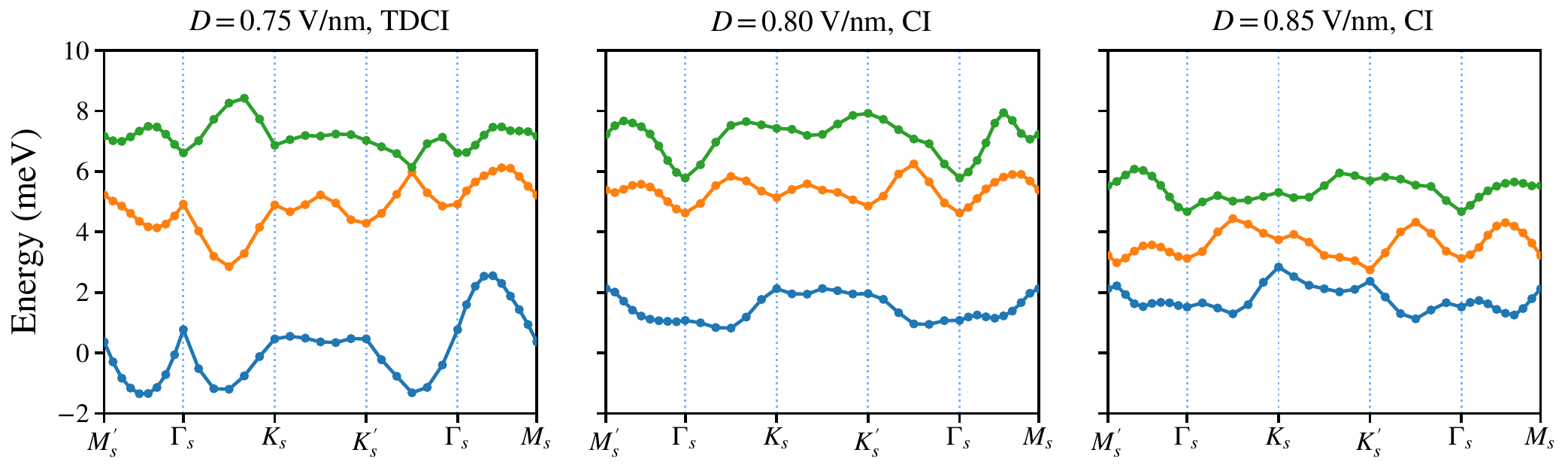}
 \caption{Comparison of charge collective modes in heptalayer RMG/hBN at $\theta=0.356^\circ$. From left to right, $D=0.75$, $0.80$, and $0.85\,\mathrm{V/nm}$. The first state is a TDCI and the latter two are ordinary Chern insulators (CI). The vertical axis gives the charge-mode excitation energy $\Omega_\nu(\mathbf Q)$ in meV. At $D=0.75\,\mathrm{V/nm}$, the positive-norm charge branch becomes negative at finite momentum, signaling an energetic instability toward charge-density-wave order. At $D=0.80$ and $0.85\,\mathrm{V/nm}$, the displayed branches remain positive along the plotted path.}
 \label{fig:charge_modes_comparison}
\end{figure}

For comparison, Fig.~\ref{fig:charge_modes_comparison} shows the charge spectra at $\theta=0.356^\circ$ for the TDCI at $D=0.75\,\mathrm{V/nm}$ and ordinary Chern insulators at $D=0.80$ and $0.85\,\mathrm{V/nm}$. The displayed charge branches of both ordinary Chern insulators remain positive along the same high-symmetry path.

\clearpage
\subsection{Calculations of in-plane orbital magnetizations}
\label{sec:sm-inplane-orbital-hf}

We report the in-plane orbital magnetization $M_y$ per moir\'e unit cell,
in $\mu_B$ per cell, for symmetry-broken Hartree--Fock (HF) states
including TDCI. At $\nu=1$, one additional electron occupies each cell, so the same value gives the moment per added carrier.
We use a zero-field expectation value and a weak-field one-shot energy
difference. With all energies defined per cell, $M_y=-d\mathcal
E_{\mathrm{HF}}/dB_y$.

\paragraph{Zero-field method.}
At zero temperature and fixed particle number, the HF energy is
$\mathcal E_{\mathrm{HF}}=\mathcal E_0+\mathcal E_{\mathrm{int}}[\rho]$,
with $\mathcal E_0=N_s^{-1}\sum_{\bm{k}}
\operatorname{Tr}[\rho_{\bm{k}}H_0(\bm{k},B_y)]$.
The Hartree and Fock energy $\mathcal E_{\mathrm{int}}$ includes the
appropriate $1/2$ factors and has no explicit field dependence
in a fixed orbital basis\cite{SI-Kang2026OrbitalHF}.
Its density variation defines $H_{\mathrm{HF}}=H_0+\Sigma_{\mathrm{HF}}[\rho]$.
Along a differentiable self-consistent branch,
\begin{equation}
 \frac{d\mathcal E_{\mathrm{HF}}}{dB_y}
 =\frac{1}{N_s}\sum_{\bm{k}}
 \operatorname{Tr}[\rho_{\bm{k}}\partial_{B_y}H_0]
 +\frac{1}{N_s}\sum_{\bm{k}}
 \operatorname{Tr}[H_{\mathrm{HF}}\dot\rho_{\bm{k}}],
 \label{eq:sm-orb-chain-rule}
\end{equation}
where a dot denotes a field derivative and $E_F$ is the chemical potential.
The normalized HF eigenvectors satisfy
$\operatorname{Re}\langle u|\dot u\rangle=0$, while occupation changes
at the Fermi surface give
$N_s^{-1}\sum_{\bm{k},n}\varepsilon_{n\bm{k}}^{\mathrm{HF}}\dot f_{n\bm{k}}
=E_F\dot{\mathcal N}=0$.
Thus the second term vanishes, leaving the explicit orbital coupling.

For electron charge $-e$ ($e>0$), we use
$\bm A(z)=\bm B_\parallel\times(z\bm e_z)$ and
$Z_0=\sum_{\ell=1}^{N_L}(\ell-1)d_0\Pi_\ell$,
where $\Pi_\ell$ is a layer projector and $d_0=3.35\,\text{\AA}$.
The corresponding layer-dependent Peierls coupling is
\begin{equation}
 H_{0,\ell\ell'}(\bm{k},\bm B_\parallel)
 =H_{0,\ell\ell'}\!\left[
 \bm{k}+\frac{e}{\hbar}\frac{(\ell+\ell'-2)d_0}{2}
 \bm B_\parallel\times\bm e_z\right].
 \label{eq:sm-orb-peierls}
\end{equation}
Differentiation gives
$m_y^{(0)}=-\partial_{B_y}H_0|_0
=-(e/2\hbar)\{Z_0,\partial_{k_x}H_0\}$\cite{SI-Hu2026InPlane},
with $\{A,B\}=AB+BA$.
Therefore,
\begin{equation}
 \boxed{
 M_y^{\mathrm{ZF}}
 =-\frac{e}{2\hbar N_s}
 \sum_{\bm{k},\,n\in\mathrm{occ}}
 \langle u_{n\bm{k}}^{\mathrm{HF}}|
 \{Z_0,\partial_{k_x}H_0\}
 |u_{n\bm{k}}^{\mathrm{HF}}\rangle .}
 \label{eq:sm-orb-hf-band-sum}
\end{equation}
Here $|u_{n\bm{k}}^{\mathrm{HF}}\rangle$ is an eigenstate of the
self-consistent $H_{\mathrm{HF}}(0)$, all quantities are evaluated at
$\bm B_\parallel=0$, and $N_s$ counts uniformly sampled momenta
in one moir\'e Brillouin zone.

\paragraph{Weak-field one-shot method.}
Starting from the same zero-field HF solution at fixed filling, we
apply a small $\delta B_y$ through the Peierls coupling above and evaluate the
orbital energy change in a one-shot calculation, without additional
HF self-consistency iterations.
Writing $\mathcal E_{\mathrm{1shot}}(\delta B_y)$ for the energy per cell
evaluated with the same HF functional, we estimate
\begin{equation}
 M_y^{\mathrm{1shot}}
 =-\frac{\mathcal E_{\mathrm{1shot}}(\delta B_y)
 -\mathcal E_{\mathrm{HF}}(0)}
 {\delta B_y}.
 \label{eq:sm-orb-finite-difference}
\end{equation}
The linear orbital energy shift is
$\delta\mathcal E_{\mathrm{orb}}
=-M_y\,\delta B_y+O(\delta B_y^2)$.
The one-shot difference probes the same linear orbital coupling as
the zero-field formula.
In our calculations, the two methods yield quantitatively consistent
values of the in-plane orbital magnetization.

\clearpage
\subsection{Single-band projection and exact diagonalization}
\label{sec:ed_methods}

We study fractional filling of the isolated spin--valley-polarized Hartree--Fock band using the single-band projection and hole-doping approach of Ref.~\cite{SI-guo-prb24}. The reference state is the $\nu=1$ TDCI in octalayer RMG/hBN at $\theta=0.621^\circ$ and $D=0.65\,\mathrm{V/nm}$ discussed in Fig.~4 of the main text. We retain its Hartree--Fock band dispersion and project the screened interaction in Eq.~\eqref{eq:bse_screening} onto the same Bloch eigenstates. The occupied bands outside this isolated band remain frozen at their integer-filling Hartree--Fock values.

Fixing the spin, valley, and band labels to $(\sigma_0,\mu_0,n_0)$, we write $\hat d_{\btk}=\hat d_{\sigma_0\mu_0,n_0}(\btk)$ and $\varepsilon^{\rm HF}(\btk)=\varepsilon^{\rm HF}_{\sigma_0\mu_0,n_0}(\btk)$. The projected density vertex is
\begin{equation}
 \Lambda(\btk,\btq,\mathbf G_0)
 =\sum_{\alpha\mathbf G}
 C^{{\rm HF}*}_{\sigma_0\mu_0\alpha(\mathbf G+\mathbf G_0),n_0}(\btk+\btq)
 C^{\rm HF}_{\sigma_0\mu_0\alpha\mathbf G,n_0}(\btk),
 \label{eq:ed_density_vertex}
\end{equation}
so that $\hat\rho_e(\btq,\mathbf G_0)=\sum_{\btk}\Lambda(\btk,\btq,\mathbf G_0)\hat d^\dagger_{\btk+\btq}\hat d_{\btk}$. This vertex uses the same coefficients as the Hartree--Fock interaction form factor; restricting all four band labels to $n_0$ gives
\begin{equation}
 \Omega^{{\rm HF};\sigma_0\mu_0,\sigma_0\mu_0}_{n_0n_0,n_0n_0}
 (\btk,\btk',\btq,\mathbf G_0)
 =\Lambda(\btk,\btq,\mathbf G_0)
 \Lambda(\btk',-\btq,-\mathbf G_0).
 \label{eq:ed_form_factor}
\end{equation}

The completely filled isolated band defines the hole vacuum $\ket{F}$. We introduce $\hat h^\dagger_{\btk}=\hat d_{-\btk}$, so that the hole density has vertex
\begin{equation}
 \Lambda_h(\btk,\btq,\mathbf G_0)
 =\Lambda(-\btk-\btq,\btq,\mathbf G_0),\qquad
 \hat\rho_h(\btq,\mathbf G_0)
 =\sum_{\btk}\Lambda_h(\btk,\btq,\mathbf G_0)
 \hat h^\dagger_{\btk+\btq}\hat h_{\btk}.
 \label{eq:ed_hole_density}
\end{equation}
Apart from the filled-band density, the electron-density fluctuation is $-\hat\rho_h$. Normal ordering about the integer Hartree--Fock reference yields, up to the constant energy of that state,
\begin{equation}
 \hat H_{\rm ED}^{(h)}
 =-\sum_{\btk}\varepsilon^{\rm HF}(-\btk)
 \hat h^\dagger_{\btk}\hat h_{\btk}
 +\frac{1}{2N_s}\sum_{\btq,\mathbf G_0}V(\btq+\mathbf G_0)
 :\!\hat\rho_h(\btq,\mathbf G_0)
 \hat\rho_h(-\btq,-\mathbf G_0)\!:_h.
 \label{eq:ed_hole_hamiltonian}
\end{equation}
Here $:\cdots:_h$ denotes normal ordering with respect to the hole vacuum. The one-body term is the Hartree--Fock electron-removal energy, including the interaction with the filled reference state; the normal-ordered two-body term describes the residual interaction between the doped holes. Thus the integer-filling Hartree--Fock background is included once. At fixed hole number, a common shift of $\varepsilon^{\rm HF}$ changes only the energy origin. The single-band Hamiltonian is diagonalized within sectors of fixed particle number and total moir\'e crystal momentum.

For a torus with periods $N_1\mathbf L_1$ and $N_2\mathbf L_2$, the allowed momenta are $\btk=(j_1/N_1)\mathbf G_1+(j_2/N_2)\mathbf G_2$, with $j_i=0,\ldots,N_i-1$ and $N_s=N_1N_2$. Electron filling $\nu_{\rm band}=1/3$ corresponds to hole filling $\nu_h=1-\nu_{\rm band}=2/3$. The $6\times6$ spectrum in Fig.~4(b) of the main text therefore has $N_s=36$ and $N_h=24$, equivalent to 12 electrons in the isolated band. The $4\times6$ spectral-flow calculation has $N_s=24$ and $N_h=16$, equivalent to eight electrons.

Flux insertion is implemented through twisted boundary conditions, $\hat h_{\mathbf R+N_i\mathbf L_i}=e^{i\varphi_i}\hat h_{\mathbf R}$, giving
\begin{equation}
 \btk(\boldsymbol\varphi)
 =\sum_{i=1}^2\frac{j_i+\varphi_i/(2\pi)}{N_i}\mathbf G_i.
 \label{eq:ed_twisted_momenta}
\end{equation}
One flux quantum corresponds to a $2\pi$ change of a boundary phase. The band energies and projected interaction vertices are evaluated on this shifted grid, and the low-energy spectrum is followed as the boundary phase varies. The three low-energy states shown in the main text remain separated from higher levels and cycle into one another under flux insertion. Their threefold low-energy manifold and spectral flow are consistent with a Laughlin-type $1/3$ FCI in the $C=1$ band~\cite{SI-neupert-fci-prl11,SI-regnault-fci-prx11}.

\clearpage
\section{Two-band effective description of the transdimensional Chern insulator}
\label{sec:two_band}

\subsection{Lattice Hamiltonian and symmetry-related domains}
We introduce a minimal lattice model that captures the coexistence of a Chern band, non-reciprocal dispersion, and in-plane orbital magnetization. This model describes a fixed spin and valley $\mu=\pm1$ and retains two effective layer orbitals at heights $z_{1,2}=\pm\ell/2$. It is a qualitative effective description rather than a quantitative downfolding of the self-consistent Hartree--Fock Hamiltonian. In particular, $\ell$ is the separation of the effective orbital centers and need not equal the spacing between adjacent graphene layers.

Following the notation of the main text, the Hamiltonian is
\begin{equation}
 \begin{aligned}
 h_{\mu i}(\bk)&=d_{0,i}(\bk)\sigma_0+\bd_\mu(\bk)\cdot\bm\sigma,
 & d_{0,i}(\bk)&=t\hat{\mathbf n}_i\cdot\mathbf F(\bk),\\
 \bd_\mu(\bk)&=(-\mu vF_x,vF_y,m(\bk)),
 & m(\bk)&=m+rG(\bk).
 \end{aligned}
 \label{eq:tb_hamiltonian}
\end{equation}
Here $\sigma_{x,y,z}$ act on the two effective layer orbitals, $v$ is the interorbital hybridization amplitude, $t$ is the signed tilt amplitude, and $m$ and $r$ determine the band-inversion mass. All four parameters have units of energy. The tilt axis is specified by $\hat{\mathbf n}_i=\hat{\mathbf M}_i\times\hat{\mathbf z}$ as in the main text, where $\hat{\mathbf M}_i$ labels the symmetry-allowed in-plane orbital axis. The signed magnetization along $\hat{\mathbf z}\times\hat{\mathbf n}_i$ is determined by the parameters and occupation. For the representative domain with $\hat{\mathbf n}_i=\hat{\mathbf x}$, $d_{0,i}=tF_x$ and
\begin{equation}
 h_{\mu x}(\bk)=
 \begin{pmatrix}
 tF_x+m(\bk)&-\mu vF_x-ivF_y\\
 -\mu vF_x+ivF_y&tF_x-m(\bk)
 \end{pmatrix},\qquad \hat z=\frac{\ell}{2}\sigma_z.
 \label{eq:tb_layer_matrix}
\end{equation}

The hexagonal moir\'e lattice is represented by a triangular Bravais lattice with
\begin{equation}
 \boldsymbol\delta_j=L_s(\cos\theta_j,\sin\theta_j),\qquad
 \theta_j=\frac{2\pi(j-1)}{3},\qquad
 \Omega_0=\frac{\sqrt3}{2}L_s^2,
\end{equation}
where $\Omega_0$ is the moir\'e unit-cell area. The periodic, dimensionless lattice functions are
\begin{equation}
 \mathbf F(\bk)=\frac23\sum_{j=1}^3\frac{\boldsymbol\delta_j}{L_s}
 \sin(\bk\cdot\boldsymbol\delta_j),\qquad
 G(\bk)=\sum_{j=1}^3[1-\cos(\bk\cdot\boldsymbol\delta_j)].
 \label{eq:tb_lattice_functions}
\end{equation}
Explicitly,
\begin{align}
 F_x&=\frac23\left[\sin(k_xL_s)+\sin\left(\frac{k_xL_s}{2}\right)
 \cos\left(\frac{\sqrt3k_yL_s}{2}\right)\right],\\
 F_y&=\frac{2\sqrt3}{3}\cos\left(\frac{k_xL_s}{2}\right)
 \sin\left(\frac{\sqrt3k_yL_s}{2}\right),\\
 G&=3-\cos(k_xL_s)-2\cos\left(\frac{k_xL_s}{2}\right)
 \cos\left(\frac{\sqrt3k_yL_s}{2}\right).
 \label{eq:tb_explicit_functions}
\end{align}
Near $\Gamma$, $\mathbf F=L_s\bk+O[(kL_s)^3]$ and $G=3L_s^2|\bk|^2/4+O[(kL_s)^4]$. The lattice regularization makes the Chern number well defined over the full moir\'e Brillouin zone (mBZ).

The three tilt axes related by $C_3$, together with their time-reversed partners, describe the six orbital domains. Time reversal sends $(\mu,\hat{\mathbf n}_i)$ to $(-\mu,-\hat{\mathbf n}_i)$ at fixed $t$. Selecting one valley breaks both time reversal and the valley-exchanging mirror $\mathcal M_y:(x,y,z)\mapsto(-x,y,z)$; selecting a nonzero tilt breaks $C_3$. For the $x$-tilted domain,
\begin{equation}
 h_{\mu x}^*(k_x,-k_y)=h_{\mu x}(k_x,k_y).
\end{equation}
This is the orbital representation of the residual $\mathcal M_y\mathcal T$ symmetry. It permits $M_y$ and $M_z$, while forcing $M_x=0$. These statements concern the orbital sector; the spin orientation is not fixed by the direction of orbital magnetization.

\subsection{Band topology and the insulating condition}
The band energies and layer polarizations are
\begin{equation}
 \varepsilon_{b,i}=d_{0,i}+b\mathcal E,\qquad
 \mathcal E(\bk)=\sqrt{v^2|\mathbf F(\bk)|^2+m(\bk)^2},\qquad
 \langle\sigma_z\rangle_{b,i}=b\frac{m(\bk)}{\mathcal E(\bk)},\qquad b=\pm1.
 \label{eq:tb_spectrum}
\end{equation}
The scalar tilt changes the energies but not the eigenvectors. Consequently, the minimal model retains a $C_3$-symmetric Berry-curvature distribution even though its dispersion breaks $C_3$. With the Berry-connection convention $\mathbf A_b=i\langle u_b|\nabla_{\bk}u_b\rangle$, the Berry curvature is
\begin{equation}
 \Omega_{b,z}(\bk)=-\frac{b}{2\mathcal E^3}
 \bd_\mu\cdot(\partial_{k_x}\bd_\mu\times\partial_{k_y}\bd_\mu),
 \qquad C_b=\frac{1}{2\pi}\int_{\rm mBZ}d^2k\,\Omega_{b,z}.
 \label{eq:tb_curvature}
\end{equation}
For $v\ne0$, the direct gap can close only where $\mathbf F=0$. There are six inequivalent zeros on the Brillouin-zone torus: $\Gamma$, three $M$ points, and two $K$ points. The windings of $(F_x,F_y)$ are $+1,-1,+1$, respectively, and the corresponding masses are $m,m+4r,m+9r/2$. Summing their massive-Dirac contributions gives the lower-band Chern number
\begin{equation}
 C_-=\mu c(m),\qquad
 c(m)=-\frac12\left[\operatorname{sgn}m-3\operatorname{sgn}(m+4r)
 +2\operatorname{sgn}(m+9r/2)\right].
 \label{eq:tb_chern}
\end{equation}
For $r>0$ this yields
\begin{equation}
 C_-=\begin{cases}
 0,&m>0,\\
 \mu,&-4r<m<0,\\
 -2\mu,&-9r/2<m<-4r,\\
 0,&m<-9r/2.
 \end{cases}
\end{equation}
The Chern number is undefined at the direct-gap closing points. To realize a Chern \emph{insulator}, the band topology must additionally be accompanied by a positive indirect gap,
\begin{equation}
 \Delta_{\rm ind}=\min_{\bk}\varepsilon_{+,i}(\bk)
 -\max_{\bk}\varepsilon_{-,i}(\bk)>0,\qquad
 \max\varepsilon_{-,i}<E_F<\min\varepsilon_{+,i}.
 \label{eq:tb_indirect_gap}
\end{equation}
Here $E_F$ denotes the chemical potential, distinct from the valley index $\mu$. A sufficiently strong tilt can close the indirect gap without changing $C_-$.

\subsection{Non-reciprocal velocity and in-plane orbital magnetization}
Since $d_{0,i}$ is odd and $\mathcal E$ is even under $\bk\mapsto-\bk$, the even part of the lower-band group velocity is
\begin{equation}
 \delta\mathbf v_i(\bk)\equiv\frac{\mathbf v_{-,i}(\bk)+\mathbf v_{-,i}(-\bk)}{2}
 =\frac{1}{\hbar}\nabla_{\bk}d_{0,i}(\bk),\qquad
 \mathbf v_{-,i}=\frac{1}{\hbar}\nabla_{\bk}\varepsilon_{-,i}.
 \label{eq:tb_nonreciprocal}
\end{equation}
To derive its connection to orbital magnetization, take $e>0$ and use the gauge
\begin{equation}
 \mathbf A(z)=z(B_y,-B_x,0)
\end{equation}
for $\mathbf B_\parallel=(B_x,B_y,0)$. With straight-bond Peierls coupling, a hopping between orbitals at heights $z_a$ and $z_b$ acquires the momentum shift
\begin{equation}
 h_{ab}(\bk,\mathbf B_\parallel)=h_{ab}\left(\bk+
 \frac{e(z_a+z_b)}{2\hbar}(B_y,-B_x)\right).
 \label{eq:tb_peierls}
\end{equation}
Thus the intralayer matrix elements are shifted with their respective heights, while the interorbital elements have zero shift in the chosen height origin. Expanding to first order gives
\begin{equation}
 \delta h=\frac{e}{2\hbar}\left[B_y\{\hat z,\partial_{k_x}h\}
 -B_x\{\hat z,\partial_{k_y}h\}\right]
 =-\hat{\mathfrak m}_xB_x-\hat{\mathfrak m}_yB_y,
\end{equation}
where $\{A,B\}=AB+BA$. The magnetic-moment operators are therefore
\begin{equation}
 \hat{\mathfrak m}_x=\frac{e}{2\hbar}\{\hat z,\partial_{k_y}h\},\qquad
 \hat{\mathfrak m}_y=-\frac{e}{2\hbar}\{\hat z,\partial_{k_x}h\}.
 \label{eq:tb_moment_operator}
\end{equation}
Using $\hat z=\ell\sigma_z/2$, the moment in band $b$ is
\begin{align}
 \mathfrak m_{x,b}&=\frac{e\ell}{2\hbar}
 \left[\partial_{k_y}m(\bk)+b\frac{m(\bk)}{\mathcal E}\partial_{k_y}d_{0,i}\right],\\
 \mathfrak m_{y,b}&=-\frac{e\ell}{2\hbar}
 \left[\partial_{k_x}m(\bk)+b\frac{m(\bk)}{\mathcal E}\partial_{k_x}d_{0,i}\right].
 \label{eq:tb_band_moment}
\end{align}
For a completely filled lower band, the integral of $\nabla_{\bk}m(\bk)$ over the periodic mBZ vanishes. Consequently,
\begin{equation}
 \mathbf M_{\parallel,i}=
 \frac{e\ell}{2}\int_{\rm mBZ}\frac{d^2k}{(2\pi)^2}
 \frac{m(\bk)}{\mathcal E(\bk)}\,
 \hat{\mathbf z}\times\delta\mathbf v_i(\bk).
 \label{eq:tb_magnetization_velocity}
\end{equation}
For the representative $x$-tilted domain this becomes
\begin{equation}
 M_y=\frac{e\ell t}{2\hbar}\int_{\rm mBZ}\frac{d^2k}{(2\pi)^2}
 \frac{m(\bk)}{\mathcal E(\bk)}\partial_{k_x}F_x(\bk),\qquad M_x=0.
 \label{eq:tb_my}
\end{equation}
The vanishing of $M_x$ follows from $k_y$ parity. The $M_y$ integrand is even and need not cancel. The magnetization is linear in the signed tilt and in $\ell$ at fixed remaining parameters, and it vanishes when either is zero.

All magnetizations in this section are defined per unit area, as in the main text. The in-plane moment per moir\'e cell is $\mathbf m_{{\rm cell},\parallel,i}=\Omega_0\mathbf M_{\parallel,i}$. Its $y$ component corresponds to the per-cell orbital moment reported in the Hartree--Fock subsection; at $\nu=1$, it is also the moment per added carrier.

The finite orbital moment does not imply a net equilibrium transport current. For a filled band,
\begin{equation}
 \mathbf j_{\rm tot}=-\frac{e}{\hbar}\int_{\rm mBZ}\frac{d^2k}{(2\pi)^2}
 \nabla_{\bk}\varepsilon_{-,i}=0.
\end{equation}
Nevertheless, defining layer current densities with $P_{1,2}=(\sigma_0\pm\sigma_z)/2$ as
$j_{a,x}=-e\int_{\rm mBZ}d^2k/(2\pi)^2\,\langle\{P_a,\hat v_x\}/2\rangle_-$, one obtains
$M_y=\ell(j_{1,x}-j_{2,x})/2$ for $z_1=+\ell/2$. Layer counterflow can therefore generate an orbital moment, with return currents closing the loops at boundaries. Changing the height origin adds a term proportional to the total current and leaves the equilibrium magnetization unchanged.

\begin{figure}[tb]
 \centering
 \includegraphics[width=0.95\textwidth]{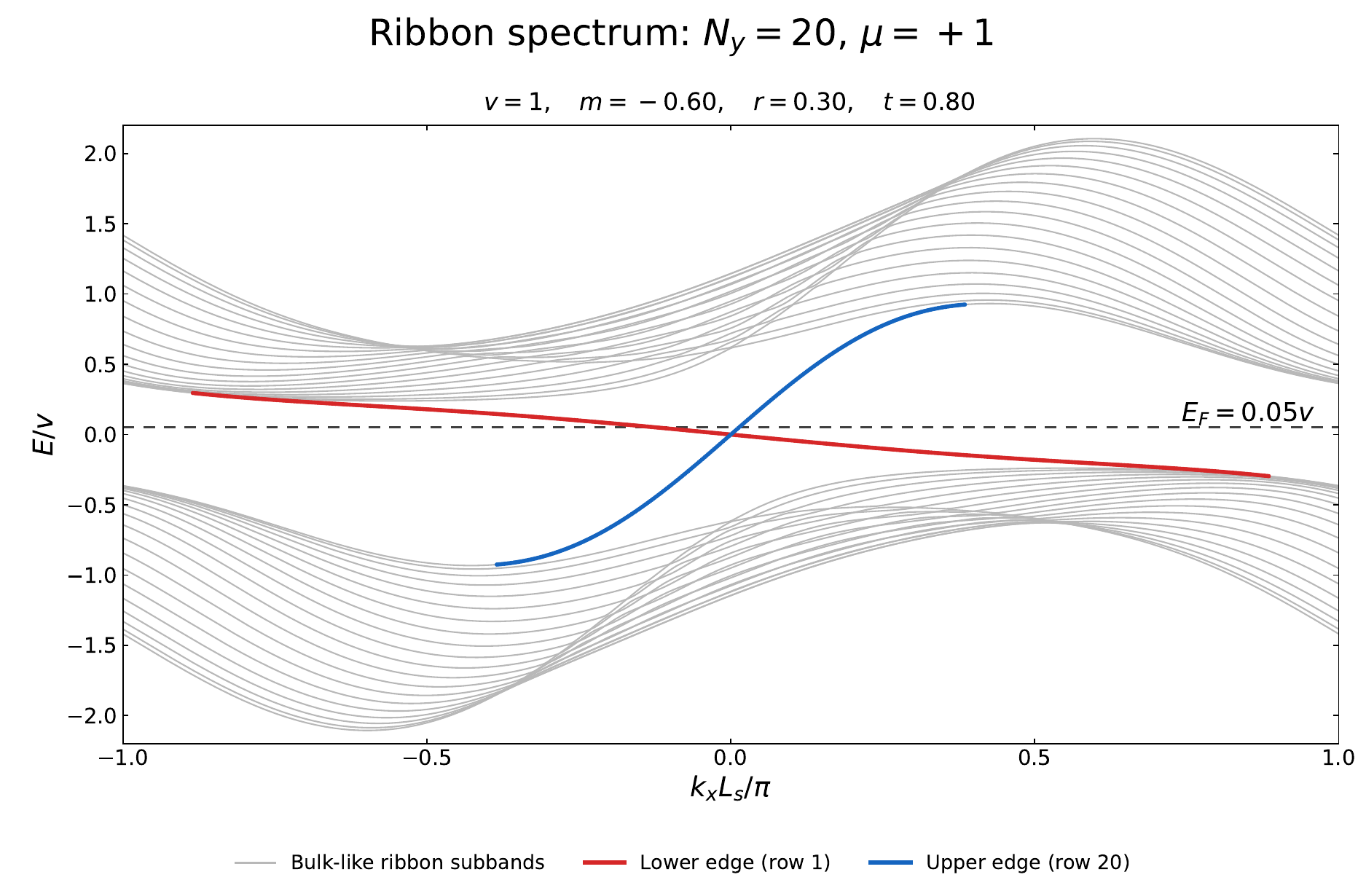}
 \caption{\textbf{Ribbon spectrum of the two-band model.} The ribbon has $N_y=20$ rows, valley $\mu=+1$, and domain axis $\hat{\mathbf n}_i=\hat{\mathbf x}$. The hybridization amplitude $v$ sets the energy unit, with $t/v=0.80$, $m/v=-0.60$, and $r/v=0.30$. Gray curves denote bulk-like subbands; red and blue curves highlight states within the projected bulk gap, assigned to the edge with the larger weight on row 1 or row 20, respectively. The dashed line marks $E_F/v=0.05$. The opposite edges show counterpropagating branches crossing the bulk gap, with markedly different dispersions in the selected domain.}
 \label{fig:tb_ribbon}
\end{figure}

As a numerical consistency check, take $L_s=1$ and an energy unit $E_0$, with $v=r=E_0$, $m=-0.30E_0$, $t=0.15\mu E_0$, and $E_F=0.10E_0$. These are time-reversed parameter sets when $\mu$ is reversed. Direct integration of Eq.~\eqref{eq:tb_curvature} on $60\times60$, $120\times120$, and $180\times180$ periodic meshes gives $C_-=\mu$. The moment per cell obeys
\begin{equation}
 \frac{m_{{\rm cell},y}}{e\ell L_sE_0/(2\hbar)}=-0.0140656454\,\mu.
\end{equation}
On the $120\times120$ mesh, a centered finite-field difference of the filled-band energy using Eq.~\eqref{eq:tb_peierls}, with $e\ell B_yL_s/(2\hbar)=\pm10^{-5}$, agrees with Eq.~\eqref{eq:tb_my} to within $3\times10^{-11}$ in these normalized units. For the parameters of Fig.~\ref{fig:tb_ribbon}, a $240\times240$ bulk mesh gives $C_-=1$ and $\Delta_{\rm ind}\simeq0.46762v$; the displayed $E_F=0.05v$ lies inside this gap. These checks concern the effective model and do not constitute a fit to the multilayer Hartree--Fock states.

\subsection{Out-of-plane orbital magnetization}
For completeness, the wave-packet self-rotation moment and the Berry-phase contribution must both be included in the out-of-plane magnetization~\cite{SI-xiao_rmp2010}. In the present two-band model,
\begin{equation}
 \mathfrak m_{b,z}=\frac{e}{\hbar}(\varepsilon_{b,i}-d_{0,i})\Omega_{b,z},\qquad
 M_z=\sum_b\int_{\rm mBZ}\frac{d^2k}{(2\pi)^2}f_b
 \left[\mathfrak m_{b,z}+\frac{e}{\hbar}(E_F-\varepsilon_{b,i})\Omega_{b,z}\right]
 \label{eq:tb_mz_general}
\end{equation}
at zero temperature. If only the lower band is filled, the two terms combine to give
\begin{equation}
 M_z=\frac{e}{\hbar}\int_{\rm mBZ}\frac{d^2k}{(2\pi)^2}
 [E_F-d_{0,i}(\bk)]\Omega_{-,z}(\bk)=\frac{eE_F}{h}C_-.
 \label{eq:tb_mz}
\end{equation}
The last equality uses the even parity of $\Omega_{-,z}$ and the odd parity of the tilt. The energy origin in this expression is the center of the two-band Hamiltonian in Eq.~\eqref{eq:tb_hamiltonian}. At its special symmetric point $E_F=0$, $M_z$ vanishes even when $C_-\ne0$; a Chern number alone does not require a nonzero magnetization at every chemical potential. For generic $E_F$ in the gap, the model supports simultaneous in-plane and out-of-plane orbital magnetizations.

\subsection{Ribbon spectrum and chiral edge states}
The bulk-boundary correspondence can be illustrated by a ribbon that is periodic along $\boldsymbol\delta_1$ and open in the transverse direction. Let $n=1,\ldots,N_y$ label the rows, with positions $\mathbf R_{p,n}=p\boldsymbol\delta_1+(n-1)\boldsymbol\delta_2$, and retain two layer orbitals per row. A real-space representation of the $x$-tilted Hamiltonian is
\begin{equation}
 H=\sum_{\mathbf R}c_{\mathbf R}^\dagger(m+3r)\sigma_z c_{\mathbf R}
 +\sum_{\mathbf R,j}\left(c_{\mathbf R}^\dagger T_jc_{\mathbf R+\boldsymbol\delta_j}+\mathrm{h.c.}\right),\qquad
 T_j=-\frac r2\sigma_z+\frac{1}{3i}
 \left[t\hat\delta_{j,x}\sigma_0-\mu v\hat\delta_{j,x}\sigma_x+v\hat\delta_{j,y}\sigma_y\right],
 \label{eq:tb_real_space}
\end{equation}
where $\hat{\boldsymbol\delta}_j=\boldsymbol\delta_j/L_s$, and the Fourier convention is $c_{\mathbf R}=N_s^{-1/2}\sum_{\bk}e^{i\bk\cdot\mathbf R}c_{\bk}$. Removing all hoppings whose endpoints lie outside the strip gives a $2N_y\times2N_y$ Hermitian matrix at each $k_x$.

For a normalized ribbon eigenstate $\psi_{\lambda,k_x}$, its row weight is
\begin{equation}
 P_n(\lambda,k_x)=\sum_{a=1}^2|\psi_{\lambda,k_x}(n,a)|^2,\qquad
 \sum_{n=1}^{N_y}P_n=1.
\end{equation}
This distinguishes the two boundary-localized branches from bulk-like subbands. In the $|C_-|=1$ regime, one net chiral branch appears at each boundary, with opposite propagation directions on the two opposite edges. Tilt removes any requirement that their dispersions be related by $k_x\mapsto-k_x$ within one domain. Time reversal reverses both the Chern number and the edge propagation directions.

\subsection{Finite-momentum particle--hole excitations and field response}
The noninteracting particle--hole energy for an electron promoted from the lower band at $\bk$ to the upper band at $\bk+\mathbf Q$ is
\begin{equation}
 E^{\rm ph}_i(\bk,\mathbf Q)=\varepsilon_{+,i}(\bk+\mathbf Q)-\varepsilon_{-,i}(\bk)
 =\mathcal E(\bk+\mathbf Q)+\mathcal E(\bk)
 +t\hat{\mathbf n}_i\cdot[\mathbf F(\bk+\mathbf Q)-\mathbf F(\bk)].
 \label{eq:tb_particle_hole}
\end{equation}
The continuum threshold is $E_{\rm cont}(\mathbf Q)=\min_{\bk}E^{\rm ph}_i(\bk,\mathbf Q)$. At $\mathbf Q=0$ the tilt cancels and the threshold is the minimum direct gap. At finite momentum it contributes already at first order,
\begin{equation}
 d_{0,i}(\bk+\mathbf Q)-d_{0,i}(\bk)
 =\hbar\mathbf Q\cdot\delta\mathbf v_i(\bk)+O(Q^2).
\end{equation}
Tilt can move the conduction-band minimum and valence-band maximum to different momenta. In fact,
\begin{equation*}
 \min_{\mathbf Q}E_{\rm cont}(\mathbf Q)=\Delta_{\rm ind}.
\end{equation*}
Thus a finite-momentum continuum minimum can occur while the system remains insulating.

An interacting collective mode additionally depends on the particle--hole kernel. In the resonant particle--hole subspace, a schematic projected equation has the form
\begin{equation}
 \sum_{\bk'}\left[E^{\rm ph}_i(\bk,\mathbf Q)\delta_{\bk\bk'}
 +\mathcal K_i(\bk,\bk';\mathbf Q)\right]\Phi_i^{\nu}(\bk';\mathbf Q)
 =\Omega_{\nu,i}(\mathbf Q)\Phi_i^{\nu}(\bk;\mathbf Q),
 \label{eq:tb_collective}
\end{equation}
where $\nu$ labels the collective branch and $\mathcal K_i$ includes the interaction and orbital form factors. An attractive channel can bind a mode below the continuum and drive its energy to zero at finite $\mathbf Q$ while $\Delta_{\rm ind}>0$. The two-band model therefore supplies a kinematic explanation for finite-momentum softening; it does not by itself determine the critical interaction, the ordering wavevector, or the spin-channel spectrum of the full Hartree--Fock state.

The same Peierls coupling fixes the linear orbital response to an in-plane field. For the $x$-tilted domain,
\begin{equation}
 h_{\mu x}(\bk,B_y)=h_{\mu x}(\bk,0)
 +\frac{e\ell B_y}{2\hbar}
 \left[(\partial_{k_x}m(\bk))\sigma_0+t(\partial_{k_x}F_x)\sigma_z\right]+O(B_y^2).
 \label{eq:tb_field_response}
\end{equation}
The corresponding band shift is $\delta\varepsilon_b=-\mathfrak m_{y,b}B_y$, so the change in an unbound particle--hole energy is
\begin{equation}
 \delta E^{\rm ph}_i(\bk,\mathbf Q)
 =-B_y[\mathfrak m_{y,+}(\bk+\mathbf Q)-\mathfrak m_{y,-}(\bk)].
\end{equation}
This establishes a coupling of the excitation spectrum to the in-plane orbital field, although the sign of a collective-mode shift also depends on its wavefunction and the field dependence of $\mathcal K_i$. It does not imply universal hardening for every domain. Separately, the free-energy bias $-\Omega_0\mathbf M_{\parallel,i}\cdot\mathbf B_\parallel$ selects between orbital domains. Time-reversed domains have opposite orbital magnetizations and opposite Chern numbers; a field-induced domain switch can therefore reverse the Hall and edge chiralities. Merely reversing a weak field within a fixed metastable domain need not switch its topology.

\clearpage

\twocolumngrid
\def\bibsection{\section*{Supplementary References}}

% Independent Supplementary references, embedded in this file.
\begingroup
\makeatletter
\let\thebibliography\NAT@thebibliography
\let\endthebibliography\endNAT@thebibliography
\makeatother
\makeatother
\endgroup


\begin{thebibliography}{56}%
\makeatletter
\providecommand \@ifxundefined [1]{%
 \@ifx{#1\undefined}
}%
\providecommand \@ifnum [1]{%
 \ifnum #1\expandafter \@firstoftwo
 \else \expandafter \@secondoftwo
 \fi
}%
\providecommand \@ifx [1]{%
 \ifx #1\expandafter \@firstoftwo
 \else \expandafter \@secondoftwo
 \fi
}%
\providecommand \natexlab [1]{#1}%
\providecommand \enquote  [1]{``#1''}%
\providecommand \bibnamefont  [1]{#1}%
\providecommand \bibfnamefont [1]{#1}%
\providecommand \citenamefont [1]{#1}%
\providecommand \href@noop [0]{\@secondoftwo}%
\providecommand \href [0]{\begingroup \@sanitize@url \@href}%
\providecommand \@href[1]{\@@startlink{#1}\@@href}%
\providecommand \@@href[1]{\endgroup#1\@@endlink}%
\providecommand \@sanitize@url [0]{\catcode `\\12\catcode `\$12\catcode
  `\&12\catcode `\#12\catcode `\^12\catcode `\_12\catcode `\%12\relax}%
\providecommand \@@startlink[1]{}%
\providecommand \@@endlink[0]{}%
\providecommand \url  [0]{\begingroup\@sanitize@url \@url }%
\providecommand \@url [1]{\endgroup\@href {#1}{\urlprefix }}%
\providecommand \urlprefix  [0]{URL }%
\providecommand \Eprint [0]{\href }%
\providecommand \doibase [0]{https://doi.org/}%
\providecommand \selectlanguage [0]{\@gobble}%
\providecommand \bibinfo  [0]{\@secondoftwo}%
\providecommand \bibfield  [0]{\@secondoftwo}%
\providecommand \translation [1]{[#1]}%
\providecommand \BibitemOpen [0]{}%
\providecommand \bibitemStop [0]{}%
\providecommand \bibitemNoStop [0]{.\EOS\space}%
\providecommand \EOS [0]{\spacefactor3000\relax}%
\providecommand \BibitemShut  [1]{\csname bibitem#1\endcsname}%
\let\auto@bib@innerbib\@empty
%</preamble>
\bibitem [{\citenamefont {Min}\ and\ \citenamefont
  {MacDonald}(2008)}]{min-chiral-prb08}%
  \BibitemOpen
  \bibfield  {author} {\bibinfo {author} {\bibfnamefont {H.}~\bibnamefont
  {Min}}\ and\ \bibinfo {author} {\bibfnamefont {A.~H.}\ \bibnamefont
  {MacDonald}},\ }\href {https://doi.org/10.1103/PhysRevB.77.155416} {\bibfield
   {journal} {\bibinfo  {journal} {Phys. Rev. B}\ }\textbf {\bibinfo {volume}
  {77}},\ \bibinfo {pages} {155416} (\bibinfo {year} {2008})}\BibitemShut
  {NoStop}%
\bibitem [{\citenamefont {Koshino}\ and\ \citenamefont
  {McCann}(2009)}]{koshino-abc-prb09}%
  \BibitemOpen
  \bibfield  {author} {\bibinfo {author} {\bibfnamefont {M.}~\bibnamefont
  {Koshino}}\ and\ \bibinfo {author} {\bibfnamefont {E.}~\bibnamefont
  {McCann}},\ }\href {https://doi.org/10.1103/PhysRevB.80.165409} {\bibfield
  {journal} {\bibinfo  {journal} {Phys. Rev. B}\ }\textbf {\bibinfo {volume}
  {80}},\ \bibinfo {pages} {165409} (\bibinfo {year} {2009})}\BibitemShut
  {NoStop}%
\bibitem [{\citenamefont {Zhang}\ \emph {et~al.}(2010)\citenamefont {Zhang},
  \citenamefont {Sahu}, \citenamefont {Min},\ and\ \citenamefont
  {MacDonald}}]{zhang-abc-prb10}%
  \BibitemOpen
  \bibfield  {author} {\bibinfo {author} {\bibfnamefont {F.}~\bibnamefont
  {Zhang}}, \bibinfo {author} {\bibfnamefont {B.}~\bibnamefont {Sahu}},
  \bibinfo {author} {\bibfnamefont {H.}~\bibnamefont {Min}},\ and\ \bibinfo
  {author} {\bibfnamefont {A.~H.}\ \bibnamefont {MacDonald}},\ }\href
  {https://doi.org/10.1103/PhysRevB.82.035409} {\bibfield  {journal} {\bibinfo
  {journal} {Phys. Rev. B}\ }\textbf {\bibinfo {volume} {82}},\ \bibinfo
  {pages} {035409} (\bibinfo {year} {2010})}\BibitemShut {NoStop}%
\bibitem [{\citenamefont {Pierucci}\ \emph {et~al.}(2015)\citenamefont
  {Pierucci}, \citenamefont {Sediri}, \citenamefont {Hajlaoui}, \citenamefont
  {Girard}, \citenamefont {Brumme}, \citenamefont {Calandra}, \citenamefont
  {Velez-Fort}, \citenamefont {Patriarche}, \citenamefont {Silly},
  \citenamefont {Ferro}, \citenamefont {Souliere}, \citenamefont {Marangolo},
  \citenamefont {Sirotti}, \citenamefont {Mauri},\ and\ \citenamefont
  {Ouerghi}}]{pierucci-flat-acsnano15}%
  \BibitemOpen
  \bibfield  {author} {\bibinfo {author} {\bibfnamefont {D.}~\bibnamefont
  {Pierucci}}, \bibinfo {author} {\bibfnamefont {H.}~\bibnamefont {Sediri}},
  \bibinfo {author} {\bibfnamefont {M.}~\bibnamefont {Hajlaoui}}, \bibinfo
  {author} {\bibfnamefont {J.-C.}\ \bibnamefont {Girard}}, \bibinfo {author}
  {\bibfnamefont {T.}~\bibnamefont {Brumme}}, \bibinfo {author} {\bibfnamefont
  {M.}~\bibnamefont {Calandra}}, \bibinfo {author} {\bibfnamefont
  {E.}~\bibnamefont {Velez-Fort}}, \bibinfo {author} {\bibfnamefont
  {G.}~\bibnamefont {Patriarche}}, \bibinfo {author} {\bibfnamefont {M.~G.}\
  \bibnamefont {Silly}}, \bibinfo {author} {\bibfnamefont {G.}~\bibnamefont
  {Ferro}}, \bibinfo {author} {\bibfnamefont {V.}~\bibnamefont {Souliere}},
  \bibinfo {author} {\bibfnamefont {M.}~\bibnamefont {Marangolo}}, \bibinfo
  {author} {\bibfnamefont {F.}~\bibnamefont {Sirotti}}, \bibinfo {author}
  {\bibfnamefont {F.}~\bibnamefont {Mauri}},\ and\ \bibinfo {author}
  {\bibfnamefont {A.}~\bibnamefont {Ouerghi}},\ }\href
  {https://doi.org/10.1021/acsnano.5b01239} {\bibfield  {journal} {\bibinfo
  {journal} {ACS Nano}\ }\textbf {\bibinfo {volume} {9}},\ \bibinfo {pages}
  {5432} (\bibinfo {year} {2015})}\BibitemShut {NoStop}%
\bibitem [{\citenamefont {Zhou}\ \emph
  {et~al.}(2021{\natexlab{a}})\citenamefont {Zhou}, \citenamefont {Xie},
  \citenamefont {Ghazaryan}, \citenamefont {Holder}, \citenamefont {Ehrets},
  \citenamefont {Spanton}, \citenamefont {Taniguchi}, \citenamefont {Watanabe},
  \citenamefont {Berg}, \citenamefont {Serbyn} \emph
  {et~al.}}]{zhou-halfmetal_R3G-nature-2021}%
  \BibitemOpen
  \bibfield  {author} {\bibinfo {author} {\bibfnamefont {H.}~\bibnamefont
  {Zhou}}, \bibinfo {author} {\bibfnamefont {T.}~\bibnamefont {Xie}}, \bibinfo
  {author} {\bibfnamefont {A.}~\bibnamefont {Ghazaryan}}, \bibinfo {author}
  {\bibfnamefont {T.}~\bibnamefont {Holder}}, \bibinfo {author} {\bibfnamefont
  {J.~R.}\ \bibnamefont {Ehrets}}, \bibinfo {author} {\bibfnamefont {E.~M.}\
  \bibnamefont {Spanton}}, \bibinfo {author} {\bibfnamefont {T.}~\bibnamefont
  {Taniguchi}}, \bibinfo {author} {\bibfnamefont {K.}~\bibnamefont {Watanabe}},
  \bibinfo {author} {\bibfnamefont {E.}~\bibnamefont {Berg}}, \bibinfo {author}
  {\bibfnamefont {M.}~\bibnamefont {Serbyn}}, \emph {et~al.},\ }\href@noop {}
  {\bibfield  {journal} {\bibinfo  {journal} {Nature}\ }\textbf {\bibinfo
  {volume} {598}},\ \bibinfo {pages} {429} (\bibinfo {year}
  {2021}{\natexlab{a}})}\BibitemShut {NoStop}%
\bibitem [{\citenamefont {Liu}\ \emph {et~al.}(2024)\citenamefont {Liu},
  \citenamefont {Zheng}, \citenamefont {Sha}, \citenamefont {Lyu},
  \citenamefont {Li}, \citenamefont {Park}, \citenamefont {Ren}, \citenamefont
  {Watanabe}, \citenamefont {Taniguchi}, \citenamefont {Jia}, \citenamefont
  {Luo}, \citenamefont {Shi}, \citenamefont {Jung},\ and\ \citenamefont
  {Chen}}]{liu-symbreakR4G-natnano-2024}%
  \BibitemOpen
  \bibfield  {author} {\bibinfo {author} {\bibfnamefont {K.}~\bibnamefont
  {Liu}}, \bibinfo {author} {\bibfnamefont {J.}~\bibnamefont {Zheng}}, \bibinfo
  {author} {\bibfnamefont {Y.}~\bibnamefont {Sha}}, \bibinfo {author}
  {\bibfnamefont {B.}~\bibnamefont {Lyu}}, \bibinfo {author} {\bibfnamefont
  {F.}~\bibnamefont {Li}}, \bibinfo {author} {\bibfnamefont {Y.}~\bibnamefont
  {Park}}, \bibinfo {author} {\bibfnamefont {Y.}~\bibnamefont {Ren}}, \bibinfo
  {author} {\bibfnamefont {K.}~\bibnamefont {Watanabe}}, \bibinfo {author}
  {\bibfnamefont {T.}~\bibnamefont {Taniguchi}}, \bibinfo {author}
  {\bibfnamefont {J.}~\bibnamefont {Jia}}, \bibinfo {author} {\bibfnamefont
  {W.}~\bibnamefont {Luo}}, \bibinfo {author} {\bibfnamefont {Z.}~\bibnamefont
  {Shi}}, \bibinfo {author} {\bibfnamefont {J.}~\bibnamefont {Jung}},\ and\
  \bibinfo {author} {\bibfnamefont {G.}~\bibnamefont {Chen}},\ }\href@noop {}
  {\bibfield  {journal} {\bibinfo  {journal} {Nature nanotechnology}\ }\textbf
  {\bibinfo {volume} {19}},\ \bibinfo {pages} {188} (\bibinfo {year}
  {2024})}\BibitemShut {NoStop}%
\bibitem [{\citenamefont {Arp}\ \emph {et~al.}(2024)\citenamefont {Arp},
  \citenamefont {Sheekey}, \citenamefont {Zhou}, \citenamefont {Tschirhart},
  \citenamefont {Patterson}, \citenamefont {Yoo}, \citenamefont {Holleis},
  \citenamefont {Redekop}, \citenamefont {Babikyan}, \citenamefont {Xie},
  \citenamefont {Xiao}, \citenamefont {Vituri}, \citenamefont {Holder},
  \citenamefont {Taniguchi}, \citenamefont {Watanabe}, \citenamefont {Huber},
  \citenamefont {Berg},\ and\ \citenamefont {Young}}]{arp-ivc-natphys24}%
  \BibitemOpen
  \bibfield  {author} {\bibinfo {author} {\bibfnamefont {T.}~\bibnamefont
  {Arp}}, \bibinfo {author} {\bibfnamefont {O.}~\bibnamefont {Sheekey}},
  \bibinfo {author} {\bibfnamefont {H.}~\bibnamefont {Zhou}}, \bibinfo {author}
  {\bibfnamefont {C.~L.}\ \bibnamefont {Tschirhart}}, \bibinfo {author}
  {\bibfnamefont {C.~L.}\ \bibnamefont {Patterson}}, \bibinfo {author}
  {\bibfnamefont {H.~M.}\ \bibnamefont {Yoo}}, \bibinfo {author} {\bibfnamefont
  {L.}~\bibnamefont {Holleis}}, \bibinfo {author} {\bibfnamefont
  {E.}~\bibnamefont {Redekop}}, \bibinfo {author} {\bibfnamefont
  {G.}~\bibnamefont {Babikyan}}, \bibinfo {author} {\bibfnamefont
  {T.}~\bibnamefont {Xie}}, \bibinfo {author} {\bibfnamefont {J.}~\bibnamefont
  {Xiao}}, \bibinfo {author} {\bibfnamefont {Y.}~\bibnamefont {Vituri}},
  \bibinfo {author} {\bibfnamefont {T.}~\bibnamefont {Holder}}, \bibinfo
  {author} {\bibfnamefont {T.}~\bibnamefont {Taniguchi}}, \bibinfo {author}
  {\bibfnamefont {K.}~\bibnamefont {Watanabe}}, \bibinfo {author}
  {\bibfnamefont {M.~E.}\ \bibnamefont {Huber}}, \bibinfo {author}
  {\bibfnamefont {E.}~\bibnamefont {Berg}},\ and\ \bibinfo {author}
  {\bibfnamefont {A.~F.}\ \bibnamefont {Young}},\ }\href
  {https://doi.org/10.1038/s41567-024-02560-7} {\bibfield  {journal} {\bibinfo
  {journal} {Nat. Phys.}\ }\textbf {\bibinfo {volume} {20}},\ \bibinfo {pages}
  {1413} (\bibinfo {year} {2024})}\BibitemShut {NoStop}%
\bibitem [{\citenamefont {Huang}\ \emph {et~al.}(2023)\citenamefont {Huang},
  \citenamefont {Wolf}, \citenamefont {Qin}, \citenamefont {Wei}, \citenamefont
  {Blinov},\ and\ \citenamefont {MacDonald}}]{huang-spin-orbital-prb23}%
  \BibitemOpen
  \bibfield  {author} {\bibinfo {author} {\bibfnamefont {C.}~\bibnamefont
  {Huang}}, \bibinfo {author} {\bibfnamefont {T.~M.~R.}\ \bibnamefont {Wolf}},
  \bibinfo {author} {\bibfnamefont {W.}~\bibnamefont {Qin}}, \bibinfo {author}
  {\bibfnamefont {N.}~\bibnamefont {Wei}}, \bibinfo {author} {\bibfnamefont
  {I.~V.}\ \bibnamefont {Blinov}},\ and\ \bibinfo {author} {\bibfnamefont
  {A.~H.}\ \bibnamefont {MacDonald}},\ }\href
  {https://doi.org/10.1103/physrevb.107.l121405} {\bibfield  {journal}
  {\bibinfo  {journal} {Phys. Rev. B}\ }\textbf {\bibinfo {volume} {107}},\
  \bibinfo {pages} {L121405} (\bibinfo {year} {2023})}\BibitemShut {NoStop}%
\bibitem [{\citenamefont {Jang}\ \emph {et~al.}(2023)\citenamefont {Jang},
  \citenamefont {Park}, \citenamefont {Jung},\ and\ \citenamefont
  {Min}}]{kpgra}%
  \BibitemOpen
  \bibfield  {author} {\bibinfo {author} {\bibfnamefont {Y.}~\bibnamefont
  {Jang}}, \bibinfo {author} {\bibfnamefont {Y.}~\bibnamefont {Park}}, \bibinfo
  {author} {\bibfnamefont {J.}~\bibnamefont {Jung}},\ and\ \bibinfo {author}
  {\bibfnamefont {H.}~\bibnamefont {Min}},\ }\href
  {https://doi.org/10.1103/PhysRevB.108.L041101} {\bibfield  {journal}
  {\bibinfo  {journal} {Phys. Rev. B}\ }\textbf {\bibinfo {volume} {108}},\
  \bibinfo {pages} {L041101} (\bibinfo {year} {2023})}\BibitemShut {NoStop}%
\bibitem [{\citenamefont {Zhou}\ \emph
  {et~al.}(2021{\natexlab{b}})\citenamefont {Zhou}, \citenamefont {Xie},
  \citenamefont {Taniguchi}, \citenamefont {Watanabe},\ and\ \citenamefont
  {Young}}]{zhou-superconductivity_R3G-nature-2021}%
  \BibitemOpen
  \bibfield  {author} {\bibinfo {author} {\bibfnamefont {H.}~\bibnamefont
  {Zhou}}, \bibinfo {author} {\bibfnamefont {T.}~\bibnamefont {Xie}}, \bibinfo
  {author} {\bibfnamefont {T.}~\bibnamefont {Taniguchi}}, \bibinfo {author}
  {\bibfnamefont {K.}~\bibnamefont {Watanabe}},\ and\ \bibinfo {author}
  {\bibfnamefont {A.~F.}\ \bibnamefont {Young}},\ }\href
  {https://doi.org/10.1038/s41586-021-03926-0} {\bibfield  {journal} {\bibinfo
  {journal} {Nature}\ }\textbf {\bibinfo {volume} {598}},\ \bibinfo {pages}
  {434} (\bibinfo {year} {2021}{\natexlab{b}})}\BibitemShut {NoStop}%
\bibitem [{\citenamefont {Han}\ \emph {et~al.}(2025)\citenamefont {Han},
  \citenamefont {Lu}, \citenamefont {Hadjri}, \citenamefont {Shi},
  \citenamefont {Wu}, \citenamefont {Xu}, \citenamefont {Yao}, \citenamefont
  {Cotten}, \citenamefont {Sharifi~Sedeh}, \citenamefont {Weldeyesus},
  \citenamefont {Yang}, \citenamefont {Seo}, \citenamefont {Ye}, \citenamefont
  {Zhou}, \citenamefont {Liu}, \citenamefont {Shi}, \citenamefont {Hua},
  \citenamefont {Watanabe}, \citenamefont {Taniguchi}, \citenamefont {Xiong},
  \citenamefont {Zumb{\"u}hl}, \citenamefont {Fu},\ and\ \citenamefont
  {Ju}}]{han-chiral-sc-nature25}%
  \BibitemOpen
  \bibfield  {author} {\bibinfo {author} {\bibfnamefont {T.}~\bibnamefont
  {Han}}, \bibinfo {author} {\bibfnamefont {Z.}~\bibnamefont {Lu}}, \bibinfo
  {author} {\bibfnamefont {Z.}~\bibnamefont {Hadjri}}, \bibinfo {author}
  {\bibfnamefont {L.}~\bibnamefont {Shi}}, \bibinfo {author} {\bibfnamefont
  {Z.}~\bibnamefont {Wu}}, \bibinfo {author} {\bibfnamefont {W.}~\bibnamefont
  {Xu}}, \bibinfo {author} {\bibfnamefont {Y.}~\bibnamefont {Yao}}, \bibinfo
  {author} {\bibfnamefont {A.~A.}\ \bibnamefont {Cotten}}, \bibinfo {author}
  {\bibfnamefont {O.}~\bibnamefont {Sharifi~Sedeh}}, \bibinfo {author}
  {\bibfnamefont {H.}~\bibnamefont {Weldeyesus}}, \bibinfo {author}
  {\bibfnamefont {J.}~\bibnamefont {Yang}}, \bibinfo {author} {\bibfnamefont
  {J.}~\bibnamefont {Seo}}, \bibinfo {author} {\bibfnamefont {S.}~\bibnamefont
  {Ye}}, \bibinfo {author} {\bibfnamefont {M.}~\bibnamefont {Zhou}}, \bibinfo
  {author} {\bibfnamefont {H.}~\bibnamefont {Liu}}, \bibinfo {author}
  {\bibfnamefont {G.}~\bibnamefont {Shi}}, \bibinfo {author} {\bibfnamefont
  {Z.}~\bibnamefont {Hua}}, \bibinfo {author} {\bibfnamefont {K.}~\bibnamefont
  {Watanabe}}, \bibinfo {author} {\bibfnamefont {T.}~\bibnamefont {Taniguchi}},
  \bibinfo {author} {\bibfnamefont {P.}~\bibnamefont {Xiong}}, \bibinfo
  {author} {\bibfnamefont {D.~M.}\ \bibnamefont {Zumb{\"u}hl}}, \bibinfo
  {author} {\bibfnamefont {L.}~\bibnamefont {Fu}},\ and\ \bibinfo {author}
  {\bibfnamefont {L.}~\bibnamefont {Ju}},\ }\href
  {https://doi.org/10.1038/s41586-025-09169-7} {\bibfield  {journal} {\bibinfo
  {journal} {Nature}\ }\textbf {\bibinfo {volume} {643}},\ \bibinfo {pages}
  {654} (\bibinfo {year} {2025})}\BibitemShut {NoStop}%
\bibitem [{\citenamefont {Ghazaryan}\ \emph {et~al.}(2021)\citenamefont
  {Ghazaryan}, \citenamefont {Holder}, \citenamefont {Serbyn},\ and\
  \citenamefont {Berg}}]{ghazaryan-annular-sc-prl21}%
  \BibitemOpen
  \bibfield  {author} {\bibinfo {author} {\bibfnamefont {A.}~\bibnamefont
  {Ghazaryan}}, \bibinfo {author} {\bibfnamefont {T.}~\bibnamefont {Holder}},
  \bibinfo {author} {\bibfnamefont {M.}~\bibnamefont {Serbyn}},\ and\ \bibinfo
  {author} {\bibfnamefont {E.}~\bibnamefont {Berg}},\ }\href
  {https://doi.org/10.1103/PhysRevLett.127.247001} {\bibfield  {journal}
  {\bibinfo  {journal} {Phys. Rev. Lett.}\ }\textbf {\bibinfo {volume} {127}},\
  \bibinfo {pages} {247001} (\bibinfo {year} {2021})}\BibitemShut {NoStop}%
\bibitem [{\citenamefont {Qin}\ \emph {et~al.}(2023)\citenamefont {Qin},
  \citenamefont {Huang}, \citenamefont {Wolf}, \citenamefont {Wei},
  \citenamefont {Blinov},\ and\ \citenamefont {MacDonald}}]{qin-frg-sc-prl23}%
  \BibitemOpen
  \bibfield  {author} {\bibinfo {author} {\bibfnamefont {W.}~\bibnamefont
  {Qin}}, \bibinfo {author} {\bibfnamefont {C.}~\bibnamefont {Huang}}, \bibinfo
  {author} {\bibfnamefont {T.}~\bibnamefont {Wolf}}, \bibinfo {author}
  {\bibfnamefont {N.}~\bibnamefont {Wei}}, \bibinfo {author} {\bibfnamefont
  {I.}~\bibnamefont {Blinov}},\ and\ \bibinfo {author} {\bibfnamefont {A.~H.}\
  \bibnamefont {MacDonald}},\ }\href
  {https://doi.org/10.1103/PhysRevLett.130.146001} {\bibfield  {journal}
  {\bibinfo  {journal} {Phys. Rev. Lett.}\ }\textbf {\bibinfo {volume} {130}},\
  \bibinfo {pages} {146001} (\bibinfo {year} {2023})}\BibitemShut {NoStop}%
\bibitem [{\citenamefont {You}\ and\ \citenamefont
  {Vishwanath}(2022)}]{you-ivc-sc-prb22}%
  \BibitemOpen
  \bibfield  {author} {\bibinfo {author} {\bibfnamefont {Y.-Z.}\ \bibnamefont
  {You}}\ and\ \bibinfo {author} {\bibfnamefont {A.}~\bibnamefont
  {Vishwanath}},\ }\href {https://doi.org/10.1103/PhysRevB.105.134524}
  {\bibfield  {journal} {\bibinfo  {journal} {Phys. Rev. B}\ }\textbf {\bibinfo
  {volume} {105}},\ \bibinfo {pages} {134524} (\bibinfo {year}
  {2022})}\BibitemShut {NoStop}%
\bibitem [{\citenamefont {Patterson}\ \emph {et~al.}(2025)\citenamefont
  {Patterson}, \citenamefont {Sheekey}, \citenamefont {Arp}, \citenamefont
  {Holleis}, \citenamefont {Koh}, \citenamefont {Choi}, \citenamefont {Xie},
  \citenamefont {Xu}, \citenamefont {Guo}, \citenamefont {Stoyanov},
  \citenamefont {Redekop}, \citenamefont {Zhang}, \citenamefont {Babikyan},
  \citenamefont {Gong}, \citenamefont {Zhou}, \citenamefont {Cheng},
  \citenamefont {Taniguchi}, \citenamefont {Watanabe}, \citenamefont {Huber},
  \citenamefont {Jin}, \citenamefont {Lantagne-Hurtubise}, \citenamefont
  {Alicea},\ and\ \citenamefont {Young}}]{patterson-spin-canting-nature25}%
  \BibitemOpen
  \bibfield  {author} {\bibinfo {author} {\bibfnamefont {C.~L.}\ \bibnamefont
  {Patterson}}, \bibinfo {author} {\bibfnamefont {O.~I.}\ \bibnamefont
  {Sheekey}}, \bibinfo {author} {\bibfnamefont {T.~B.}\ \bibnamefont {Arp}},
  \bibinfo {author} {\bibfnamefont {L.~F.~W.}\ \bibnamefont {Holleis}},
  \bibinfo {author} {\bibfnamefont {J.~M.}\ \bibnamefont {Koh}}, \bibinfo
  {author} {\bibfnamefont {Y.}~\bibnamefont {Choi}}, \bibinfo {author}
  {\bibfnamefont {T.}~\bibnamefont {Xie}}, \bibinfo {author} {\bibfnamefont
  {S.}~\bibnamefont {Xu}}, \bibinfo {author} {\bibfnamefont {Y.}~\bibnamefont
  {Guo}}, \bibinfo {author} {\bibfnamefont {H.}~\bibnamefont {Stoyanov}},
  \bibinfo {author} {\bibfnamefont {E.}~\bibnamefont {Redekop}}, \bibinfo
  {author} {\bibfnamefont {C.}~\bibnamefont {Zhang}}, \bibinfo {author}
  {\bibfnamefont {G.}~\bibnamefont {Babikyan}}, \bibinfo {author}
  {\bibfnamefont {D.}~\bibnamefont {Gong}}, \bibinfo {author} {\bibfnamefont
  {H.}~\bibnamefont {Zhou}}, \bibinfo {author} {\bibfnamefont {X.}~\bibnamefont
  {Cheng}}, \bibinfo {author} {\bibfnamefont {T.}~\bibnamefont {Taniguchi}},
  \bibinfo {author} {\bibfnamefont {K.}~\bibnamefont {Watanabe}}, \bibinfo
  {author} {\bibfnamefont {M.~E.}\ \bibnamefont {Huber}}, \bibinfo {author}
  {\bibfnamefont {C.}~\bibnamefont {Jin}}, \bibinfo {author} {\bibfnamefont
  {{\'E}.}~\bibnamefont {Lantagne-Hurtubise}}, \bibinfo {author} {\bibfnamefont
  {J.}~\bibnamefont {Alicea}},\ and\ \bibinfo {author} {\bibfnamefont {A.~F.}\
  \bibnamefont {Young}},\ }\href {https://doi.org/10.1038/s41586-025-08863-w}
  {\bibfield  {journal} {\bibinfo  {journal} {Nature}\ }\textbf {\bibinfo
  {volume} {641}},\ \bibinfo {pages} {632} (\bibinfo {year}
  {2025})}\BibitemShut {NoStop}%
\bibitem [{\citenamefont {Yang}\ \emph {et~al.}(2025)\citenamefont {Yang},
  \citenamefont {Shi}, \citenamefont {Ye}, \citenamefont {Yoon}, \citenamefont
  {Lu}, \citenamefont {Kakani}, \citenamefont {Han}, \citenamefont {Seo},
  \citenamefont {Shi}, \citenamefont {Watanabe}, \citenamefont {Taniguchi},
  \citenamefont {Zhang},\ and\ \citenamefont {Ju}}]{yang-soc-sc-natmat25}%
  \BibitemOpen
  \bibfield  {author} {\bibinfo {author} {\bibfnamefont {J.}~\bibnamefont
  {Yang}}, \bibinfo {author} {\bibfnamefont {X.}~\bibnamefont {Shi}}, \bibinfo
  {author} {\bibfnamefont {S.}~\bibnamefont {Ye}}, \bibinfo {author}
  {\bibfnamefont {C.}~\bibnamefont {Yoon}}, \bibinfo {author} {\bibfnamefont
  {Z.}~\bibnamefont {Lu}}, \bibinfo {author} {\bibfnamefont {V.}~\bibnamefont
  {Kakani}}, \bibinfo {author} {\bibfnamefont {T.}~\bibnamefont {Han}},
  \bibinfo {author} {\bibfnamefont {J.}~\bibnamefont {Seo}}, \bibinfo {author}
  {\bibfnamefont {L.}~\bibnamefont {Shi}}, \bibinfo {author} {\bibfnamefont
  {K.}~\bibnamefont {Watanabe}}, \bibinfo {author} {\bibfnamefont
  {T.}~\bibnamefont {Taniguchi}}, \bibinfo {author} {\bibfnamefont
  {F.}~\bibnamefont {Zhang}},\ and\ \bibinfo {author} {\bibfnamefont
  {L.}~\bibnamefont {Ju}},\ }\href {https://doi.org/10.1038/s41563-025-02156-3}
  {\bibfield  {journal} {\bibinfo  {journal} {Nat. Mater.}\ }\textbf {\bibinfo
  {volume} {24}},\ \bibinfo {pages} {1058} (\bibinfo {year}
  {2025})}\BibitemShut {NoStop}%
\bibitem [{\citenamefont {Seo}\ \emph {et~al.}(2026)\citenamefont {Seo},
  \citenamefont {Cotten}, \citenamefont {Ye}, \citenamefont {Xu}, \citenamefont
  {Sharifi~Sedeh}, \citenamefont {Weldeyesus}, \citenamefont {Han},
  \citenamefont {Lu}, \citenamefont {Wu}, \citenamefont {Xu}, \citenamefont
  {Yang}, \citenamefont {Aitken}, \citenamefont {Liong}, \citenamefont
  {Pattanakanvijit}, \citenamefont {Hadjri}, \citenamefont {Gazizulin},
  \citenamefont {Watanabe}, \citenamefont {Taniguchi}, \citenamefont {Li},
  \citenamefont {Zumb{\"u}hl},\ and\ \citenamefont
  {Ju}}]{seo-field-boosted-sc-nature26}%
  \BibitemOpen
  \bibfield  {author} {\bibinfo {author} {\bibfnamefont {J.}~\bibnamefont
  {Seo}}, \bibinfo {author} {\bibfnamefont {A.~A.}\ \bibnamefont {Cotten}},
  \bibinfo {author} {\bibfnamefont {S.}~\bibnamefont {Ye}}, \bibinfo {author}
  {\bibfnamefont {M.}~\bibnamefont {Xu}}, \bibinfo {author} {\bibfnamefont
  {O.}~\bibnamefont {Sharifi~Sedeh}}, \bibinfo {author} {\bibfnamefont
  {H.}~\bibnamefont {Weldeyesus}}, \bibinfo {author} {\bibfnamefont
  {T.}~\bibnamefont {Han}}, \bibinfo {author} {\bibfnamefont {Z.}~\bibnamefont
  {Lu}}, \bibinfo {author} {\bibfnamefont {Z.}~\bibnamefont {Wu}}, \bibinfo
  {author} {\bibfnamefont {W.}~\bibnamefont {Xu}}, \bibinfo {author}
  {\bibfnamefont {J.}~\bibnamefont {Yang}}, \bibinfo {author} {\bibfnamefont
  {E.}~\bibnamefont {Aitken}}, \bibinfo {author} {\bibfnamefont {P.~P.}\
  \bibnamefont {Liong}}, \bibinfo {author} {\bibfnamefont {P.}~\bibnamefont
  {Pattanakanvijit}}, \bibinfo {author} {\bibfnamefont {Z.}~\bibnamefont
  {Hadjri}}, \bibinfo {author} {\bibfnamefont {R.}~\bibnamefont {Gazizulin}},
  \bibinfo {author} {\bibfnamefont {K.}~\bibnamefont {Watanabe}}, \bibinfo
  {author} {\bibfnamefont {T.}~\bibnamefont {Taniguchi}}, \bibinfo {author}
  {\bibfnamefont {M.}~\bibnamefont {Li}}, \bibinfo {author} {\bibfnamefont
  {D.~M.}\ \bibnamefont {Zumb{\"u}hl}},\ and\ \bibinfo {author} {\bibfnamefont
  {L.}~\bibnamefont {Ju}},\ }\href {https://doi.org/10.1038/s41586-026-10815-x}
  {\bibfield  {journal} {\bibinfo  {journal} {Nature}\ }\textbf {\bibinfo
  {volume} {656}},\ \bibinfo {pages} {60} (\bibinfo {year} {2026})}\BibitemShut
  {NoStop}%
\bibitem [{\citenamefont {Okounkova}\ \emph {et~al.}(2026)\citenamefont
  {Okounkova}, \citenamefont {Sohm}, \citenamefont {Faehndrich}, \citenamefont
  {Kumar}, \citenamefont {Waleffe}, \citenamefont {Yan}, \citenamefont
  {Watanabe}, \citenamefont {Taniguchi}, \citenamefont {Folk},\ and\
  \citenamefont {Yankowitz}}]{okounkova-anomalous-metal-nature26}%
  \BibitemOpen
  \bibfield  {author} {\bibinfo {author} {\bibfnamefont {A.}~\bibnamefont
  {Okounkova}}, \bibinfo {author} {\bibfnamefont {A.}~\bibnamefont {Sohm}},
  \bibinfo {author} {\bibfnamefont {T.}~\bibnamefont {Faehndrich}}, \bibinfo
  {author} {\bibfnamefont {M.}~\bibnamefont {Kumar}}, \bibinfo {author}
  {\bibfnamefont {D.}~\bibnamefont {Waleffe}}, \bibinfo {author} {\bibfnamefont
  {J.}~\bibnamefont {Yan}}, \bibinfo {author} {\bibfnamefont {K.}~\bibnamefont
  {Watanabe}}, \bibinfo {author} {\bibfnamefont {T.}~\bibnamefont {Taniguchi}},
  \bibinfo {author} {\bibfnamefont {J.}~\bibnamefont {Folk}},\ and\ \bibinfo
  {author} {\bibfnamefont {M.}~\bibnamefont {Yankowitz}},\ }\href
  {https://doi.org/10.1038/s41586-026-11033-1} {\bibfield  {journal} {\bibinfo
  {journal} {Nature}\ }\textbf {\bibinfo {volume} {658}},\ \bibinfo {pages}
  {65} (\bibinfo {year} {2026})}\BibitemShut {NoStop}%
\bibitem [{\citenamefont {Deng}\ \emph {et~al.}(2025)\citenamefont {Deng},
  \citenamefont {Xie}, \citenamefont {Li}, \citenamefont {Taniguchi},
  \citenamefont {Watanabe}, \citenamefont {Shan}, \citenamefont {Mak},\ and\
  \citenamefont {Liu}}]{deng-hexalayer-sc-arxiv25}%
  \BibitemOpen
  \bibfield  {author} {\bibinfo {author} {\bibfnamefont {J.}~\bibnamefont
  {Deng}}, \bibinfo {author} {\bibfnamefont {J.}~\bibnamefont {Xie}}, \bibinfo
  {author} {\bibfnamefont {H.}~\bibnamefont {Li}}, \bibinfo {author}
  {\bibfnamefont {T.}~\bibnamefont {Taniguchi}}, \bibinfo {author}
  {\bibfnamefont {K.}~\bibnamefont {Watanabe}}, \bibinfo {author}
  {\bibfnamefont {J.}~\bibnamefont {Shan}}, \bibinfo {author} {\bibfnamefont
  {K.~F.}\ \bibnamefont {Mak}},\ and\ \bibinfo {author} {\bibfnamefont
  {X.}~\bibnamefont {Liu}},\ }\href {https://arxiv.org/abs/2508.15909}
  {\bibinfo {title} {Superconductivity and ferroelectric orbital magnetism in
  semimetallic rhombohedral hexalayer graphene}} (\bibinfo {year} {2025}),\
  \Eprint {https://arxiv.org/abs/2508.15909} {arXiv:2508.15909
  [cond-mat.str-el]} \BibitemShut {NoStop}%
\bibitem [{\citenamefont {Han}\ \emph {et~al.}(2024)\citenamefont {Han},
  \citenamefont {Lu}, \citenamefont {Scuri}, \citenamefont {Sung},
  \citenamefont {Wang}, \citenamefont {Han}, \citenamefont {Watanabe},
  \citenamefont {Taniguchi}, \citenamefont {Park},\ and\ \citenamefont
  {Ju}}]{han-symbreakR5G-natnano-2024}%
  \BibitemOpen
  \bibfield  {author} {\bibinfo {author} {\bibfnamefont {T.}~\bibnamefont
  {Han}}, \bibinfo {author} {\bibfnamefont {Z.}~\bibnamefont {Lu}}, \bibinfo
  {author} {\bibfnamefont {G.}~\bibnamefont {Scuri}}, \bibinfo {author}
  {\bibfnamefont {J.}~\bibnamefont {Sung}}, \bibinfo {author} {\bibfnamefont
  {J.}~\bibnamefont {Wang}}, \bibinfo {author} {\bibfnamefont {T.}~\bibnamefont
  {Han}}, \bibinfo {author} {\bibfnamefont {K.}~\bibnamefont {Watanabe}},
  \bibinfo {author} {\bibfnamefont {T.}~\bibnamefont {Taniguchi}}, \bibinfo
  {author} {\bibfnamefont {H.}~\bibnamefont {Park}},\ and\ \bibinfo {author}
  {\bibfnamefont {L.}~\bibnamefont {Ju}},\ }\href@noop {} {\bibfield  {journal}
  {\bibinfo  {journal} {Nature Nanotechnology}\ }\textbf {\bibinfo {volume}
  {19}},\ \bibinfo {pages} {181} (\bibinfo {year} {2024})}\BibitemShut
  {NoStop}%
\bibitem [{\citenamefont {Winterer}\ \emph {et~al.}(2024)\citenamefont
  {Winterer}, \citenamefont {Geisenhof}, \citenamefont {Fernandez},
  \citenamefont {Seiler}, \citenamefont {Zhang},\ and\ \citenamefont
  {Weitz}}]{winterer-ferroelectric-natphys24}%
  \BibitemOpen
  \bibfield  {author} {\bibinfo {author} {\bibfnamefont {F.}~\bibnamefont
  {Winterer}}, \bibinfo {author} {\bibfnamefont {F.~R.}\ \bibnamefont
  {Geisenhof}}, \bibinfo {author} {\bibfnamefont {N.}~\bibnamefont
  {Fernandez}}, \bibinfo {author} {\bibfnamefont {A.~M.}\ \bibnamefont
  {Seiler}}, \bibinfo {author} {\bibfnamefont {F.}~\bibnamefont {Zhang}},\ and\
  \bibinfo {author} {\bibfnamefont {R.~T.}\ \bibnamefont {Weitz}},\ }\href
  {https://doi.org/10.1038/s41567-023-02327-6} {\bibfield  {journal} {\bibinfo
  {journal} {Nat. Phys.}\ }\textbf {\bibinfo {volume} {20}},\ \bibinfo {pages}
  {422} (\bibinfo {year} {2024})}\BibitemShut {NoStop}%
\bibitem [{\citenamefont {Zhang}\ \emph {et~al.}(2011)\citenamefont {Zhang},
  \citenamefont {Jung}, \citenamefont {Fiete}, \citenamefont {Niu},\ and\
  \citenamefont {MacDonald}}]{zhang-spontaneous-qh-prl11}%
  \BibitemOpen
  \bibfield  {author} {\bibinfo {author} {\bibfnamefont {F.}~\bibnamefont
  {Zhang}}, \bibinfo {author} {\bibfnamefont {J.}~\bibnamefont {Jung}},
  \bibinfo {author} {\bibfnamefont {G.~A.}\ \bibnamefont {Fiete}}, \bibinfo
  {author} {\bibfnamefont {Q.}~\bibnamefont {Niu}},\ and\ \bibinfo {author}
  {\bibfnamefont {A.~H.}\ \bibnamefont {MacDonald}},\ }\href
  {https://doi.org/10.1103/physrevlett.106.156801} {\bibfield  {journal}
  {\bibinfo  {journal} {Phys. Rev. Lett.}\ }\textbf {\bibinfo {volume} {106}},\
  \bibinfo {pages} {156801} (\bibinfo {year} {2011})}\BibitemShut {NoStop}%
\bibitem [{\citenamefont {Han}\ \emph {et~al.}(2023)\citenamefont {Han},
  \citenamefont {Lu}, \citenamefont {Scuri}, \citenamefont {Sung},
  \citenamefont {Wang}, \citenamefont {Han}, \citenamefont {Watanabe},
  \citenamefont {Taniguchi}, \citenamefont {Fu}, \citenamefont {Park},\ and\
  \citenamefont {Ju}}]{ju-orbitalmulti-nature23}%
  \BibitemOpen
  \bibfield  {author} {\bibinfo {author} {\bibfnamefont {T.}~\bibnamefont
  {Han}}, \bibinfo {author} {\bibfnamefont {Z.}~\bibnamefont {Lu}}, \bibinfo
  {author} {\bibfnamefont {G.}~\bibnamefont {Scuri}}, \bibinfo {author}
  {\bibfnamefont {J.}~\bibnamefont {Sung}}, \bibinfo {author} {\bibfnamefont
  {J.}~\bibnamefont {Wang}}, \bibinfo {author} {\bibfnamefont {T.}~\bibnamefont
  {Han}}, \bibinfo {author} {\bibfnamefont {K.}~\bibnamefont {Watanabe}},
  \bibinfo {author} {\bibfnamefont {T.}~\bibnamefont {Taniguchi}}, \bibinfo
  {author} {\bibfnamefont {L.}~\bibnamefont {Fu}}, \bibinfo {author}
  {\bibfnamefont {H.}~\bibnamefont {Park}},\ and\ \bibinfo {author}
  {\bibfnamefont {L.}~\bibnamefont {Ju}},\ }\href
  {https://doi.org/10.1038/s41586-023-06572-w} {\bibfield  {journal} {\bibinfo
  {journal} {Nature}\ }\textbf {\bibinfo {volume} {623}},\ \bibinfo {pages}
  {41} (\bibinfo {year} {2023})}\BibitemShut {NoStop}%
\bibitem [{\citenamefont {Li}\ \emph {et~al.}(2026)\citenamefont {Li},
  \citenamefont {Fan}, \citenamefont {Li}, \citenamefont {Xu}, \citenamefont
  {Song}, \citenamefont {Watanabe}, \citenamefont {Taniguchi}, \citenamefont
  {Jiang}, \citenamefont {Xie}, \citenamefont {Hone}, \citenamefont {Dean},
  \citenamefont {Zhao}, \citenamefont {Liu},\ and\ \citenamefont
  {Wang}}]{comment_exp}%
  \BibitemOpen
  \bibfield  {author} {\bibinfo {author} {\bibfnamefont {Q.}~\bibnamefont
  {Li}}, \bibinfo {author} {\bibfnamefont {H.}~\bibnamefont {Fan}}, \bibinfo
  {author} {\bibfnamefont {M.}~\bibnamefont {Li}}, \bibinfo {author}
  {\bibfnamefont {Y.}~\bibnamefont {Xu}}, \bibinfo {author} {\bibfnamefont
  {J.}~\bibnamefont {Song}}, \bibinfo {author} {\bibfnamefont {K.}~\bibnamefont
  {Watanabe}}, \bibinfo {author} {\bibfnamefont {T.}~\bibnamefont {Taniguchi}},
  \bibinfo {author} {\bibfnamefont {H.}~\bibnamefont {Jiang}}, \bibinfo
  {author} {\bibfnamefont {X.~C.}\ \bibnamefont {Xie}}, \bibinfo {author}
  {\bibfnamefont {J.}~\bibnamefont {Hone}}, \bibinfo {author} {\bibfnamefont
  {C.}~\bibnamefont {Dean}}, \bibinfo {author} {\bibfnamefont {Y.}~\bibnamefont
  {Zhao}}, \bibinfo {author} {\bibfnamefont {J.}~\bibnamefont {Liu}},\ and\
  \bibinfo {author} {\bibfnamefont {L.}~\bibnamefont {Wang}},\ }\href
  {https://doi.org/10.1038/s41586-026-10471-1} {\bibfield  {journal} {\bibinfo
  {journal} {Nature}\ }\textbf {\bibinfo {volume} {653}},\ \bibinfo {pages}
  {384} (\bibinfo {year} {2026})}\BibitemShut {NoStop}%
\bibitem [{\citenamefont {Das}\ and\ \citenamefont
  {Huang}(2025)}]{das-multiferroic-prb25}%
  \BibitemOpen
  \bibfield  {author} {\bibinfo {author} {\bibfnamefont {M.}~\bibnamefont
  {Das}}\ and\ \bibinfo {author} {\bibfnamefont {C.}~\bibnamefont {Huang}},\
  }\href {https://doi.org/10.1103/PhysRevB.111.L161106} {\bibfield  {journal}
  {\bibinfo  {journal} {Phys. Rev. B}\ }\textbf {\bibinfo {volume} {111}},\
  \bibinfo {pages} {L161106} (\bibinfo {year} {2025})}\BibitemShut {NoStop}%
\bibitem [{\citenamefont {Moon}\ and\ \citenamefont
  {Koshino}(2014)}]{koshino-hBN-prb14}%
  \BibitemOpen
  \bibfield  {author} {\bibinfo {author} {\bibfnamefont {P.}~\bibnamefont
  {Moon}}\ and\ \bibinfo {author} {\bibfnamefont {M.}~\bibnamefont {Koshino}},\
  }\href {https://doi.org/10.1103/PhysRevB.90.155406} {\bibfield  {journal}
  {\bibinfo  {journal} {Phys. Rev. B}\ }\textbf {\bibinfo {volume} {90}},\
  \bibinfo {pages} {155406} (\bibinfo {year} {2014})}\BibitemShut {NoStop}%
\bibitem [{\citenamefont {Jung}\ \emph {et~al.}(2014)\citenamefont {Jung},
  \citenamefont {Raoux}, \citenamefont {Qiao},\ and\ \citenamefont
  {MacDonald}}]{jung-hbn-graphene-prb14}%
  \BibitemOpen
  \bibfield  {author} {\bibinfo {author} {\bibfnamefont {J.}~\bibnamefont
  {Jung}}, \bibinfo {author} {\bibfnamefont {A.}~\bibnamefont {Raoux}},
  \bibinfo {author} {\bibfnamefont {Z.}~\bibnamefont {Qiao}},\ and\ \bibinfo
  {author} {\bibfnamefont {A.~H.}\ \bibnamefont {MacDonald}},\ }\href
  {https://doi.org/10.1103/PhysRevB.89.205414} {\bibfield  {journal} {\bibinfo
  {journal} {Phys. Rev. B}\ }\textbf {\bibinfo {volume} {89}},\ \bibinfo
  {pages} {205414} (\bibinfo {year} {2014})}\BibitemShut {NoStop}%
\bibitem [{\citenamefont {Park}\ \emph {et~al.}(2023)\citenamefont {Park},
  \citenamefont {Kim}, \citenamefont {Chittari},\ and\ \citenamefont
  {Jung}}]{jung-multilayer-arxiv23}%
  \BibitemOpen
  \bibfield  {author} {\bibinfo {author} {\bibfnamefont {Y.}~\bibnamefont
  {Park}}, \bibinfo {author} {\bibfnamefont {Y.}~\bibnamefont {Kim}}, \bibinfo
  {author} {\bibfnamefont {B.~L.}\ \bibnamefont {Chittari}},\ and\ \bibinfo
  {author} {\bibfnamefont {J.}~\bibnamefont {Jung}},\ }\href
  {https://doi.org/10.1103/PhysRevB.108.155406} {\bibfield  {journal} {\bibinfo
   {journal} {Phys. Rev. B}\ }\textbf {\bibinfo {volume} {108}},\ \bibinfo
  {pages} {155406} (\bibinfo {year} {2023})}\BibitemShut {NoStop}%
\bibitem [{\citenamefont {Herzog-Arbeitman}\ \emph {et~al.}(2024)\citenamefont
  {Herzog-Arbeitman}, \citenamefont {Wang}, \citenamefont {Liu}, \citenamefont
  {Tam}, \citenamefont {Qi}, \citenamefont {Jia}, \citenamefont {Efetov},
  \citenamefont {Vafek}, \citenamefont {Regnault}, \citenamefont {Weng},
  \citenamefont {Wu}, \citenamefont {Bernevig},\ and\ \citenamefont
  {Yu}}]{bernevig-fci-plg-arxiv23}%
  \BibitemOpen
  \bibfield  {author} {\bibinfo {author} {\bibfnamefont {J.}~\bibnamefont
  {Herzog-Arbeitman}}, \bibinfo {author} {\bibfnamefont {Y.}~\bibnamefont
  {Wang}}, \bibinfo {author} {\bibfnamefont {J.}~\bibnamefont {Liu}}, \bibinfo
  {author} {\bibfnamefont {P.~M.}\ \bibnamefont {Tam}}, \bibinfo {author}
  {\bibfnamefont {Z.}~\bibnamefont {Qi}}, \bibinfo {author} {\bibfnamefont
  {Y.}~\bibnamefont {Jia}}, \bibinfo {author} {\bibfnamefont {D.~K.}\
  \bibnamefont {Efetov}}, \bibinfo {author} {\bibfnamefont {O.}~\bibnamefont
  {Vafek}}, \bibinfo {author} {\bibfnamefont {N.}~\bibnamefont {Regnault}},
  \bibinfo {author} {\bibfnamefont {H.}~\bibnamefont {Weng}}, \bibinfo {author}
  {\bibfnamefont {Q.}~\bibnamefont {Wu}}, \bibinfo {author} {\bibfnamefont
  {B.~A.}\ \bibnamefont {Bernevig}},\ and\ \bibinfo {author} {\bibfnamefont
  {J.}~\bibnamefont {Yu}},\ }\href
  {https://doi.org/10.1103/PhysRevB.109.205122} {\bibfield  {journal} {\bibinfo
   {journal} {Phys. Rev. B}\ }\textbf {\bibinfo {volume} {109}},\ \bibinfo
  {pages} {205122} (\bibinfo {year} {2024})}\BibitemShut {NoStop}%
\bibitem [{\citenamefont {Lu}\ \emph {et~al.}(2024)\citenamefont {Lu},
  \citenamefont {Han}, \citenamefont {Yao}, \citenamefont {Reddy},
  \citenamefont {Yang}, \citenamefont {Seo}, \citenamefont {Watanabe},
  \citenamefont {Taniguchi}, \citenamefont {Fu},\ and\ \citenamefont
  {Ju}}]{fqah-ju-nature24}%
  \BibitemOpen
  \bibfield  {author} {\bibinfo {author} {\bibfnamefont {Z.}~\bibnamefont
  {Lu}}, \bibinfo {author} {\bibfnamefont {T.}~\bibnamefont {Han}}, \bibinfo
  {author} {\bibfnamefont {Y.}~\bibnamefont {Yao}}, \bibinfo {author}
  {\bibfnamefont {A.~P.}\ \bibnamefont {Reddy}}, \bibinfo {author}
  {\bibfnamefont {J.}~\bibnamefont {Yang}}, \bibinfo {author} {\bibfnamefont
  {J.}~\bibnamefont {Seo}}, \bibinfo {author} {\bibfnamefont {K.}~\bibnamefont
  {Watanabe}}, \bibinfo {author} {\bibfnamefont {T.}~\bibnamefont {Taniguchi}},
  \bibinfo {author} {\bibfnamefont {L.}~\bibnamefont {Fu}},\ and\ \bibinfo
  {author} {\bibfnamefont {L.}~\bibnamefont {Ju}},\ }\href
  {https://doi.org/10.1038/s41586-023-07010-7} {\bibfield  {journal} {\bibinfo
  {journal} {Nature}\ }\textbf {\bibinfo {volume} {626}},\ \bibinfo {pages}
  {759} (\bibinfo {year} {2024})}\BibitemShut {NoStop}%
\bibitem [{\citenamefont {Lu}\ \emph {et~al.}(2025)\citenamefont {Lu},
  \citenamefont {Han}, \citenamefont {Yao}, \citenamefont {Hadjri},
  \citenamefont {Yang}, \citenamefont {Seo}, \citenamefont {Shi}, \citenamefont
  {Ye}, \citenamefont {Watanabe}, \citenamefont {Taniguchi},\ and\
  \citenamefont {Ju}}]{ju-eqah-nature25}%
  \BibitemOpen
  \bibfield  {author} {\bibinfo {author} {\bibfnamefont {Z.}~\bibnamefont
  {Lu}}, \bibinfo {author} {\bibfnamefont {T.}~\bibnamefont {Han}}, \bibinfo
  {author} {\bibfnamefont {Y.}~\bibnamefont {Yao}}, \bibinfo {author}
  {\bibfnamefont {Z.}~\bibnamefont {Hadjri}}, \bibinfo {author} {\bibfnamefont
  {J.}~\bibnamefont {Yang}}, \bibinfo {author} {\bibfnamefont {J.}~\bibnamefont
  {Seo}}, \bibinfo {author} {\bibfnamefont {L.}~\bibnamefont {Shi}}, \bibinfo
  {author} {\bibfnamefont {S.}~\bibnamefont {Ye}}, \bibinfo {author}
  {\bibfnamefont {K.}~\bibnamefont {Watanabe}}, \bibinfo {author}
  {\bibfnamefont {T.}~\bibnamefont {Taniguchi}},\ and\ \bibinfo {author}
  {\bibfnamefont {L.}~\bibnamefont {Ju}},\ }\href
  {https://doi.org/10.1038/s41586-024-08470-1} {\bibfield  {journal} {\bibinfo
  {journal} {Nature}\ }\textbf {\bibinfo {volume} {637}},\ \bibinfo {pages}
  {1090} (\bibinfo {year} {2025})}\BibitemShut {NoStop}%
\bibitem [{\citenamefont {Aronson}\ \emph {et~al.}(2025)\citenamefont
  {Aronson}, \citenamefont {Han}, \citenamefont {Lu}, \citenamefont {Yao},
  \citenamefont {Butler}, \citenamefont {Watanabe}, \citenamefont {Taniguchi},
  \citenamefont {Ju},\ and\ \citenamefont {Ashoori}}]{aronson-fci-cdw-prx25}%
  \BibitemOpen
  \bibfield  {author} {\bibinfo {author} {\bibfnamefont {S.~H.}\ \bibnamefont
  {Aronson}}, \bibinfo {author} {\bibfnamefont {T.}~\bibnamefont {Han}},
  \bibinfo {author} {\bibfnamefont {Z.}~\bibnamefont {Lu}}, \bibinfo {author}
  {\bibfnamefont {Y.}~\bibnamefont {Yao}}, \bibinfo {author} {\bibfnamefont
  {J.~P.}\ \bibnamefont {Butler}}, \bibinfo {author} {\bibfnamefont
  {K.}~\bibnamefont {Watanabe}}, \bibinfo {author} {\bibfnamefont
  {T.}~\bibnamefont {Taniguchi}}, \bibinfo {author} {\bibfnamefont
  {L.}~\bibnamefont {Ju}},\ and\ \bibinfo {author} {\bibfnamefont {R.~C.}\
  \bibnamefont {Ashoori}},\ }\href {https://doi.org/10.1103/75gl-jzl6}
  {\bibfield  {journal} {\bibinfo  {journal} {Phys. Rev. X}\ }\textbf {\bibinfo
  {volume} {15}},\ \bibinfo {pages} {031026} (\bibinfo {year}
  {2025})}\BibitemShut {NoStop}%
\bibitem [{\citenamefont {Xie}\ \emph {et~al.}(2025)\citenamefont {Xie},
  \citenamefont {Huo}, \citenamefont {Lu}, \citenamefont {Feng}, \citenamefont
  {Zhang}, \citenamefont {Wang}, \citenamefont {Yang}, \citenamefont
  {Watanabe}, \citenamefont {Taniguchi}, \citenamefont {Liu}, \citenamefont
  {Song}, \citenamefont {Xie}, \citenamefont {Liu},\ and\ \citenamefont
  {Lu}}]{xie-tunable-fci-natmat25}%
  \BibitemOpen
  \bibfield  {author} {\bibinfo {author} {\bibfnamefont {J.}~\bibnamefont
  {Xie}}, \bibinfo {author} {\bibfnamefont {Z.}~\bibnamefont {Huo}}, \bibinfo
  {author} {\bibfnamefont {X.}~\bibnamefont {Lu}}, \bibinfo {author}
  {\bibfnamefont {Z.}~\bibnamefont {Feng}}, \bibinfo {author} {\bibfnamefont
  {Z.}~\bibnamefont {Zhang}}, \bibinfo {author} {\bibfnamefont
  {W.}~\bibnamefont {Wang}}, \bibinfo {author} {\bibfnamefont {Q.}~\bibnamefont
  {Yang}}, \bibinfo {author} {\bibfnamefont {K.}~\bibnamefont {Watanabe}},
  \bibinfo {author} {\bibfnamefont {T.}~\bibnamefont {Taniguchi}}, \bibinfo
  {author} {\bibfnamefont {K.}~\bibnamefont {Liu}}, \bibinfo {author}
  {\bibfnamefont {Z.}~\bibnamefont {Song}}, \bibinfo {author} {\bibfnamefont
  {X.~C.}\ \bibnamefont {Xie}}, \bibinfo {author} {\bibfnamefont
  {J.}~\bibnamefont {Liu}},\ and\ \bibinfo {author} {\bibfnamefont
  {X.}~\bibnamefont {Lu}},\ }\href {https://doi.org/10.1038/s41563-025-02225-7}
  {\bibfield  {journal} {\bibinfo  {journal} {Nature Materials}\ }\textbf
  {\bibinfo {volume} {24}},\ \bibinfo {pages} {1042} (\bibinfo {year}
  {2025})}\BibitemShut {NoStop}%
\bibitem [{\citenamefont {Zheng}\ \emph {et~al.}(2025)\citenamefont {Zheng},
  \citenamefont {Wu}, \citenamefont {Liu}, \citenamefont {Lyu}, \citenamefont
  {Liu}, \citenamefont {Sha}, \citenamefont {Li}, \citenamefont {Watanabe},
  \citenamefont {Taniguchi}, \citenamefont {Jia}, \citenamefont {Shi},\ and\
  \citenamefont {Chen}}]{zheng-switchable-r6g-prl25}%
  \BibitemOpen
  \bibfield  {author} {\bibinfo {author} {\bibfnamefont {J.}~\bibnamefont
  {Zheng}}, \bibinfo {author} {\bibfnamefont {S.}~\bibnamefont {Wu}}, \bibinfo
  {author} {\bibfnamefont {K.}~\bibnamefont {Liu}}, \bibinfo {author}
  {\bibfnamefont {B.}~\bibnamefont {Lyu}}, \bibinfo {author} {\bibfnamefont
  {S.}~\bibnamefont {Liu}}, \bibinfo {author} {\bibfnamefont {Y.}~\bibnamefont
  {Sha}}, \bibinfo {author} {\bibfnamefont {Z.}~\bibnamefont {Li}}, \bibinfo
  {author} {\bibfnamefont {K.}~\bibnamefont {Watanabe}}, \bibinfo {author}
  {\bibfnamefont {T.}~\bibnamefont {Taniguchi}}, \bibinfo {author}
  {\bibfnamefont {J.}~\bibnamefont {Jia}}, \bibinfo {author} {\bibfnamefont
  {Z.}~\bibnamefont {Shi}},\ and\ \bibinfo {author} {\bibfnamefont
  {G.}~\bibnamefont {Chen}},\ }\href {https://doi.org/10.1103/chsq-ndzs}
  {\bibfield  {journal} {\bibinfo  {journal} {Phys. Rev. Lett.}\ }\textbf
  {\bibinfo {volume} {135}},\ \bibinfo {pages} {136302} (\bibinfo {year}
  {2025})}\BibitemShut {NoStop}%
\bibitem [{\citenamefont {Liu}\ \emph {et~al.}(2026)\citenamefont {Liu},
  \citenamefont {Wang}, \citenamefont {Han}, \citenamefont {Li}, \citenamefont
  {Li}, \citenamefont {Zhou}, \citenamefont {Hu}, \citenamefont {Qu},
  \citenamefont {Han}, \citenamefont {Watanabe}, \citenamefont {Taniguchi},
  \citenamefont {Han}, \citenamefont {Tong}, \citenamefont {Liu}, \citenamefont
  {Lu}, \citenamefont {Wu},\ and\ \citenamefont
  {Lu}}]{liu-heptalayer-qah-prl26}%
  \BibitemOpen
  \bibfield  {author} {\bibinfo {author} {\bibfnamefont {Q.}~\bibnamefont
  {Liu}}, \bibinfo {author} {\bibfnamefont {Z.}~\bibnamefont {Wang}}, \bibinfo
  {author} {\bibfnamefont {X.}~\bibnamefont {Han}}, \bibinfo {author}
  {\bibfnamefont {Z.}~\bibnamefont {Li}}, \bibinfo {author} {\bibfnamefont
  {B.}~\bibnamefont {Li}}, \bibinfo {author} {\bibfnamefont {S.}~\bibnamefont
  {Zhou}}, \bibinfo {author} {\bibfnamefont {L.}~\bibnamefont {Hu}}, \bibinfo
  {author} {\bibfnamefont {Z.}~\bibnamefont {Qu}}, \bibinfo {author}
  {\bibfnamefont {C.}~\bibnamefont {Han}}, \bibinfo {author} {\bibfnamefont
  {K.}~\bibnamefont {Watanabe}}, \bibinfo {author} {\bibfnamefont
  {T.}~\bibnamefont {Taniguchi}}, \bibinfo {author} {\bibfnamefont {Z.~V.}\
  \bibnamefont {Han}}, \bibinfo {author} {\bibfnamefont {B.}~\bibnamefont
  {Tong}}, \bibinfo {author} {\bibfnamefont {G.}~\bibnamefont {Liu}}, \bibinfo
  {author} {\bibfnamefont {L.}~\bibnamefont {Lu}}, \bibinfo {author}
  {\bibfnamefont {F.}~\bibnamefont {Wu}},\ and\ \bibinfo {author}
  {\bibfnamefont {J.}~\bibnamefont {Lu}},\ }\href
  {https://doi.org/10.1103/gm64-vxdm} {\bibfield  {journal} {\bibinfo
  {journal} {Phys. Rev. Lett.}\ }\textbf {\bibinfo {volume} {136}},\ \bibinfo
  {pages} {016602} (\bibinfo {year} {2026})}\BibitemShut {NoStop}%
\bibitem [{\citenamefont {Guo}\ \emph {et~al.}(2024)\citenamefont {Guo},
  \citenamefont {Lu}, \citenamefont {Xie},\ and\ \citenamefont
  {Liu}}]{guo-prb24}%
  \BibitemOpen
  \bibfield  {author} {\bibinfo {author} {\bibfnamefont {Z.}~\bibnamefont
  {Guo}}, \bibinfo {author} {\bibfnamefont {X.}~\bibnamefont {Lu}}, \bibinfo
  {author} {\bibfnamefont {B.}~\bibnamefont {Xie}},\ and\ \bibinfo {author}
  {\bibfnamefont {J.}~\bibnamefont {Liu}},\ }\href
  {https://doi.org/10.1103/PhysRevB.110.075109} {\bibfield  {journal} {\bibinfo
   {journal} {Phys. Rev. B}\ }\textbf {\bibinfo {volume} {110}},\ \bibinfo
  {pages} {075109} (\bibinfo {year} {2024})}\BibitemShut {NoStop}%
\bibitem [{\citenamefont {Huang}\ \emph {et~al.}(2024)\citenamefont {Huang},
  \citenamefont {Li}, \citenamefont {Das~Sarma},\ and\ \citenamefont
  {Zhang}}]{lixiao-fci-prb24}%
  \BibitemOpen
  \bibfield  {author} {\bibinfo {author} {\bibfnamefont {K.}~\bibnamefont
  {Huang}}, \bibinfo {author} {\bibfnamefont {X.}~\bibnamefont {Li}}, \bibinfo
  {author} {\bibfnamefont {S.}~\bibnamefont {Das~Sarma}},\ and\ \bibinfo
  {author} {\bibfnamefont {F.}~\bibnamefont {Zhang}},\ }\href
  {https://doi.org/10.1103/PhysRevB.110.115146} {\bibfield  {journal} {\bibinfo
   {journal} {Phys. Rev. B}\ }\textbf {\bibinfo {volume} {110}},\ \bibinfo
  {pages} {115146} (\bibinfo {year} {2024})}\BibitemShut {NoStop}%
\bibitem [{\citenamefont {Dong}\ \emph
  {et~al.}(2024{\natexlab{a}})\citenamefont {Dong}, \citenamefont {Patri},\
  and\ \citenamefont {Senthil}}]{senthil-fci-prl24}%
  \BibitemOpen
  \bibfield  {author} {\bibinfo {author} {\bibfnamefont {Z.}~\bibnamefont
  {Dong}}, \bibinfo {author} {\bibfnamefont {A.~S.}\ \bibnamefont {Patri}},\
  and\ \bibinfo {author} {\bibfnamefont {T.}~\bibnamefont {Senthil}},\ }\href
  {https://doi.org/10.1103/PhysRevLett.133.206502} {\bibfield  {journal}
  {\bibinfo  {journal} {Phys. Rev. Lett.}\ }\textbf {\bibinfo {volume} {133}},\
  \bibinfo {pages} {206502} (\bibinfo {year} {2024}{\natexlab{a}})}\BibitemShut
  {NoStop}%
\bibitem [{\citenamefont {Dong}\ \emph
  {et~al.}(2024{\natexlab{b}})\citenamefont {Dong}, \citenamefont {Wang},
  \citenamefont {Wang}, \citenamefont {Soejima}, \citenamefont {Zaletel},
  \citenamefont {Vishwanath},\ and\ \citenamefont {Parker}}]{ashvin-fci-prl24}%
  \BibitemOpen
  \bibfield  {author} {\bibinfo {author} {\bibfnamefont {J.}~\bibnamefont
  {Dong}}, \bibinfo {author} {\bibfnamefont {T.}~\bibnamefont {Wang}}, \bibinfo
  {author} {\bibfnamefont {T.}~\bibnamefont {Wang}}, \bibinfo {author}
  {\bibfnamefont {T.}~\bibnamefont {Soejima}}, \bibinfo {author} {\bibfnamefont
  {M.~P.}\ \bibnamefont {Zaletel}}, \bibinfo {author} {\bibfnamefont
  {A.}~\bibnamefont {Vishwanath}},\ and\ \bibinfo {author} {\bibfnamefont
  {D.~E.}\ \bibnamefont {Parker}},\ }\href
  {https://doi.org/10.1103/PhysRevLett.133.206503} {\bibfield  {journal}
  {\bibinfo  {journal} {Phys. Rev. Lett.}\ }\textbf {\bibinfo {volume} {133}},\
  \bibinfo {pages} {206503} (\bibinfo {year} {2024}{\natexlab{b}})}\BibitemShut
  {NoStop}%
\bibitem [{\citenamefont {Zhou}\ \emph {et~al.}(2024)\citenamefont {Zhou},
  \citenamefont {Yang},\ and\ \citenamefont {Zhang}}]{zhou-moireless-prl24}%
  \BibitemOpen
  \bibfield  {author} {\bibinfo {author} {\bibfnamefont {B.}~\bibnamefont
  {Zhou}}, \bibinfo {author} {\bibfnamefont {H.}~\bibnamefont {Yang}},\ and\
  \bibinfo {author} {\bibfnamefont {Y.-H.}\ \bibnamefont {Zhang}},\ }\href
  {https://doi.org/10.1103/PhysRevLett.133.206504} {\bibfield  {journal}
  {\bibinfo  {journal} {Phys. Rev. Lett.}\ }\textbf {\bibinfo {volume} {133}},\
  \bibinfo {pages} {206504} (\bibinfo {year} {2024})}\BibitemShut {NoStop}%
\bibitem [{\citenamefont {Tan}\ and\ \citenamefont
  {Devakul}(2024)}]{tan-parent-berry-prx24}%
  \BibitemOpen
  \bibfield  {author} {\bibinfo {author} {\bibfnamefont {T.}~\bibnamefont
  {Tan}}\ and\ \bibinfo {author} {\bibfnamefont {T.}~\bibnamefont {Devakul}},\
  }\href {https://doi.org/10.1103/PhysRevX.14.041040} {\bibfield  {journal}
  {\bibinfo  {journal} {Phys. Rev. X}\ }\textbf {\bibinfo {volume} {14}},\
  \bibinfo {pages} {041040} (\bibinfo {year} {2024})}\BibitemShut {NoStop}%
\bibitem [{\citenamefont {Soejima}\ \emph {et~al.}(2024)\citenamefont
  {Soejima}, \citenamefont {Dong}, \citenamefont {Wang}, \citenamefont {Wang},
  \citenamefont {Zaletel}, \citenamefont {Vishwanath},\ and\ \citenamefont
  {Parker}}]{soejima-ahc-model-prb24}%
  \BibitemOpen
  \bibfield  {author} {\bibinfo {author} {\bibfnamefont {T.}~\bibnamefont
  {Soejima}}, \bibinfo {author} {\bibfnamefont {J.}~\bibnamefont {Dong}},
  \bibinfo {author} {\bibfnamefont {T.}~\bibnamefont {Wang}}, \bibinfo {author}
  {\bibfnamefont {T.}~\bibnamefont {Wang}}, \bibinfo {author} {\bibfnamefont
  {M.~P.}\ \bibnamefont {Zaletel}}, \bibinfo {author} {\bibfnamefont
  {A.}~\bibnamefont {Vishwanath}},\ and\ \bibinfo {author} {\bibfnamefont
  {D.~E.}\ \bibnamefont {Parker}},\ }\href
  {https://doi.org/10.1103/PhysRevB.110.205124} {\bibfield  {journal} {\bibinfo
   {journal} {Phys. Rev. B}\ }\textbf {\bibinfo {volume} {110}},\ \bibinfo
  {pages} {205124} (\bibinfo {year} {2024})}\BibitemShut {NoStop}%
\bibitem [{\citenamefont {Dong}\ \emph
  {et~al.}(2024{\natexlab{c}})\citenamefont {Dong}, \citenamefont {Patri},\
  and\ \citenamefont {Senthil}}]{dong-ahc-stability-prb24}%
  \BibitemOpen
  \bibfield  {author} {\bibinfo {author} {\bibfnamefont {Z.}~\bibnamefont
  {Dong}}, \bibinfo {author} {\bibfnamefont {A.~S.}\ \bibnamefont {Patri}},\
  and\ \bibinfo {author} {\bibfnamefont {T.}~\bibnamefont {Senthil}},\ }\href
  {https://doi.org/10.1103/physrevb.110.205130} {\bibfield  {journal} {\bibinfo
   {journal} {Phys. Rev. B}\ }\textbf {\bibinfo {volume} {110}},\ \bibinfo
  {pages} {205130} (\bibinfo {year} {2024}{\natexlab{c}})}\BibitemShut
  {NoStop}%
\bibitem [{\citenamefont {Li}\ \emph {et~al.}(2026)\citenamefont {Li},
  \citenamefont {Li}, \citenamefont {Lu}, \citenamefont {Fan}, \citenamefont
  {Watanabe}, \citenamefont {Taniguchi}, \citenamefont {Zhao}, \citenamefont
  {Xie}, \citenamefont {Wang},\ and\ \citenamefont {Liu}}]{fermilune}%
  \BibitemOpen
  \bibfield  {author} {\bibinfo {author} {\bibfnamefont {M.}~\bibnamefont
  {Li}}, \bibinfo {author} {\bibfnamefont {Q.}~\bibnamefont {Li}}, \bibinfo
  {author} {\bibfnamefont {X.}~\bibnamefont {Lu}}, \bibinfo {author}
  {\bibfnamefont {H.}~\bibnamefont {Fan}}, \bibinfo {author} {\bibfnamefont
  {K.}~\bibnamefont {Watanabe}}, \bibinfo {author} {\bibfnamefont
  {T.}~\bibnamefont {Taniguchi}}, \bibinfo {author} {\bibfnamefont
  {Y.}~\bibnamefont {Zhao}}, \bibinfo {author} {\bibfnamefont {X.-C.}\
  \bibnamefont {Xie}}, \bibinfo {author} {\bibfnamefont {L.}~\bibnamefont
  {Wang}},\ and\ \bibinfo {author} {\bibfnamefont {J.}~\bibnamefont {Liu}},\
  }\href {https://arxiv.org/abs/2505.05414} {\bibinfo {title} {Fermi lune and non-reciprocal transport in rhombohedral multilayer graphene}}
  (\bibinfo {year} {2026}),\ \Eprint {https://arxiv.org/abs/2505.05414}
  {arXiv:2505.05414 [cond-mat.mes-hall]} \BibitemShut {NoStop}%
\bibitem [{tdc()}]{tdci-supplemental}%
  \BibitemOpen
  \href@noop {} {}\bibinfo {note} {See Supplemental Material for: (a)
  continuum, renormalization, and Hartree--Fock methods; (b) band unfolding;
  (c) full Bethe--Salpeter kernels, normalization, and charge, spin, and valley
  spectra; (d) zero-field and weak-field orbital magnetization; (e) single-band
  exact diagonalization with Hartree--Fock dispersion in the hole
  representation; and (f) two-band topology, magnetization, edge states,
  finite-momentum excitations, and field response}\BibitemShut {NoStop}%
\bibitem [{\citenamefont {Negele}(1982)}]{negele-rmp82}%
  \BibitemOpen
  \bibfield  {author} {\bibinfo {author} {\bibfnamefont {J.~W.}\ \bibnamefont
  {Negele}},\ }\href {https://doi.org/10.1103/RevModPhys.54.913} {\bibfield
  {journal} {\bibinfo  {journal} {Rev. Mod. Phys.}\ }\textbf {\bibinfo {volume}
  {54}},\ \bibinfo {pages} {913} (\bibinfo {year} {1982})}\BibitemShut
  {NoStop}%
\bibitem [{\citenamefont {Ur{\'i}a-{\'A}lvarez}\ \emph
  {et~al.}(2024)\citenamefont {Ur{\'i}a-{\'A}lvarez}, \citenamefont
  {Esteve-Paredes}, \citenamefont {Garc{\'i}a-Bl{\'a}zquez},\ and\
  \citenamefont {Palacios}}]{uria-xatu-cpc24}%
  \BibitemOpen
  \bibfield  {author} {\bibinfo {author} {\bibfnamefont {A.~J.}\ \bibnamefont
  {Ur{\'i}a-{\'A}lvarez}}, \bibinfo {author} {\bibfnamefont {J.~J.}\
  \bibnamefont {Esteve-Paredes}}, \bibinfo {author} {\bibfnamefont {M.~A.}\
  \bibnamefont {Garc{\'i}a-Bl{\'a}zquez}},\ and\ \bibinfo {author}
  {\bibfnamefont {J.~J.}\ \bibnamefont {Palacios}},\ }\href
  {https://doi.org/10.1016/j.cpc.2023.109001} {\bibfield  {journal} {\bibinfo
  {journal} {Computer Physics Communications}\ }\textbf {\bibinfo {volume}
  {295}},\ \bibinfo {pages} {109001} (\bibinfo {year} {2024})}\BibitemShut
  {NoStop}%
\bibitem [{\citenamefont {Wu}\ and\ \citenamefont
  {Das~Sarma}(2020)}]{wu-tbg-collective-prl20}%
  \BibitemOpen
  \bibfield  {author} {\bibinfo {author} {\bibfnamefont {F.}~\bibnamefont
  {Wu}}\ and\ \bibinfo {author} {\bibfnamefont {S.}~\bibnamefont {Das~Sarma}},\
  }\href {https://doi.org/10.1103/PhysRevLett.124.046403} {\bibfield  {journal}
  {\bibinfo  {journal} {Phys. Rev. Lett.}\ }\textbf {\bibinfo {volume} {124}},\
  \bibinfo {pages} {046403} (\bibinfo {year} {2020})}\BibitemShut {NoStop}%
\bibitem [{\citenamefont {Kumar}\ \emph {et~al.}(2021)\citenamefont {Kumar},
  \citenamefont {Xie},\ and\ \citenamefont
  {MacDonald}}]{kumar-collective-prb21}%
  \BibitemOpen
  \bibfield  {author} {\bibinfo {author} {\bibfnamefont {A.}~\bibnamefont
  {Kumar}}, \bibinfo {author} {\bibfnamefont {M.}~\bibnamefont {Xie}},\ and\
  \bibinfo {author} {\bibfnamefont {A.~H.}\ \bibnamefont {MacDonald}},\ }\href
  {https://doi.org/10.1103/PhysRevB.104.035119} {\bibfield  {journal} {\bibinfo
   {journal} {Phys. Rev. B}\ }\textbf {\bibinfo {volume} {104}},\ \bibinfo
  {pages} {035119} (\bibinfo {year} {2021})}\BibitemShut {NoStop}%
\bibitem [{\citenamefont {Kwan}\ \emph {et~al.}(2025)\citenamefont {Kwan},
  \citenamefont {Yu}, \citenamefont {Herzog-Arbeitman}, \citenamefont {Efetov},
  \citenamefont {Regnault},\ and\ \citenamefont {Bernevig}}]{kwan-fci-prb25}%
  \BibitemOpen
  \bibfield  {author} {\bibinfo {author} {\bibfnamefont {Y.~H.}\ \bibnamefont
  {Kwan}}, \bibinfo {author} {\bibfnamefont {J.}~\bibnamefont {Yu}}, \bibinfo
  {author} {\bibfnamefont {J.}~\bibnamefont {Herzog-Arbeitman}}, \bibinfo
  {author} {\bibfnamefont {D.~K.}\ \bibnamefont {Efetov}}, \bibinfo {author}
  {\bibfnamefont {N.}~\bibnamefont {Regnault}},\ and\ \bibinfo {author}
  {\bibfnamefont {B.~A.}\ \bibnamefont {Bernevig}},\ }\href
  {https://doi.org/10.1103/PhysRevB.112.075109} {\bibfield  {journal} {\bibinfo
   {journal} {Phys. Rev. B}\ }\textbf {\bibinfo {volume} {112}},\ \bibinfo
  {pages} {075109} (\bibinfo {year} {2025})}\BibitemShut {NoStop}%
\bibitem [{\citenamefont {Froese}\ \emph {et~al.}(2025)\citenamefont {Froese},
  \citenamefont {Neupert},\ and\ \citenamefont
  {Wagner}}]{froese-excitons-prr25}%
  \BibitemOpen
  \bibfield  {author} {\bibinfo {author} {\bibfnamefont {P.}~\bibnamefont
  {Froese}}, \bibinfo {author} {\bibfnamefont {T.}~\bibnamefont {Neupert}},\
  and\ \bibinfo {author} {\bibfnamefont {G.}~\bibnamefont {Wagner}},\ }\href
  {https://doi.org/10.1103/PhysRevResearch.7.023047} {\bibfield  {journal}
  {\bibinfo  {journal} {Phys. Rev. Research}\ }\textbf {\bibinfo {volume}
  {7}},\ \bibinfo {pages} {023047} (\bibinfo {year} {2025})}\BibitemShut
  {NoStop}%
\bibitem [{\citenamefont {Xie}\ and\ \citenamefont
  {Das~Sarma}(2026)}]{xie-tmd-magnons-2026}%
  \BibitemOpen
  \bibfield  {author} {\bibinfo {author} {\bibfnamefont {M.}~\bibnamefont
  {Xie}}\ and\ \bibinfo {author} {\bibfnamefont {S.}~\bibnamefont
  {Das~Sarma}},\ }\href {https://doi.org/10.48550/arXiv.2603.20370} {\bibinfo
  {title} {Collective spin excitations in correlated moir\'e chern
  ferromagnets}} (\bibinfo {year} {2026}),\ \Eprint
  {https://arxiv.org/abs/2603.20370} {arXiv:2603.20370 [cond-mat.str-el]}
  \BibitemShut {NoStop}%
\bibitem [{\citenamefont {Alavirad}\ and\ \citenamefont
  {Sau}(2020)}]{alavirad-sau-prb20}%
  \BibitemOpen
  \bibfield  {author} {\bibinfo {author} {\bibfnamefont {Y.}~\bibnamefont
  {Alavirad}}\ and\ \bibinfo {author} {\bibfnamefont {J.~D.}\ \bibnamefont
  {Sau}},\ }\href {https://doi.org/10.1103/PhysRevB.102.235123} {\bibfield
  {journal} {\bibinfo  {journal} {Phys. Rev. B}\ }\textbf {\bibinfo {volume}
  {102}},\ \bibinfo {pages} {235123} (\bibinfo {year} {2020})}\BibitemShut
  {NoStop}%
\bibitem [{\citenamefont {Desrochers}\ \emph {et~al.}(2026)\citenamefont
  {Desrochers}, \citenamefont {Hirsbrunner}, \citenamefont {Huxford},
  \citenamefont {Patri}, \citenamefont {Senthil},\ and\ \citenamefont
  {Kim}}]{desrochers-ahc-instability-prl26}%
  \BibitemOpen
  \bibfield  {author} {\bibinfo {author} {\bibfnamefont {F.}~\bibnamefont
  {Desrochers}}, \bibinfo {author} {\bibfnamefont {M.~R.}\ \bibnamefont
  {Hirsbrunner}}, \bibinfo {author} {\bibfnamefont {J.}~\bibnamefont
  {Huxford}}, \bibinfo {author} {\bibfnamefont {A.~S.}\ \bibnamefont {Patri}},
  \bibinfo {author} {\bibfnamefont {T.}~\bibnamefont {Senthil}},\ and\ \bibinfo
  {author} {\bibfnamefont {Y.~B.}\ \bibnamefont {Kim}},\ }\href
  {https://doi.org/10.1103/vyss-2hkj} {\bibfield  {journal} {\bibinfo
  {journal} {Phys. Rev. Lett.}\ }\textbf {\bibinfo {volume} {136}},\ \bibinfo
  {pages} {166503} (\bibinfo {year} {2026})}\BibitemShut {NoStop}%
\bibitem [{\citenamefont {Neupert}\ \emph {et~al.}(2011)\citenamefont
  {Neupert}, \citenamefont {Santos}, \citenamefont {Chamon},\ and\
  \citenamefont {Mudry}}]{neupert-fci-prl11}%
  \BibitemOpen
  \bibfield  {author} {\bibinfo {author} {\bibfnamefont {T.}~\bibnamefont
  {Neupert}}, \bibinfo {author} {\bibfnamefont {L.}~\bibnamefont {Santos}},
  \bibinfo {author} {\bibfnamefont {C.}~\bibnamefont {Chamon}},\ and\ \bibinfo
  {author} {\bibfnamefont {C.}~\bibnamefont {Mudry}},\ }\href
  {https://doi.org/10.1103/PhysRevLett.106.236804} {\bibfield  {journal}
  {\bibinfo  {journal} {Phys. Rev. Lett.}\ }\textbf {\bibinfo {volume} {106}},\
  \bibinfo {pages} {236804} (\bibinfo {year} {2011})}\BibitemShut {NoStop}%
\bibitem [{\citenamefont {Regnault}\ and\ \citenamefont
  {Bernevig}(2011)}]{regnault-fci-prx11}%
  \BibitemOpen
  \bibfield  {author} {\bibinfo {author} {\bibfnamefont {N.}~\bibnamefont
  {Regnault}}\ and\ \bibinfo {author} {\bibfnamefont {B.~A.}\ \bibnamefont
  {Bernevig}},\ }\href {https://doi.org/10.1103/PhysRevX.1.021014} {\bibfield
  {journal} {\bibinfo  {journal} {Phys. Rev. X}\ }\textbf {\bibinfo {volume}
  {1}},\ \bibinfo {pages} {021014} (\bibinfo {year} {2011})}\BibitemShut
  {NoStop}%
\end{thebibliography}

\begin{thebibliography}{17}%
\makeatletter
\providecommand \@ifxundefined [1]{%
 \@ifx{#1\undefined}
}%
\providecommand \@ifnum [1]{%
 \ifnum #1\expandafter \@firstoftwo
 \else \expandafter \@secondoftwo
 \fi
}%
\providecommand \@ifx [1]{%
 \ifx #1\expandafter \@firstoftwo
 \else \expandafter \@secondoftwo
 \fi
}%
\providecommand \natexlab [1]{#1}%
\providecommand \enquote  [1]{``#1''}%
\providecommand \bibnamefont  [1]{#1}%
\providecommand \bibfnamefont [1]{#1}%
\providecommand \citenamefont [1]{#1}%
\providecommand \href@noop [0]{\@secondoftwo}%
\providecommand \href [0]{\begingroup \@sanitize@url \@href}%
\providecommand \@href[1]{\@@startlink{#1}\@@href}%
\providecommand \@@href[1]{\endgroup#1\@@endlink}%
\providecommand \@sanitize@url [0]{\catcode `\\12\catcode `\$12\catcode
  `\&12\catcode `\#12\catcode `\^12\catcode `\_12\catcode `\%12\relax}%
\providecommand \@@startlink[1]{}%
\providecommand \@@endlink[0]{}%
\providecommand \url  [0]{\begingroup\@sanitize@url \@url }%
\providecommand \@url [1]{\endgroup\@href {#1}{\urlprefix }}%
\providecommand \urlprefix  [0]{URL }%
\providecommand \Eprint [0]{\href }%
\providecommand \doibase [0]{https://doi.org/}%
\providecommand \selectlanguage [0]{\@gobble}%
\providecommand \bibinfo  [0]{\@secondoftwo}%
\providecommand \bibfield  [0]{\@secondoftwo}%
\providecommand \translation [1]{[#1]}%
\providecommand \BibitemOpen [0]{}%
\providecommand \bibitemStop [0]{}%
\providecommand \bibitemNoStop [0]{.\EOS\space}%
\providecommand \EOS [0]{\spacefactor3000\relax}%
\providecommand \BibitemShut  [1]{\csname bibitem#1\endcsname}%
\let\auto@bib@innerbib\@empty
%</preamble>
\bibitem [{\citenamefont {Moon}\ and\ \citenamefont
  {Koshino}(2014)}]{SI-koshino-hBN-prb14}%
  \BibitemOpen
  \bibfield  {author} {\bibinfo {author} {\bibfnamefont {P.}~\bibnamefont
  {Moon}}\ and\ \bibinfo {author} {\bibfnamefont {M.}~\bibnamefont {Koshino}},\
  }\bibfield  {title} {\bibinfo {title} {Electronic properties of
  graphene/hexagonal-boron-nitride moir\'e superlattice},\ }\href
  {https://doi.org/10.1103/PhysRevB.90.155406} {\bibfield  {journal} {\bibinfo
  {journal} {Phys. Rev. B}\ }\textbf {\bibinfo {volume} {90}},\ \bibinfo
  {pages} {155406} (\bibinfo {year} {2014})}\BibitemShut {NoStop}%
\bibitem [{\citenamefont {Jang}\ \emph {et~al.}(2023)\citenamefont {Jang},
  \citenamefont {Park}, \citenamefont {Jung},\ and\ \citenamefont
  {Min}}]{SI-kpgra}%
  \BibitemOpen
  \bibfield  {author} {\bibinfo {author} {\bibfnamefont {Y.}~\bibnamefont
  {Jang}}, \bibinfo {author} {\bibfnamefont {Y.}~\bibnamefont {Park}}, \bibinfo
  {author} {\bibfnamefont {J.}~\bibnamefont {Jung}},\ and\ \bibinfo {author}
  {\bibfnamefont {H.}~\bibnamefont {Min}},\ }\bibfield  {title} {\bibinfo
  {title} {Chirality and correlations in the spontaneous spin-valley
  polarization of rhombohedral multilayer graphene},\ }\href
  {https://doi.org/10.1103/PhysRevB.108.L041101} {\bibfield  {journal}
  {\bibinfo  {journal} {Phys. Rev. B}\ }\textbf {\bibinfo {volume} {108}},\
  \bibinfo {pages} {L041101} (\bibinfo {year} {2023})}\BibitemShut {NoStop}%
\bibitem [{\citenamefont {Lee}\ \emph {et~al.}(2019)\citenamefont {Lee},
  \citenamefont {Khalaf}, \citenamefont {Liu}, \citenamefont {Liu},
  \citenamefont {Hao}, \citenamefont {Kim},\ and\ \citenamefont
  {Vishwanath}}]{SI-lee-mag-inplane}%
  \BibitemOpen
  \bibfield  {author} {\bibinfo {author} {\bibfnamefont {J.~Y.}\ \bibnamefont
  {Lee}}, \bibinfo {author} {\bibfnamefont {E.}~\bibnamefont {Khalaf}},
  \bibinfo {author} {\bibfnamefont {S.}~\bibnamefont {Liu}}, \bibinfo {author}
  {\bibfnamefont {X.}~\bibnamefont {Liu}}, \bibinfo {author} {\bibfnamefont
  {Z.}~\bibnamefont {Hao}}, \bibinfo {author} {\bibfnamefont {P.}~\bibnamefont
  {Kim}},\ and\ \bibinfo {author} {\bibfnamefont {A.}~\bibnamefont
  {Vishwanath}},\ }\bibfield  {title} {\bibinfo {title} {Theory of correlated
  insulating behaviour and spin-triplet superconductivity in twisted double
  bilayer graphene},\ }\href@noop {} {\bibfield  {journal} {\bibinfo  {journal}
  {Nature communications}\ }\textbf {\bibinfo {volume} {10}},\ \bibinfo {pages}
  {5333} (\bibinfo {year} {2019})}\BibitemShut {NoStop}%
\bibitem [{\citenamefont {Guo}\ \emph {et~al.}(2024)\citenamefont {Guo},
  \citenamefont {Lu}, \citenamefont {Xie},\ and\ \citenamefont
  {Liu}}]{SI-guo-prb24}%
  \BibitemOpen
  \bibfield  {author} {\bibinfo {author} {\bibfnamefont {Z.}~\bibnamefont
  {Guo}}, \bibinfo {author} {\bibfnamefont {X.}~\bibnamefont {Lu}}, \bibinfo
  {author} {\bibfnamefont {B.}~\bibnamefont {Xie}},\ and\ \bibinfo {author}
  {\bibfnamefont {J.}~\bibnamefont {Liu}},\ }\bibfield  {title} {\bibinfo
  {title} {Fractional chern insulator states in multilayer graphene moir\'e
  superlattices},\ }\href {https://doi.org/10.1103/PhysRevB.110.075109}
  {\bibfield  {journal} {\bibinfo  {journal} {Phys. Rev. B}\ }\textbf {\bibinfo
  {volume} {110}},\ \bibinfo {pages} {075109} (\bibinfo {year}
  {2024})}\BibitemShut {NoStop}%
\bibitem [{\citenamefont {Ur{\'i}a-{\'A}lvarez}\ \emph
  {et~al.}(2024)\citenamefont {Ur{\'i}a-{\'A}lvarez}, \citenamefont
  {Esteve-Paredes}, \citenamefont {Garc{\'i}a-Bl{\'a}zquez},\ and\
  \citenamefont {Palacios}}]{SI-uria-xatu-cpc24}%
  \BibitemOpen
  \bibfield  {author} {\bibinfo {author} {\bibfnamefont {A.~J.}\ \bibnamefont
  {Ur{\'i}a-{\'A}lvarez}}, \bibinfo {author} {\bibfnamefont {J.~J.}\
  \bibnamefont {Esteve-Paredes}}, \bibinfo {author} {\bibfnamefont {M.~A.}\
  \bibnamefont {Garc{\'i}a-Bl{\'a}zquez}},\ and\ \bibinfo {author}
  {\bibfnamefont {J.~J.}\ \bibnamefont {Palacios}},\ }\bibfield  {title}
  {\bibinfo {title} {Efficient computation of optical excitations in
  two-dimensional materials with the {Xatu} code},\ }\href
  {https://doi.org/10.1016/j.cpc.2023.109001} {\bibfield  {journal} {\bibinfo
  {journal} {Computer Physics Communications}\ }\textbf {\bibinfo {volume}
  {295}},\ \bibinfo {pages} {109001} (\bibinfo {year} {2024})}\BibitemShut
  {NoStop}%
\bibitem [{\citenamefont {Negele}(1982)}]{SI-negele-rmp82}%
  \BibitemOpen
  \bibfield  {author} {\bibinfo {author} {\bibfnamefont {J.~W.}\ \bibnamefont
  {Negele}},\ }\bibfield  {title} {\bibinfo {title} {The mean-field theory of
  nuclear structure and dynamics},\ }\href
  {https://doi.org/10.1103/RevModPhys.54.913} {\bibfield  {journal} {\bibinfo
  {journal} {Rev. Mod. Phys.}\ }\textbf {\bibinfo {volume} {54}},\ \bibinfo
  {pages} {913} (\bibinfo {year} {1982})}\BibitemShut {NoStop}%
\bibitem [{\citenamefont {Kumar}\ \emph {et~al.}(2021)\citenamefont {Kumar},
  \citenamefont {Xie},\ and\ \citenamefont
  {MacDonald}}]{SI-kumar-collective-prb21}%
  \BibitemOpen
  \bibfield  {author} {\bibinfo {author} {\bibfnamefont {A.}~\bibnamefont
  {Kumar}}, \bibinfo {author} {\bibfnamefont {M.}~\bibnamefont {Xie}},\ and\
  \bibinfo {author} {\bibfnamefont {A.~H.}\ \bibnamefont {MacDonald}},\
  }\bibfield  {title} {\bibinfo {title} {Lattice collective modes from a
  continuum model of magic-angle twisted bilayer graphene},\ }\href
  {https://doi.org/10.1103/PhysRevB.104.035119} {\bibfield  {journal} {\bibinfo
   {journal} {Phys. Rev. B}\ }\textbf {\bibinfo {volume} {104}},\ \bibinfo
  {pages} {035119} (\bibinfo {year} {2021})}\BibitemShut {NoStop}%
\bibitem [{\citenamefont {Froese}\ \emph {et~al.}(2025)\citenamefont {Froese},
  \citenamefont {Neupert},\ and\ \citenamefont
  {Wagner}}]{SI-froese-excitons-prr25}%
  \BibitemOpen
  \bibfield  {author} {\bibinfo {author} {\bibfnamefont {P.}~\bibnamefont
  {Froese}}, \bibinfo {author} {\bibfnamefont {T.}~\bibnamefont {Neupert}},\
  and\ \bibinfo {author} {\bibfnamefont {G.}~\bibnamefont {Wagner}},\
  }\bibfield  {title} {\bibinfo {title} {Topological excitons in moir\'e
  {MoTe$_2$}/{WSe$_2$} heterobilayers},\ }\href
  {https://doi.org/10.1103/PhysRevResearch.7.023047} {\bibfield  {journal}
  {\bibinfo  {journal} {Phys. Rev. Research}\ }\textbf {\bibinfo {volume}
  {7}},\ \bibinfo {pages} {023047} (\bibinfo {year} {2025})}\BibitemShut
  {NoStop}%
\bibitem [{\citenamefont {Kwan}\ \emph {et~al.}(2025)\citenamefont {Kwan},
  \citenamefont {Yu}, \citenamefont {Herzog-Arbeitman}, \citenamefont {Efetov},
  \citenamefont {Regnault},\ and\ \citenamefont {Bernevig}}]{SI-kwan-fci-prb25}%
  \BibitemOpen
  \bibfield  {author} {\bibinfo {author} {\bibfnamefont {Y.~H.}\ \bibnamefont
  {Kwan}}, \bibinfo {author} {\bibfnamefont {J.}~\bibnamefont {Yu}}, \bibinfo
  {author} {\bibfnamefont {J.}~\bibnamefont {Herzog-Arbeitman}}, \bibinfo
  {author} {\bibfnamefont {D.~K.}\ \bibnamefont {Efetov}}, \bibinfo {author}
  {\bibfnamefont {N.}~\bibnamefont {Regnault}},\ and\ \bibinfo {author}
  {\bibfnamefont {B.~A.}\ \bibnamefont {Bernevig}},\ }\bibfield  {title}
  {\bibinfo {title} {Moir\'e fractional chern insulators. iii. hartree-fock
  phase diagram, magic angle regime for chern insulator states, role of moir\'e
  potential, and goldstone gaps in rhombohedral graphene superlattices},\
  }\href {https://doi.org/10.1103/PhysRevB.112.075109} {\bibfield  {journal}
  {\bibinfo  {journal} {Phys. Rev. B}\ }\textbf {\bibinfo {volume} {112}},\
  \bibinfo {pages} {075109} (\bibinfo {year} {2025})}\BibitemShut {NoStop}%
\bibitem [{\citenamefont {Wu}\ and\ \citenamefont
  {Das~Sarma}(2020)}]{SI-wu-tbg-collective-prl20}%
  \BibitemOpen
  \bibfield  {author} {\bibinfo {author} {\bibfnamefont {F.}~\bibnamefont
  {Wu}}\ and\ \bibinfo {author} {\bibfnamefont {S.}~\bibnamefont {Das~Sarma}},\
  }\bibfield  {title} {\bibinfo {title} {Collective excitations of quantum
  anomalous hall ferromagnets in twisted bilayer graphene},\ }\href
  {https://doi.org/10.1103/PhysRevLett.124.046403} {\bibfield  {journal}
  {\bibinfo  {journal} {Phys. Rev. Lett.}\ }\textbf {\bibinfo {volume} {124}},\
  \bibinfo {pages} {046403} (\bibinfo {year} {2020})}\BibitemShut {NoStop}%
\bibitem [{\citenamefont {Xie}\ and\ \citenamefont
  {Das~Sarma}(2026)}]{SI-xie-tmd-magnons-2026}%
  \BibitemOpen
  \bibfield  {author} {\bibinfo {author} {\bibfnamefont {M.}~\bibnamefont
  {Xie}}\ and\ \bibinfo {author} {\bibfnamefont {S.}~\bibnamefont
  {Das~Sarma}},\ }\href {https://doi.org/10.48550/arXiv.2603.20370} {\bibinfo
  {title} {Collective spin excitations in correlated moir\'e chern
  ferromagnets}} (\bibinfo {year} {2026}),\ \Eprint
  {https://arxiv.org/abs/2603.20370} {arXiv:2603.20370 [cond-mat.str-el]}
  \BibitemShut {NoStop}%
\bibitem [{\citenamefont {Alavirad}\ and\ \citenamefont
  {Sau}(2020)}]{SI-alavirad-sau-prb20}%
  \BibitemOpen
  \bibfield  {author} {\bibinfo {author} {\bibfnamefont {Y.}~\bibnamefont
  {Alavirad}}\ and\ \bibinfo {author} {\bibfnamefont {J.}~\bibnamefont {Sau}},\
  }\bibfield  {title} {\bibinfo {title} {Ferromagnetism and its stability from
  the one-magnon spectrum in twisted bilayer graphene},\ }\href
  {https://doi.org/10.1103/PhysRevB.102.235123} {\bibfield  {journal} {\bibinfo
   {journal} {Phys. Rev. B}\ }\textbf {\bibinfo {volume} {102}},\ \bibinfo
  {pages} {235123} (\bibinfo {year} {2020})}\BibitemShut {NoStop}%
\bibitem [{\citenamefont {Kang}\ \emph {et~al.}(2026)\citenamefont {Kang},
  \citenamefont {Wang},\ and\ \citenamefont {Vafek}}]{SI-Kang2026OrbitalHF}%
  \BibitemOpen
  \bibfield  {author} {\bibinfo {author} {\bibfnamefont {J.}~\bibnamefont
  {Kang}}, \bibinfo {author} {\bibfnamefont {M.}~\bibnamefont {Wang}},\ and\
  \bibinfo {author} {\bibfnamefont {O.}~\bibnamefont {Vafek}},\ }\bibfield
  {title} {\bibinfo {title} {Orbital magnetization and magnetic susceptibility
  of interacting electrons},\ }\href {https://doi.org/10.1103/9l9c-mp3t}
  {\bibfield  {journal} {\bibinfo  {journal} {Phys. Rev. Lett.}\ }\textbf
  {\bibinfo {volume} {137}},\ \bibinfo {pages} {106703} (\bibinfo {year}
  {2026})}\BibitemShut {NoStop}%
\bibitem [{\citenamefont {Hu}\ \emph {et~al.}(2026)\citenamefont {Hu},
  \citenamefont {Sun},\ and\ \citenamefont {Yao}}]{SI-Hu2026InPlane}%
  \BibitemOpen
  \bibfield  {author} {\bibinfo {author} {\bibfnamefont {J.-X.}\ \bibnamefont
  {Hu}}, \bibinfo {author} {\bibfnamefont {Z.-T.}\ \bibnamefont {Sun}},\ and\
  \bibinfo {author} {\bibfnamefont {Y.}~\bibnamefont {Yao}},\ }\href@noop {}
  {\bibinfo {title} {Theory of in-plane orbital magnetization with layer
  hybridization}} (\bibinfo {year} {2026}),\ \Eprint
  {https://arxiv.org/abs/2606.18891} {arXiv:2606.18891} \BibitemShut {NoStop}%
\bibitem [{\citenamefont {Neupert}\ \emph {et~al.}(2011)\citenamefont
  {Neupert}, \citenamefont {Santos}, \citenamefont {Chamon},\ and\
  \citenamefont {Mudry}}]{SI-neupert-fci-prl11}%
  \BibitemOpen
  \bibfield  {author} {\bibinfo {author} {\bibfnamefont {T.}~\bibnamefont
  {Neupert}}, \bibinfo {author} {\bibfnamefont {L.}~\bibnamefont {Santos}},
  \bibinfo {author} {\bibfnamefont {C.}~\bibnamefont {Chamon}},\ and\ \bibinfo
  {author} {\bibfnamefont {C.}~\bibnamefont {Mudry}},\ }\bibfield  {title}
  {\bibinfo {title} {Fractional quantum hall states at zero magnetic field},\
  }\href {https://doi.org/10.1103/PhysRevLett.106.236804} {\bibfield  {journal}
  {\bibinfo  {journal} {Phys. Rev. Lett.}\ }\textbf {\bibinfo {volume} {106}},\
  \bibinfo {pages} {236804} (\bibinfo {year} {2011})}\BibitemShut {NoStop}%
\bibitem [{\citenamefont {Regnault}\ and\ \citenamefont
  {Bernevig}(2011)}]{SI-regnault-fci-prx11}%
  \BibitemOpen
  \bibfield  {author} {\bibinfo {author} {\bibfnamefont {N.}~\bibnamefont
  {Regnault}}\ and\ \bibinfo {author} {\bibfnamefont {B.~A.}\ \bibnamefont
  {Bernevig}},\ }\bibfield  {title} {\bibinfo {title} {Fractional chern
  insulator},\ }\href {https://doi.org/10.1103/PhysRevX.1.021014} {\bibfield
  {journal} {\bibinfo  {journal} {Phys. Rev. X}\ }\textbf {\bibinfo {volume}
  {1}},\ \bibinfo {pages} {021014} (\bibinfo {year} {2011})}\BibitemShut
  {NoStop}%
\bibitem [{\citenamefont {Xiao}\ \emph {et~al.}(2010)\citenamefont {Xiao},
  \citenamefont {Chang},\ and\ \citenamefont {Niu}}]{SI-xiao_rmp2010}%
  \BibitemOpen
  \bibfield  {author} {\bibinfo {author} {\bibfnamefont {D.}~\bibnamefont
  {Xiao}}, \bibinfo {author} {\bibfnamefont {M.-C.}\ \bibnamefont {Chang}},\
  and\ \bibinfo {author} {\bibfnamefont {Q.}~\bibnamefont {Niu}},\ }\bibfield
  {title} {\bibinfo {title} {Berry phase effects on electronic properties},\
  }\href {https://doi.org/10.1103/RevModPhys.82.1959} {\bibfield  {journal}
  {\bibinfo  {journal} {Rev. Mod. Phys.}\ }\textbf {\bibinfo {volume} {82}},\
  \bibinfo {pages} {1959} (\bibinfo {year} {2010})}\BibitemShut {NoStop}%
\end{thebibliography}
\end{document}